\documentclass{article}
\usepackage{geometry}
\usepackage{array}
\usepackage{amsmath}
\usepackage{amsfonts}
\usepackage{array}
\usepackage{multirow}
\usepackage{graphics}
\usepackage{booktabs}
\usepackage{graphicx}
\usepackage{placeins}
\usepackage{subcaption}
\usepackage[document]{ragged2e}
\usepackage{adjustbox}
\usepackage[font=large,labelfont=bf]{caption}
\usepackage[table,xcdraw]{xcolor}
\usepackage{amsmath}
\usepackage[utf8]{inputenc}
\usepackage[english]{babel}
\usepackage{todonotes}
\usepackage{natbib}
\usepackage{lscape}
\usepackage{float}
\usepackage{setspace}
\usepackage{comment}
\usepackage{diagbox}
\usepackage{hyperref}
\hypersetup{
    colorlinks=true,
    allcolors=blue,
}

\spacing{1.8}
\author{Laura Blattner \\ \small{Stanford University} \\ \and Scott Nelson \\ \small{Chicago Booth}}
\title{How Costly is Noise? \\ Data and Disparities in Consumer Credit\thanks{We thank Mario Curiki, David Du, Pranav Garg, Jacob Hartwig, Xuechen Hong, and Michael Yip for their excellent research assistance on the project. We thank Susan Athey, Stefania Albanesi, Bob Adams, Juliane Begenau, Neil Bhutta, Ken Brevoort, Greg Buchak, Lisa Cook, Will Dobbie, Anders Humlum, Paul Goldsmith-Pinkham, Erik Hurst, Dan Ringo, Claudia Robles-Garcia, Ashesh Rambachan, Amit Seru,  Ken Singleton, Amir Sufi, Jann Spiess, Johannes Stroebel, Paul Willen, Francis Wong, and seminar participants at the University Of Chicago, UC Davis, Dartmouth, the Federal Reserve Board, the Federal Reserve Bank of St. Louis, NBER Household Finance, NBER Economics of AI, Stanford SITE, and Stanford University for helpful discussions and comments, as well as the Becker Friedman Institute and the Fama-Miller center at Chicago Booth for generous support. An earlier version of this paper circulated under the title ``How Costly is Noise? Data and Disparities in the US Mortgage Market." Any errors or omissions are the responsibility of the authors.}}
\begin{document}

\date{\vspace*{0.75cm} May 5, 2021}
\maketitle
\justify

\begin{spacing}{1.1}
	\begin{abstract}
		\noindent
		
		We show that lenders face more uncertainty when assessing default risk of historically under-served groups in US credit markets and that this information disparity is a quantitatively important driver of inefficient and unequal credit market outcomes. We first document that widely used credit scores are statistically noisier indicators of default risk for historically under-served groups. This noise emerges primarily through the explanatory power of the underlying credit report data (e.g., thin credit files), not through issues with model fit (e.g., the inability to include protected class in the scoring model). Estimating a structural model of lending with heterogeneity in information, we quantify the gains from addressing these information disparities for the US mortgage market. We find that equalizing the precision of credit scores can reduce disparities in approval rates and in credit misallocation for disadvantaged groups by approximately half.

		
		\vskip 10 pt

	\end{abstract}
	
	\noindent
\end{spacing}
\vfill 
\pagebreak \newpage


\section{Introduction} \label{sec:intro}

Credit scores are used as a signal of applicant quality in many high-stakes screening decisions, including lending, rental housing, insurance underwriting, and hiring. Reliance on credit scores is becoming even more widespread with the advent of algorithmic decision-making, which automates the screening decision based on data inputs such as a credit score. As credit scores guide allocation in a growing share of the economy, it is crucial to understand how properties of credit scores affect allocative efficiency and who is able to borrow, rent a home, be insured, or obtain a job offer. There is particular concern as to whether credit scores can contribute to disparities in these outcomes across social groups such as minority or low-income applicants. Credit markets, specifically the US mortgage market, present a natural setting to study this question because reliance on commercial credit scores is high, the shift towards algorithmic underwriting is already underway, and resulting disparities can translate into inequality in home ownership and wealth accumulation.

This paper argues that lenders face higher uncertainty when assessing default risk of traditionally under-served groups and that this information disparity is a quantitatively important driver of inequality in credit market outcomes. We first show that widely used credit scores, which are designed to be predictors of default risk, are statistically \emph{noisier} indicators of default risk for traditionally under-served groups.\footnote{Throughout the paper, we will be studying the VantageScore 3.0 credit score.\label{foot:vantagethroughout}} We then show that this noise emerges primarily through the explanatory power of the underlying credit report data (e.g., thin credit files), not through issues with model fit (e.g., regulations that limit the flexibility of credit scoring models for disadvantaged groups), and thus cannot be fixed by more advanced prediction technology. 

We present evidence that the gains from addressing these information disparities can be substantial. Estimating a structural model of the US mortgage market, we use a series of counterfactual exercises to show that equalizing the precision of credit scores can shrink disparities in efficiency and in loan approval rates for disadvantaged groups by approximately 50\%. These estimates provide a stark illustrative example of how promoting disadvantaged groups' ability to \emph{show} their quality as borrowers can advance the goals of efficient and equitable credit access that are central in much of US policy.

New data help make these findings possible. We merge a large panel of consumer credit report data from TransUnion for 50 million anonymized consumers with three complementary datasets: first, public data from property deeds and related mortgage transactions for these consumers; second, a marketing dataset containing socio-economic characteristics of the same consumers; and third, public data on the mortgage lenders who originate loans for these consumers. The marketing dataset is sourced from the firm Infutor, as in \cite{diamond2019effects}. This unique merge across datasets makes possible both an instrumental variables strategy, which we use to identify unobserved default risk of marginally rejected applicants, and our analysis of various consumer groups, such as minority and low-income applicants.\footnote{To our knowledge, neither lender-specific information nor similarly detailed demographic data have previously been merged with credit report data in academic work, with the notable exception of the demographic variables in \cite{avery2009credit} and \cite{avery2012does}. The results in those two papers are consistent with the motivational evidence we build on in this paper.} 

We use these rich data to overcome the key challenge of measuring the default risk of unsuccessful loan applicants; this challenge is similar to assessing unobserved potential outcomes in other settings, such as judges' bail decisions \citep{arnold2018racial} or employers' hiring choices \citep{autor2008does}. We take two broad approaches in overcoming this challenge. In our structural work, we use an instrumental variables strategy based on lender-by-time-by-geography credit supply shocks that allows us to recover the default risk of marginally approved borrowers from various groups of applicants, which, together with other empirical moments, helps us estimate the distribution of unobserved default risk in different applicant populations. We complement this with a second approach in our reduced-form work, where we use consumer credit report data for rejected mortgage applicants and measure default risk using non-mortgage loans' default outcomes (e.g., a rejected mortgage applicant's auto loan). We show that this non-mortgage loan default measure is highly correlated with mortgage default in the sample of approved mortgages.

Starting with these reduced-form results, we find 7\% to 9\% lower fit of widely used credit scores (VantageScore 3.0 credit score) for minority or low-income mortgage applicants, in the sense of how well scores at the time of mortgage application predict later default. Our preferred performance metric is the  Area Under the Curve (AUC) – a simple summary measure of the difference in predictive power exhibited in receiver operating characteristic curves, a widely used tool in machine learning to evaluate model performance. We confirm these results using the R2 metric, which suggests 16 to 18\% lower model fit for the same disadvantaged groups. 

To understand the sources of these precision differences, we next characterize two reasons why statistical risk prediction models can be less informative for different consumers, which we term modeling bias and data bias.\footnote{We note that these frictions do not necessarily introduce mean-nonzero bias in the statistical sense; we nevertheless follow the terminology of the computer science literature in terming such algorithmic frictions as biases.} The first relates to whether prediction models successfully capture heterogeneity across groups in the data generating process for default, while the second relates to differences across groups in the data with which prediction models are estimated. The most obvious example of modeling bias is lower model fit for minority groups due to legal restrictions on the inclusion of protected class information in credit scoring models. The most obvious example of data bias are so-called thin-file consumers who have sparser credit report data. While modeling bias can be resolved by changing how we train scoring models, e.g., fitting separate models by group, data bias cannot be solved simply by improving the predictive models, but rather requires undoing the underlying data property that drives the poorer fit.  

We provide two sets of evidence that the differential informativeness of credit scores for disadvantaged consumer groups stems more from data bias -- that is, the underlying credit report data -- as opposed to modeling bias. First, ``fixing" modeling bias has little effect on precision differences across groups. In particular, neither training separate risk-prediction models for each group nor re-weighting groups' influence in the model's loss function reduces meaningfully the difference in predictive accuracy across groups. This finding holds both for logit models as well as advanced machine learning models that we estimate using hundreds of credit report features. Second, we find evidence in favor of data bias. Differences in the composition of credit report data -- such as sparsity (in the sense of having data from few loans or few years of history), past default history,  and account diversity (in the sense of only holding multiple types of loans) -- explain roughly half of the between-group gap in credit scores' predictive power. The residual gap is concentrated in ``clean" credit files (those without a history of default), which we argue can be explained both by differences in default reporting as well as by between-group differences in the inherent predictability of default.

To complement our reduced-form work, and to quantify the importance of these precision differences in a high-stakes context, we build a structural model of a mortgage lender making loan approval decisions based on various information sources. The goal of the model is threefold. First, the model allows us to quantify differences in credit score precision while accounting flexibly for other differences across groups, both in terms of the precision of other, non-credit-score information sources and in terms of the underlying distribution of unobserved default risk. Second, the model provides a complementary and independent empirical strategy for estimating the precision differences emphasized in our reduced-form work. Third, and most importantly, the model allows us to quantify the importance of these precision differences in economic terms: that is, given the model's estimates of between-group differences in other factors such as the distribution of actual default risk or the precision of other available information sources, how much would improvements in credit score precision for disadvantaged groups reduce disparities in credit allocation?

In the model, a mortgage applicant's true default  risk is unobserved by the lender but the lender receives both a credit score signal and other signals that she uses to screen applicants. We assume normality in both signal noise and in underlying default risk types, which allows us to express the lender's posterior belief (after observing the signals) in tractable terms. 
The lender can observe either signal precision or, equivalently, group membership, knowing how precision differs across groups, and forms her belief about default risk accordingly. We deliberately abstract from modeling pricing decisions or markups from imperfect competition in order to focus on the changes in the information structure which are the emphasis of this paper. 

Model identification relies on an instrumental variables strategy that recovers the characteristics of marginally approved loan applicants, building on a classic \cite{Becker1957} test of marginal borrowers' characteristics. Our instrument exploits Community Reinvestment Act (CRA) exams, following \citet{Agarwal2012}.  The Community Reinvestment Act, passed in 1977, mandates US financial regulators ``to encourage insured depository institutions to help meet credit needs of all segments of their local communities" \citep{FedCRA2018}. Consistent with \citet{Agarwal2012}, we find that banks increase lending, in particular in CRA-eligible locations, just prior to CRA exams.\footnote{We discuss the validity of the instrument and the broader literature on CRA-induced lending extensively in Sections \ref{sec:estimation_results:instrumentconstruction} and \ref{sec:estimation_results:instrumentvalidity}.} This plausibly exogenous variation in approval leniency of mortgage lenders allows us to identify marginal loan applicants and their default outcomes. Intuitively, holding approval rates constant, the difference between the marginal and average default rate tells us how much information lenders obtain through their screening tools -- with a perfectly uninformative screening tool implying no difference between marginal and average default rates, and informative screening tools implying a large difference. Additional data moments include the mean of credit scores of rejected and approved applicants,
approval rates, and the slopes of forward and ``reverse'' regressions relating credit score with default outcomes -- a standard way of assessing measurement error in a regressor \citep{black2000bounding}, which helps identify how much of lenders' total information across all signals is coming from credit scores as opposed to other information sources. 

The structural model yields estimates of credit score noise disparities that are similar to our reduced-form approach. Concretely, we estimate that the standard deviation of credit score noise of minority applicants is 2.2 times higher than that of non-minority applicants. Simulating out default rates and credit scores, these differences translate to a 5\% difference in terms of AUC (cf. 7\% in our reduced-form evidence). Low-income applicants similarly have greater credit score noise than higher-income applicants, which translates into a 10\% difference in terms of AUC (cf. 9\% in our reduced-form evidence). 

Finally, we use the structural model to quantify the importance of these precision differences in economic terms, by way of solving for counterfactual mortgage approval decisions under alternative information structures that reduce or change the precision differences faced by various groups. These exercises hold constant other features of the mortgage market, such as differences in the distribution of underlying default risk across groups, in order to speak precisely to the costs of credit score noise. We find that equalizing credit score signal noise can shrink efficiency differences and disparities in approval rates by up to 50\%. For minority mortgage applicants, nearly half of the counterfactual increase in approval rates is due to a reduction in inefficient rejections -- rejections for borrowers to whom it would be ex ante efficient to lend.

\paragraph{Related literature}
Our paper contributes to a large literature on the causes and consequences of credit market disparities. Much existing research focuses on human frictions such as discrimination, bias, or agency conflict \emph{after} controlling for observables such as credit score \citep{Dobbie2019, Bartlett2019, bhutta2020minorities, heimer2021using, bayer2018drives, zhang2020lenders, Hanson2016, Ross2008, butler2020racial}. We instead study how these observables themselves can drive credit misallocation and disparities. We focus on the mortgage market which plays a prominent role in the persistence of wealth gaps across generations \citep{charleshurstwealth, kuhn2020income}, with historically disadvantaged groups being less likely to transition into home ownership and build home equity \citep{charleshurstReStat}. Our results highlight a mechanism by which a credit scoring system can perpetuate credit misallocation over time: disparities in credit access at one point in time translate into disparities in credit report information in the future, a persistence that might be termed ``credit score hysteresis'' in the tradition of \cite{blanchard1986hysteresis}.\footnote{In the \cite{blanchard1986hysteresis} context, long-run unemployment becomes endemic to a labor market through a hysteresis effect. Other recent work on hysteresis at the intersection of wealth inequality and credit markets includes \cite{mian2020indebted}.}

We also contribute to a literature that studies the role of credit scores for market efficiency and distributional outcomes. Relative to this literature, we identify differences in credit score precision as a key channel through which credit market disparities arise, and we study how these disparities appear not just in credit access but also in credit misallocation. In addition, we show that realizing improvements in these dimensions requires more than changes in statistical technology alone but rather a change in the underlying credit report data. Prior work on credit scoring or statistical technology in credit markets  in contrast has emphasized credit scores' role in overcoming asymmetric information among new borrowers \citep{einav2013impact, adams2009liquidity}, incentivizing loan repayment \citep{CCDR}, and facilitating loan securitization while discouraging lenders' use of soft information \citep{keys2012lender,keys2010}. Recent work has also warned that more flexible statistical technology such as machine learning can reduce overall loan approval rates for disadvantaged groups \citep{Fuster2020}, and that modern, FinTech underwriting continues to generate cross-group disparities in loan terms \citep{Bartlett2019}; credit scores likewise are seen to play a role in geographic misallocation in the US mortgage market \citep{KeysSeru2016}. Much of this work echoes persistent policy concerns about equity across consumers in credit scoring \citep{avery2009credit, avery2012does, traub2013discredited}.

By studying the effects of changes in the information structure in lending markets, we also contribute to a literature that studies how regulation or innovation in information sources can affect different groups' outcomes in markets with screening. An active labor literature studies the consequences of using screening tools such as criminal records or credit reports,\footnote{Examples include \cite{autor2008does, agan2018ban, bartik2020deleting, Doleac2017, Wozniak2015, corbae2018employer}.} and recent work has emphasized the importance of heterogeneity in signal precision in these or related settings especially \citep{arnold2020measuring, chan2020selection,bartik2020deleting}. This draws on a longer tradition of work concerned with signal precision heterogeneity (e.g., \cite{Aigner1977}). Likewise, there is extensive work in finance that studies changes in lenders' ability to use public credit report data, their own private information, or other lenders' assessment of default risk.\footnote{While far from an exhaustive list, examples of such research include \cite{liberman2018equilibrium, nelson2018private, hertzberg2011public}.}

More broadly, we contribute to a rapidly growing literature on Fairness, Accountability, and Transparency (FAT) in computer science.\footnote{For  summaries see e.g. \cite{barocas2020, RambachanLudwig2018,barocas2016, Zafar2017, RambachanLudwig2018, lakkaraju2017selective, KearnsRoth2019, rambachan2019, cowgill2019economics}.} While this literature has taken a largely axiomatic approach to evaluating disparities in algorithmic decision-making based on various fairness metrics, for example equal error rates across demographic subgroups \citep{Zafar2017}, we offer an alternative approach grounded in economic theory.\footnote{We share this goal with \cite{rambachan_economic_2020} who use a principal-agent approach to characterize optimal fair lending policy.} The type of quantification exercises our structural model provides are key for welfare considerations which have largely been absent from the computer science or machine learning literature. A key focus of this literature has been what type of restrictions to impose on algorithms with regard to the use of protected class.  While protected class is typically considered a prohibited input in credit scoring models \citep{bartlett_algorithmic_2020, yang_equal_2019}, recent research has suggested that such restrictions are no longer feasible in a world of complex algorithms \citep{gillis_big_2019, gillis_false_2020} and less disparate outcomes might result from allowing the inclusion of protected class. 
We show that the precision gains from including protected class are limited in practice and that challenges with the underlying data, not restrictions on algorithmic inputs, present the binding constraint for improving allocative efficiency in US mortgage markets.

This paper is organized as follows: Section \ref{sec:data_motivation} presents our data. Section \ref{sec:data_motivation:motivation} presents our reduced-form evidence and Section \ref{sec:sources} develops our analysis of the sources of credit score noise. Section \ref{sec:model} describes our structural model of lender underwriting in the presence of noisy signals, and Section \ref{sec:estimation_results} presents our results from the instrumental variables strategy and then from model estimation. Section \ref{conclusion} concludes.


\section{Data} \label{sec:data_motivation}

Our dataset is based on a sample drawn from a marketing dataset with broad coverage across the US, merged to publicly available data drawn from property records from CoreLogic. These data are then linked with publicly available, bank-specific information for use in our instrumental variables strategy, and, finally, linked with a panel of credit bureau data.

The marketing dataset is acquired from Infutor, as in \cite{diamond2019effects}, \cite{bernstein2019contribution}, \cite{diamond2020effect}, and \cite{qian2020effects}. The data include a near-universe of address histories and socioeconomic information that allow us to identify the various groups studied in our analysis. We probabilistically sampled from this near-universe, allowing us both to recover a representative sample for the population and to ensure sufficient sample size in several sub-populations of interest.\footnote{We probabilistically oversampled several disadvantaged groups and geographies in order to ensure sufficient power. 
} Based on name and address, we merge these data with publicly available property-specific data for US homeowners, sourced from CoreLogic. The latter data include property-specific characteristics including the history of home prices at each sale date, and also mortgage-relevant characteristics, including the amount, date, and lender for each mortgage loan, and information on whether the loan was a refinance or purchase loan.

Based on lender names and property location, these marketing data and property-specific data are then linked with publicly available, bank-specific data for 6,954 US mortgage lenders. 
This merge allows us to include several potential shifters of bank loan supply in our analysis, including details on the timing of bank regulators' examinations under the Community Reinvestment Act (CRA), which we detail more below. We find that the idiosyncratic timing of these exams shifts bank loan supply, as also evidenced in \citet{Agarwal2012}, which we exploit in an instrumental variables strategy to identify the characteristics of marginally rejected loan applicants. Our sample includes  2,324  banks that undergo a CRA exam in our sample period. Lenders have an average of 1.2 CRA exams in our sample period with a maximum of four exams.

Finally, our credit report data from TransUnion are available for the period between 2009-2017. TransUnion performed a merge between the Infutor data and the credit report data based on social security numbers, first and last names, and the most recent address. TransUnion returned only the de-identified data. The credit report data allow us to construct mortgage applications, mortgage approvals, as well as loan performance in the 24 months following the mortgage application.  We infer applications from a hard credit inquiry by a mortgage lender \citep{avery2003overview}. We infer a rejection in any instance where we see one or more applications for a new mortgage and no new mortgage origination in the subsequent 3 quarters.\footnote{Our preferred horizon of 3 quarters is based on advice from credit risk model building teams at large financial institutions.} We show robustness to our choice of 3 quarters in Table \ref{table_orig_horizon} in Appendix \ref{appendix_more}.\footnote{Beyond those included in this section, additional details about data construction are reported in Appendix \ref{appendix_datamerge}.}

The credit report data usefully allow us to observe rejected loan applications, and to observe the performance of rejected applicants on their other, non-mortgage loans in order to learn about their underlying default risk. The availability of credit scores (VantageScore 3.0 credit score) in the credit report data also allows us to measure how the mean, variance, and default covariance of credit scores differ across accepted and rejected applicants, which we use to help discipline our model estimates about the amount of statistical noise in credit scores relative to other information sources used in approval decisions.

Table \ref{table_summarystats} shows summary statistics for our sample of \emph{originated mortgages} across the data merges. The first column reflects our primary sample, individuals for whom we successfully link credit bureau data with Infutor. Subsequent columns reflect increasingly restrictive subsamples conditioned on various merges, including our initial merge with CoreLogic (column 2), screens for high-quality matches between CoreLogic and the credit bureau data (columns 3 and 4), and a merge with mortgage lender information based on bank names (column 5). While merge rates are considerably below 100\%, given data limitations and the need for merges on inexact name strings, we are reassured that the sample summary statistics are quite comparable across columns. For example, the share of minority mortgage holders ranges from 10\% to 12\% across columns, the share of conventional mortgages ranges from 68\% to 74\%, and mortgage amounts stay within a \$7,000 range, relative to a mean of slightly over \$200,000.\footnote{Table \ref{table_summarystats_applicants} in Appendix \ref{appendix_more} shows summary statistics for our sample of \emph{mortgage applicants}. This sample only relies on the successful merge with Infutor which we require to estimate minority status. The first column shows all initial matches with Infutor, and the second column shows a subsample conditioned on various screens for high-quality matches, such as having a current address (in Infutor). As in the origination sample, the distribution of observables remains stable across the two samples. The latter sample is a superset of the sample in column (1) of Table \ref{table_summarystats}, which further conditions on observing an originated mortgage.}

We classify a mortgage origination as a refinance loan if the borrower has at least one open mortgage prior to the application date; the Infutor address indicates that the borrower does not move in a 1-year window around the origination;\footnote{Our Infutor data end in 2017. Hence for transactions in late 2016, we classify them as a refinance if there is no move before and on the date of the origination.} and the number of open mortgage does not increase (to rule out second homes). A closely similar approach is followed in \cite{mian2018credit}. We validate the refinance flag using data both from CoreLogic and HMDA. See Appendix \ref{appendix_refi} for details. 

We define two indicators of economic disadvantage: low income, and racial or ethnic minority. To classify income status, we use a measure of estimated income based on a proprietary algorithm provided by the credit bureau. The algorithm was developed based on income reported on tax filings between 2008 to 2012 and estimates income  to the nearest thousands. We define low-income applicants as those that fall in the bottom quartile of the income distribution in our sample (cut-off is defined year-by-year).  Second, we infer minority/non-minority status using a standard Bayesian Improved Surname Geocoding (BISG) approach, which uses name and geographic information to predict race and ethnicity (see Appendix \ref{appendix_race} for details).\footnote{We use a two-step procedure to assign race and ethnicity in the marketing dataset that provides names and addresses, which we merge to both the CoreLogic and credit bureau data. First, we follow an approach developed by \cite{sood2018}, which uses Voter Registration data to predict the relationship between the sequence of characters in a name and race and ethnicity. After applying their approach, we update each individual’s baseline racial/ethnic probabilities with the racial and ethnic characteristics of the census block associated with her or his place of residence in 2000 using Bayes’ Rule, and we then update the Bayesian posterior again using an individual's 2010 address and the 2010 census data. An individual is assigned to a racial/ethnic category if this category has the highest posterior probability for that individual. This two-step method is similar to methods used by the CFPB to construct race and ethnicity in fair lending analysis. Work by the \cite{CFPB2014ethn} and \cite{census2009} shows that combining geographic and name-based information outperforms methods using either of these sources of information alone. } We aggregate all ethnic and racial minorities into a single minority category and define its complement as the non-minority group.\footnote{Table \ref{tab:minority_income_crosstab} shows the cross-tab between minority and income status.} Since both our measures of economic disadvantage rely on an imputation strategy, we conduct extensive validation in Appendix \ref{appendix_hmda_income}. Our validation relies on matching originated mortgages in our sample to HMDA, which contains the borrower's self-reported income and race/ethnicity. The match is performed on mortgage amount, geography, year, and lender name. We successfully match 16m purchase loans and 24m refinance loans. Given data limitations of the publicly available HMDA data, we cannot perform the merge for loan applications.  We find that both our low-income and minority indicators are highly correlated with the corresponding indicators in HMDA.


\section{Quantifying Noise in Credit Scores} \label{sec:data_motivation:motivation}

This section presents our reduced-form work quantifying noise in credit scores. The key challenge we address in this section is unobserved potential default outcomes for rejected applicants --  also known as \textit{reject inference} in credit scoring.\footnote{See \cite{siddiqi2017intelligent} for an extended discussion. Reject inference is also analogous to what \cite{lakkaraju2017selective} call the ``selective labels problem" and more broadly to the classic econometric issue of unobserved potential outcomes. See also \cite{arnold2020algodiscrim} drawing the analogy between selective labels and unobserved potential outcomes, and \cite{rambachan2019} for a discussion of how selection into training data can have subtle effects on the predictive power of algorithms trained on those data.} A frequent solution is to simply drop rejected applicants and compute credit score performance in the sample of approved applicants. However, differences in credit score performance for approved applicants may not reflect the differences in the overall applicant sample, as loans are not randomly approved.  We propose two different solutions to the reject inference challenge: first, a reduced-form approach described below, where we measure unobserved default risk using outcomes on other credit products, and second, a structural model described in Section \ref{sec:model}, where we identify unobserved risk based on moments from approved, rejected, and marginal applicants.  

\subsection{Measuring Risk Using Non-Mortgage Loan Products}

We infer outcomes for rejected applicants using delinquency on any non-mortgage product 24 months after the application date, which is available for nearly all the mortgage rejects. This approach 
exploits the fact that we can follow the credit report of both rejected and approved applicants over time.\footnote{This is a type of reject inference by extrapolation. (See also \cite{coston_characterizing_2021} for an application of this type of reject inference to credit data.) Unlike most applications of this type of reject inference, we leverage that we observe other applicant credit behavior after the application regardless of the outcome of the application.}

Table \ref{table_default_summary} shows default rates across different loan products for the sample of applicants by loan approval status and loan type. More than 90\% of applicants in our sample have at least one loan product on their credit report.  A similarly large fraction have a credit card, around 50\% of applicants have an auto loan,  and around 10\% have a home-equity loan. We exclude information on student and personal loans due to data limitations.\footnote{We conduct robustness tests using an all-encompassing any-default variable provided by TransUnion that includes all loan products reported to the credit bureau. Our results are very similar when using this measure.} As expected, rejected applicants have higher default rates across products.

We construct a summary default measure on non-mortgage products from information on credit cards, auto loans, and HELOCs (hereafter ``non-mortgage default" measure). Table \ref{tab:default_confusionmatrix} shows how well the non-mortgage default measure predicts mortgage default in the sample of approved mortgage applicants. The table shows confusion matrices as well as precision and recall as we vary the severity of the non-mortgage measure.\footnote{Precision in this context refers to the ratio of correct positive predictions to the total \textit{predicted} positives. (This is distinct from precision in the sense of the inverse variance of signal noise, as we use in Sections \ref{sec:model} and \ref{sec:estimation_results}.) Recall refers to the  ratio of correct positive predictions to the total positive cases. A positive is a default and a true positive is a correctly predicted default.} Our strategy has to balance precision and recall.  With the most encompassing default definition, which includes a one-off missed credit card payment, we maximize recall but will sacrifice precision, as missing a card payment once is unlikely to be a good predictor of mortgage default.  As we move to more stringent default definitions (e.g., requiring delinquency of at least 60 days), we improve precision but sacrifice some recall. Our preferred measure is the non-mortgage default measure of 90 days or more past-due. This measure also corresponds to the default measure typically used in the construction of consumer credit scores \citep{board2007report}. In Appendix \ref{appendix_default}, we provide further information on these default measures. 

We use the non-mortgage default outcomes to construct receiver operating characteristic (ROC) curves, which are a popular tool in machine learning to assess model fit. ROC curves show the ability of a prediction algorithm, such as a credit score, to predict non-defaults separately from defaults as the (credit score) approval threshold is varied. An efficient predictor maximizes the fraction of non-defaults approved (referred to as a true positive rate) and minimizes the number of defaults mistakenly approved (referred to as a false positive rate). The predictive performance visualized in a ROC curve is summarized in the Area Under the Curve (AUC) measure. An alternative approach relies on the (observable) relationship between mortgage and non-mortgage default to directly compute the mortgage ROC and leads to a similar conclusion (see Appendix \ref{appendix_default} for details).

\subsection{Noise in Credit Scores: Results} \label{sec:noiseresults}

\begin{figure}[H]
	\caption{Receiver Operating Characteristic Curves by Consumer Type \label{rocgraph}}
	\centering
	\begin{subfigure}[t]{0.45\textwidth}
		\includegraphics[width=1\textwidth] {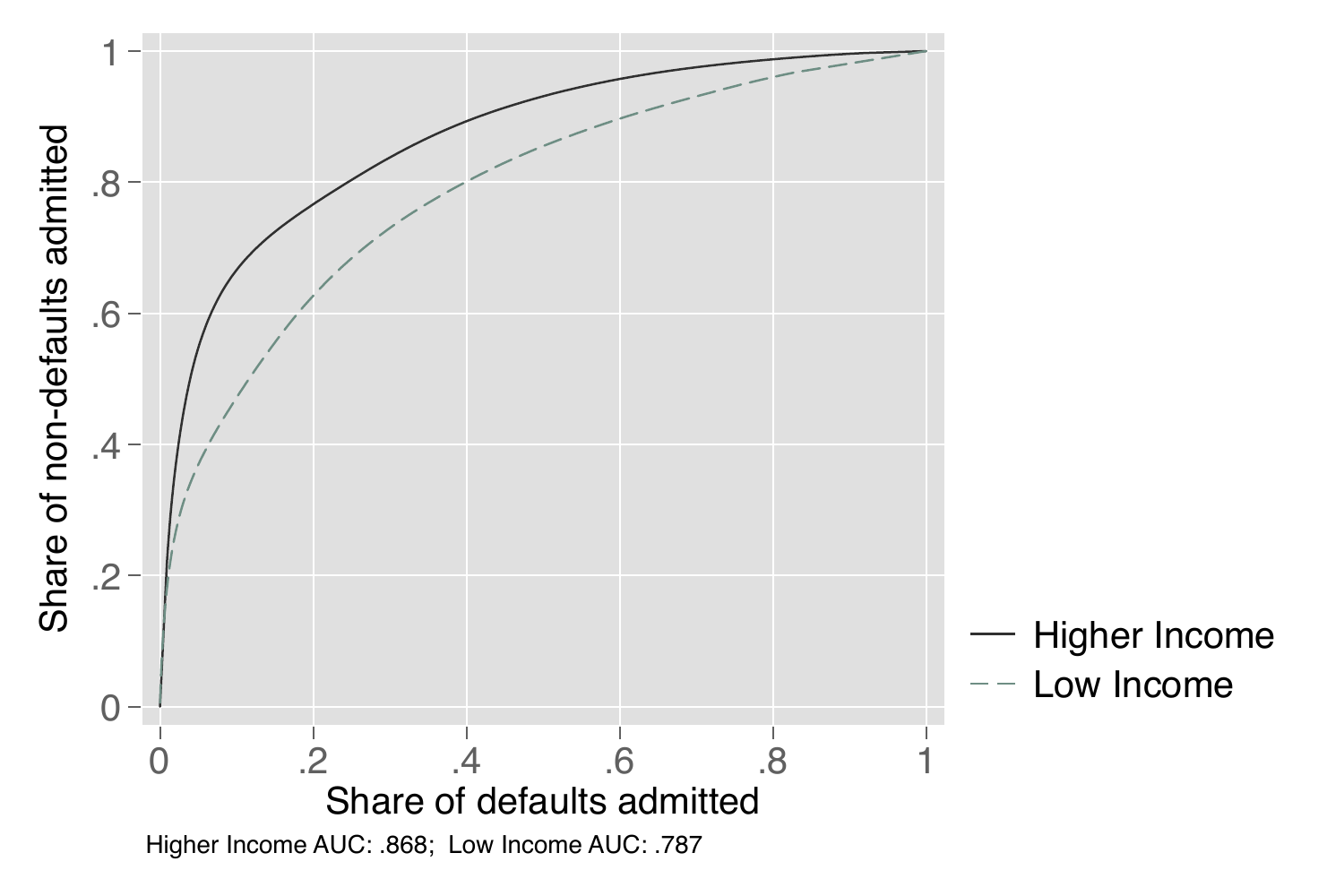}\caption{Non-Mortgage Default: Income}
	\end{subfigure} 
	\begin{subfigure}[t]{0.45\textwidth}
		\includegraphics[width=\textwidth]{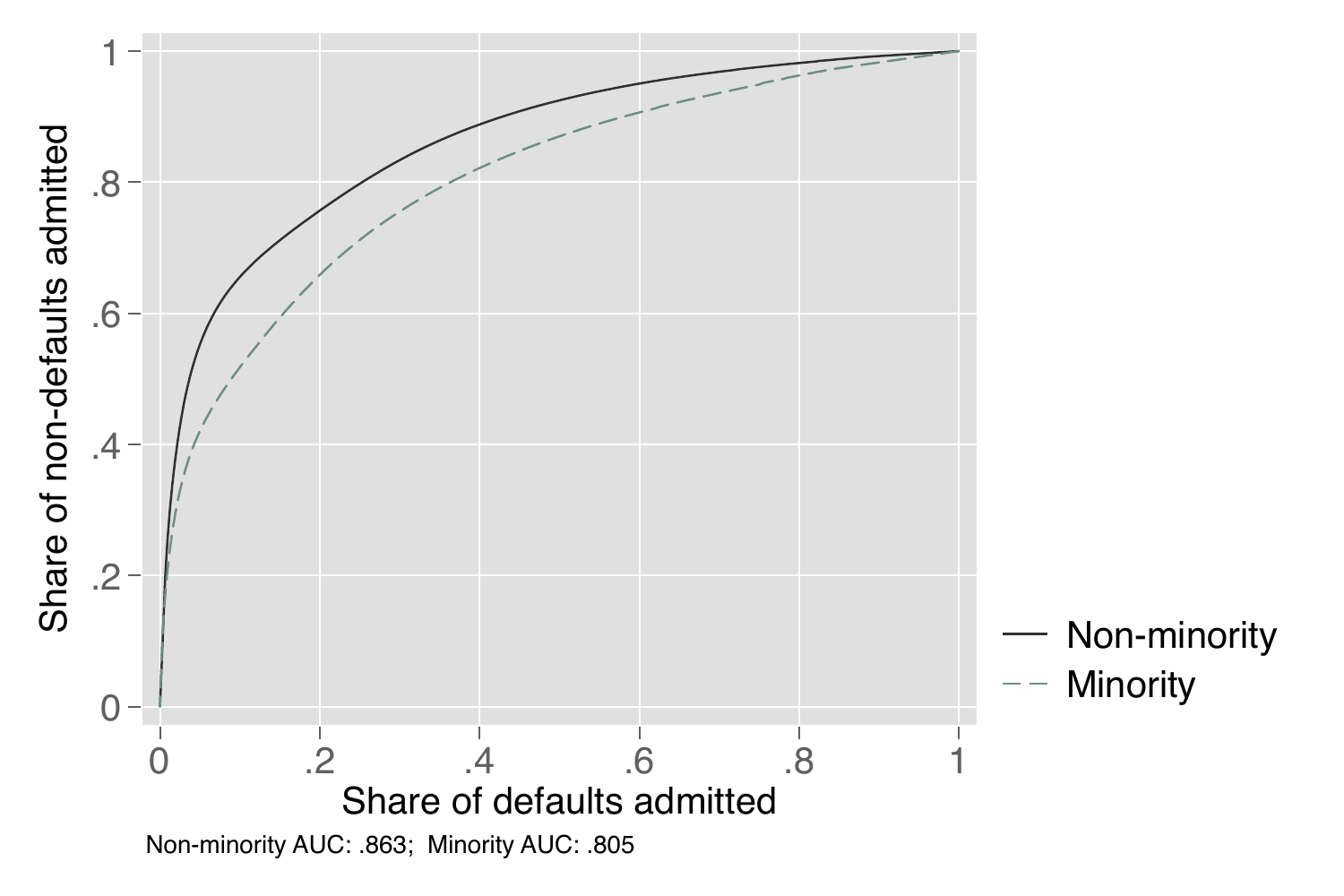}\caption{Non-Mortgage Default: Minority }
	\end{subfigure}\\
	\begin{flushleft}
		\begin{spacing}{1.1}
			\small{Note: This figure shows Receiver Operating Characteristic (ROC) curves for the VantageScore 3.0 credit score in a sample of mortgage applicants between 2009-2016. ROC curves plot the fraction of non-defaults (True Positive Rate) admitted for a given score cutoff against the fraction of defaults admitted (False Positive Rate). The Area Under the Curve (AUC) is the integral below each curve. Higher AUC, and ROC curves that are farther to the upper-left of the plot, both indicate higher predictive power. The measure of default is delinquency of at least 90+ days on any non-mortgage product 24 months after the application date. The group classifications are described in the main text. }
		\end{spacing}
	\end{flushleft}
\end{figure}

We show that the VantageScore 3.0 credit score has lower predictive accuracy -- based on AUC and R2 -- for low-income and minority groups among our sample of mortgage applicants. 
The ROC curves in Figure \ref{rocgraph} show this result visually, with minority and low-income consumers' ROC curves falling markedly below ROC curves for non-minority and higher income consumers. Table \ref{table_auc_summary_mc} tabulates the AUC and R2 metrics corresponding to the ROC curves in Figure \ref{rocgraph} as well as several alternatives.\footnote{We show results for the full sample of applicants. We obtain essentially identical results when computing predictive accuracy in a sample where both groups are equally represented to ensure that sample sizes are not affecting our performance measures.} According to the Area Under the Curve (AUC) measure, which summarizes the differential model fit across all possible credit scores, model fit is 9\% lower for the lowest income quartile relative to higher-income consumers.   The credit score at time of mortgage application has a  16\% lower R2 for default for borrowers in the lowest income quartile relative to applicants in the other three income quartiles.  For minorities, the R2 difference is 18\% while the AUC difference is 7\%. The R2 is obtained by first computing the average log odds of default in each one-point score bin and then regressing the log odds on the credit score, weighting by the number of observations in each bin.\footnote{An alternative specification would be to run a logistic regression of default on the credit score at the individual level. Given the size of our data, this aggregation approach is computationally easier. When we compare our results to R2 obtained at the individual level in a random sample of our data, we obtain very similar results on between-group gaps.}

We provide a second type of visual evidence for our finding of differential informativeness. In Figure \ref{attenuation} in Appendix \ref{appendix_more}, we plot the relationship between default and credit score separately for various consumer types. We show that this relationship is flatter for the consumer types we found to have lower R2 and AUCs, consistent with higher noise -- that is, higher measurement error -- in scores attenuating the relationship towards zero.\footnote{To understand this result, consider regressing the log odds ratio of default on a  signal $a$  of default risk (credit score) for each group $g$, as $\log(\text{pd}_g/(1-\text{pd}_g)) = \alpha + \beta \text{a}_g + \epsilon_g$, where $\text{pd}_g$ is probability of default for individuals in group $g$, conditional on a realization of the signal or score $a$. Assume the signal is measured with noise $\text{a}_g = \theta_g + v_g$. If the signal has more measurement error for a consumer group, classical errors-in-variable theory indicates that the coefficient for that group is more heavily attenuated towards zero. Table \ref{attenuationtable} in Appendix \ref{appendix_more} provides the point estimates for this specification.}

Our findings have important implications for how to interpret prior evidence that commercial credit scores are not calibrated across groups (e.g., \cite{avery1996}); that is, the probability of default conditional on a given credit score is not identical across groups. Figure \ref{attenuation} in Appendix \ref{appendix_more} reveals a similar result in our sample. This miscalibration is often interpreted as driven by some sort of ``bias," whereby the credit score signal is not centered around a consumer's true default risk but around the true default risk plus some bias term. However,  simulations in Appendix \ref{appendix_more} Figure \ref{bias_noise_sim} show that the patterns observed in the data arise even if credit scores feature \emph{only} noise (but no bias), as long as the underlying true default processes are different. The simulations also show that any potential bias would have little impact on AUC differences across groups, since AUC focuses on the ability of an algorithm to correctly \emph{rank} consumers according to risk. In other words, an algorithm can correctly rank applicants even if the point estimate of the probability of default is on average too high or too low.


\section{Sources of Noise} \label{sec:sources}

In this section, we investigate the possible sources of noise. We propose two potential sources of noise, explain how to test for sources of noise in the data, and present results.

\subsection{Framework} \label{sec:sources:framework}

We propose two potential sources of noise: \textit{modeling bias} and \textit{data bias}.\footnote{As emphasized in the introduction, these frictions do not necessarily introduce mean-nonzero bias in the statistical sense; we nevertheless follow the terminology of the computer science literature in terming such frictions as biases.}  We further sub-divide modeling bias into aggregation bias and majority bias. Recall that a credit score is a mapping from observables $X$ (e.g., credit report data) into a predicted default outcome $\hat{y}$ such that $\hat{y} = f(X)$, where $f(\cdot)$ can be viewed as a conditional expectation function. 

\textit{Aggregation bias} arises when the true conditional expectation function for two groups differs but the algorithm is prevented from fitting separate models for each group.\footnote{The terminology is based on \cite{sureshgattag2020}.} For example, the Equal Credit Opportunity Act's prohibitions against ``disparate treatment" forbid using data on protected class membership (for example, minority status) in credit risk model building. Similarly, income data are not used in credit scoring due to regulatory restrictions.\footnote{Income is typically evaluated in a separate ability-to-pay exercise that considers debt-to-income constraints. However, it is not used directly in the credit score, given concerns about disparate impact arising through income's correlation with membership in various protected classes \citep{Mays2004}.} Aggregation bias can be resolved by either fitting separate models for each group, or allowing the algorithm access to the group membership variable and sufficient flexibility to effectively fit the differences in $f(\cdot)$.

Second, \textit{majority bias} compounds the aggregation bias problem since a predictive model will be able to achieve higher precision when given more data. The fact that there are likely to be fewer minority and low-income individuals in the training data would imply that the models are inherently less able to achieve the same precision for these groups. Majority bias should disappear if the training data is re-weighted to represent both groups equally.\footnote{Some recent papers in machine learning refer to majority/aggregation biases as a type of data bias, e.g. \cite{BlumStangel2020}.}

Finally, we use \textit{data bias} to refer to sources of noise that come from the underlying data as opposed to the mapping from the data into predicted default outcomes. There are three forces that can drive data bias in our setting: compositional differences; differences in how default is or is not reported to credit bureaus for different groups; and differences in outcome predictability. Compositional differences matter if some types of credit report data are less informative about future default and disadvantaged groups are more likely to have these types of credit report data. Examples are sparse data (e.g., thin or young credit files), ``dirty" files  (e.g., showing past default), or non-diverse files (e.g., only holding one type of loan). Differences in default reporting can lead to a mismatch between the true outcome we would like to observe -- did a borrower miss a payment -- and the outcome we observe in the data -- is a missed payment reported to the credit bureau. This can be driven by, for example, differential use of lenders who do not report to traditional credit bureaus. Finally, available observables might just be less informative for predicting default of disadvantaged groups, for example because the  shock process that drives default could be more volatile for minority or low-income applicants.

\subsection{Empirical Approach and Results}  \label{sec:sources:approach}

In order to test these hypotheses on the sources of noise, we proceed in two steps. First, we build our own set of default prediction models to test whether we can reduce precision differences across groups through two approaches: training group-specific models to avoid aggregation bias, or re-weighting training data to avoid majority bias. A significant reduction in the precision gap across groups when using these approaches would point to modeling bias being a primary source of noise. Second, we investigate differences in the underlying credit report data between groups. In particular, we investigate whether there are significant differences in the composition of credit report data, and whether any differences remain once we condition on similar credit report observables. 

\subsubsection{Modeling Bias} \label{sec:modelbias}

We build a suite of default prediction models based on a logit, a random forest, and a XGBoost algorithm. We train the models on a sample of 364,979 mortgage applicants. Our baseline models are trained on a pooled sample that has 22.9\% minority applicants and 28.3\% low-income applicants. We use the same non-mortgage default outcome as in our reduced-form reject inference exercise in Section \ref{sec:data_motivation:motivation}. We select 219 credit report variables that include information on delinquencies across all credit products, evolution of the consumer's debt balance over time, public records related to debt collection (for example, a court judgment), credit availability, credit file ``thickness" (for example, the number of accounts or items on the report), applications for new credit, as well as a range of mortgage-specific variables. We deliberately do not include existing credit scores in our set of predictors, in order to focus on the explanatory power of the underlying credit report data itself. We use this full set of features for our machine learning models, whereas for the logit model, we hand-select 35 features. All models are estimated using the Python skicit-learn library.\footnote{We tune the models using a random grid search using 5-fold cross-validation. Table  \ref{table_ml_params_mc} in Appendix \ref{appendix_more} reports search grids and chosen hyperparameters for each model.}

\begin{figure}[H]
	\caption{Plot of VantageScore 3.0 Credit Score and Predicted Default (XGBoost) \label{correlation_xgboost_mc}}
	\centering
	\includegraphics[width=.8\textwidth]{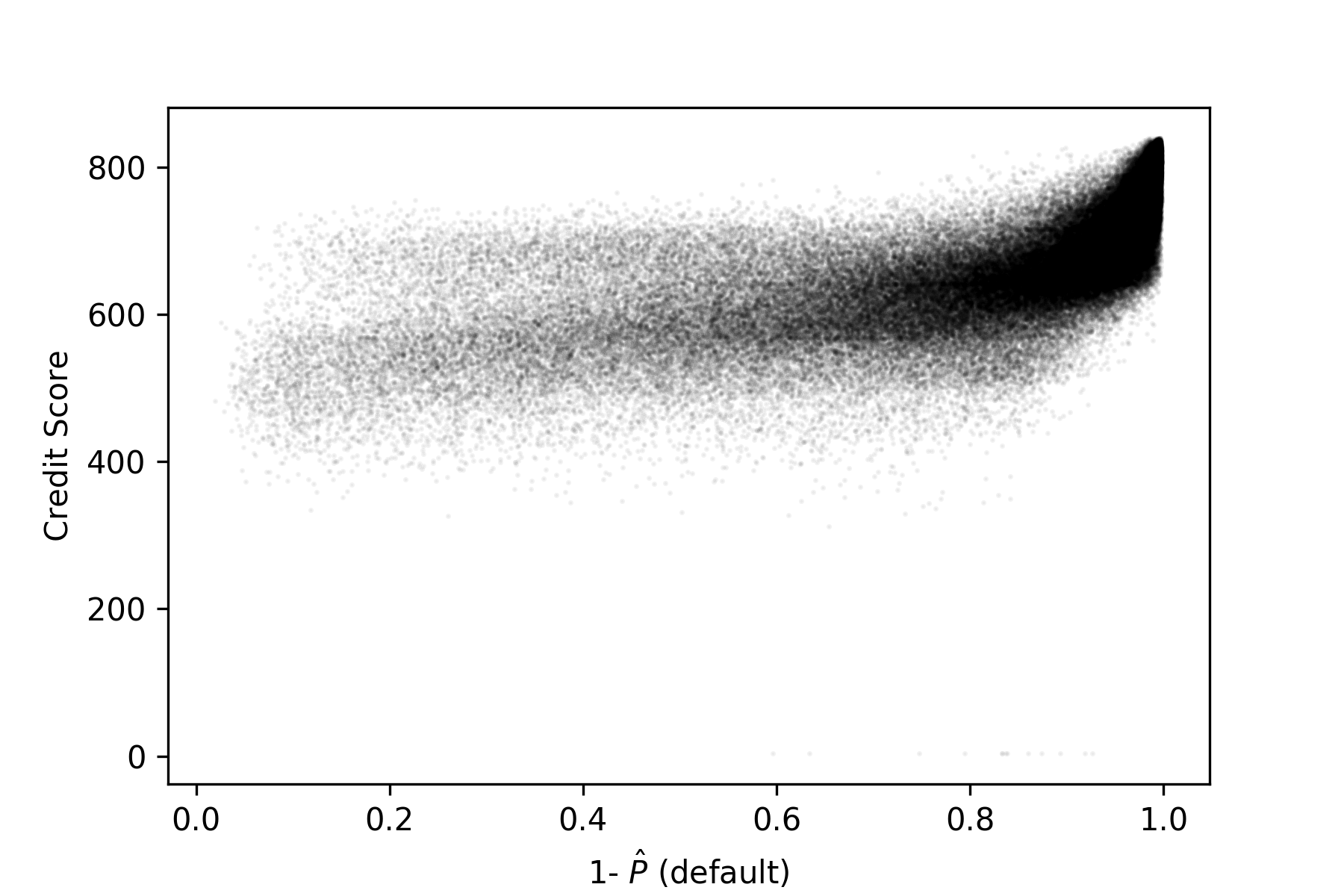}
	\begin{flushleft}
		\begin{spacing}{1.1}
			\small{Note: This figure shows the correlation of the VantageScore 3.0 credit score and our predicted default probabilities of the XGBoost machine learning model in the test data set of mortgage applicants. Our measure of default is a non-mortgage deliquency of at least 90+ days 24 months after the application date -- the same as in Figure \ref{rocgraph}. See text for more details on the test data set.  }
		\end{spacing}
	\end{flushleft}
\end{figure}

Table \ref{table_ml_applicant_mc} shows results from the estimated models. Our baseline models attain predictive performance that is comparable to the VantageScore 3.0 credit score, giving us confidence that we have built reliable and representative risk prediction models.\footnote{All performance statistics are calculated on the test (or holdout) data set.} Moreover, Figure \ref{correlation_xgboost_mc} shows the correlation between commercially available credit scores (Vantage Score 3.0) and predicted default probabilities from our XGBoost model. The correlation coefficient is 0.760, suggesting that our models generate output similar to commercially available models.

Our models also have lower precision for the disadvantaged groups we study, suggesting that our models are subject to the same challenges that drive noise  we illustrate for commercially available credit scores. In particular, column-wise comparisons in Table \ref{table_ml_applicant_mc} indicate overall lower model performance for low-income and minority applicants.

We then use our prediction models to test the effect of fitting group-specific models (addressing potential aggregation bias) and re-weighting the training data (addressing potential majority bias). Table \ref{table_ml_applicant_mc} shows the results. Across performance metrics (in table panels), sampling or re-weighting strategies (in table sub-panels), and models (in table rows), we find that allowing credit scoring models to fit more flexibly has virtually no effect on precision differentials. In particular, the \textit{Different Models} sub-panels, where we fit group-specific models, show similar precision differentials as the \textit{Baseline} sub-panels, both for a random forest (RF) model, an XGBoost model, and a 35-feature logit model. These provide evidence against aggregation bias playing a meaningful role in generating credit score precision disparities.\footnote{The changes in model performance metrics that we calculate from fitting group-specific models are similar to analogous estimates in \cite{Fuster2020}; we find these changes are small relative to the the effects of data bias, as shown in the following section.} Turning to majority bias, we similarly see in the \textit{Re-weight Training Data} sub-panels that by re-weighting, which equalizes sample size across groups to undo majority bias, we again produce similar precision differentials as in the \textit{Baseline} sub-panels. These findings suggest that majority bias is unlikely to account for the credit score precision differentials observed in the data.\footnote{One potential concern is that flexible machine-learning models were already able to re-construct group status from the data and effectively fit the differences in the prediction function. However, we also find no precision gains for a 35-feature logit model, as shown in the Logit rows of the table.} 

To understand this null result, it is helpful to consider when we would expect training group-specific models  to have the largest benefit. \cite{wang_split_2020} show theoretically that group-specific models (``model-splitting") yield the largest benefit if both groups have similar joint distribution of predictors but different conditional expectation function (see the top left panel of Figure \ref{model_bias_intuition} in Appendix \ref{appendix_more}). Intuitively, if the two groups have very distinct clustering of predictors, then a sufficiently complex algorithm would likely already have reconstructed group membership (see the bottom right panel of Figure \ref{model_bias_intuition} in Appendix \ref{appendix_more}). Conversely, there is only a benefit to model-splitting if the conditional expectation function differs across groups (see the top right panel of Figure \ref{model_bias_intuition} in Appendix \ref{appendix_more}). 

Our finding that model-splitting does not result in precision gains can then be explained either by (i) significant differences in the joint distribution of predictors across groups which are already captured by existing algorithms, or (ii) similar conditional expectation functions across group which make model-splitting unnecessary. There are several reasons why the first explanation is unlikely to apply. If already-reconstructed differences  were the main driver of our results, we would expect more complex algorithms, which are presumably better at re-constructing group membership, to feature smaller precision differentials and to have \emph{smaller} precision gains for the disadvantaged groups when training separate models. In Table \ref{tab:predict_group} in Appendix \ref{appendix_more}, we show that the machine learning algorithms are in fact better than a logit model at predicting minority and low-income status. However, Table \ref{table_ml_applicant_mc} shows that a simple 20-variable logit model has similar precision differences across groups and does not have larger precision gains for disadvantaged groups as we fit separate models. This explanation is also inconsistent with existing work that has found that the variables used in typical credit scoring models do \emph{not} serve as proxies for race and ethnicity \citep{avery2012does}.

\begin{figure}[h]
	\caption{SHAP Values for XGBoost Models by Consumer Type \label{fig:shap_xgboost_groups_mc}}
	\centering
	\begin{subfigure}[t]{0.45\textwidth}
		\includegraphics[width=1\textwidth] {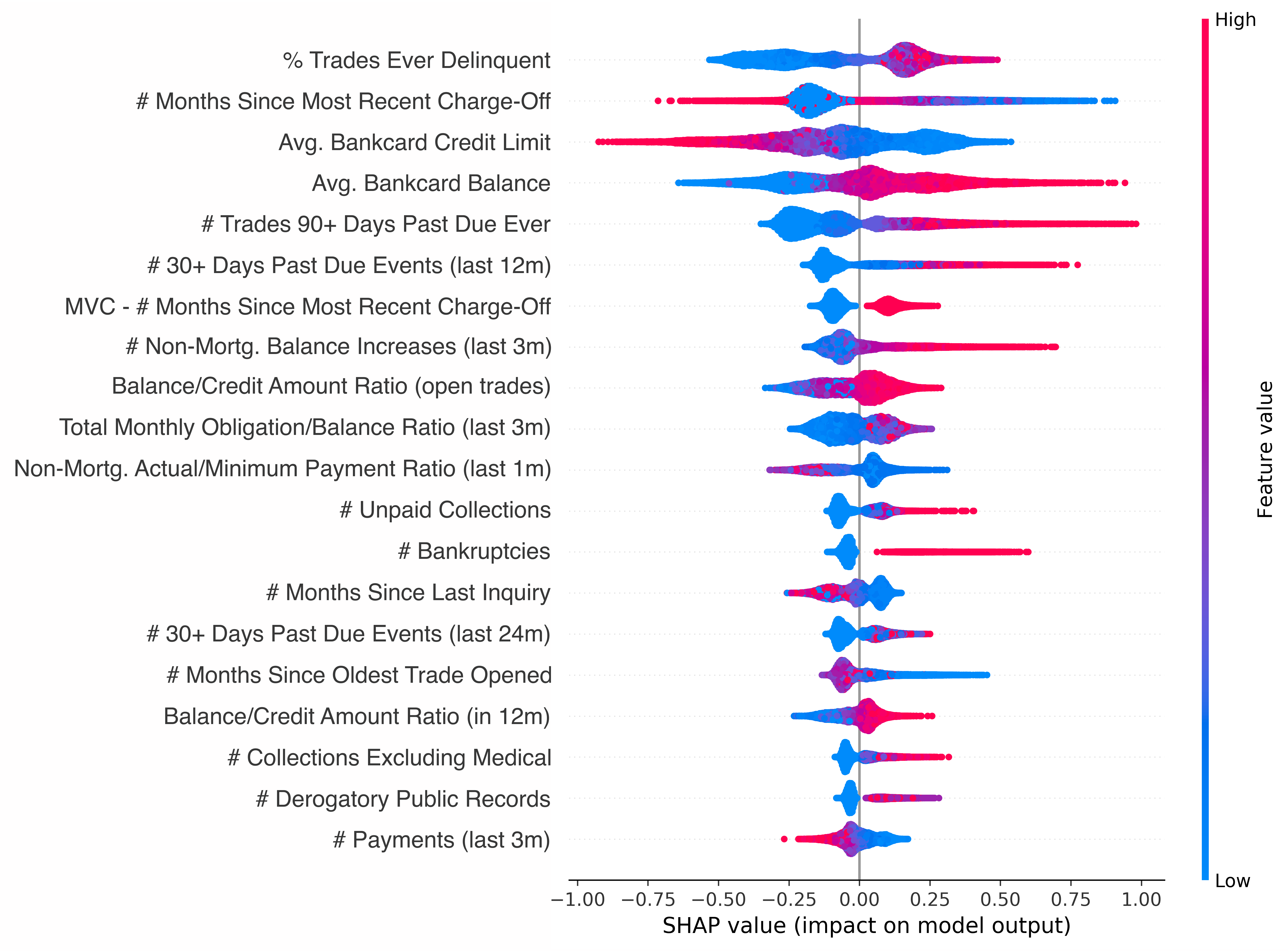}\caption{Minority}
	\end{subfigure} 
	\begin{subfigure}[t]{0.45\textwidth}
		\includegraphics[width=\textwidth]{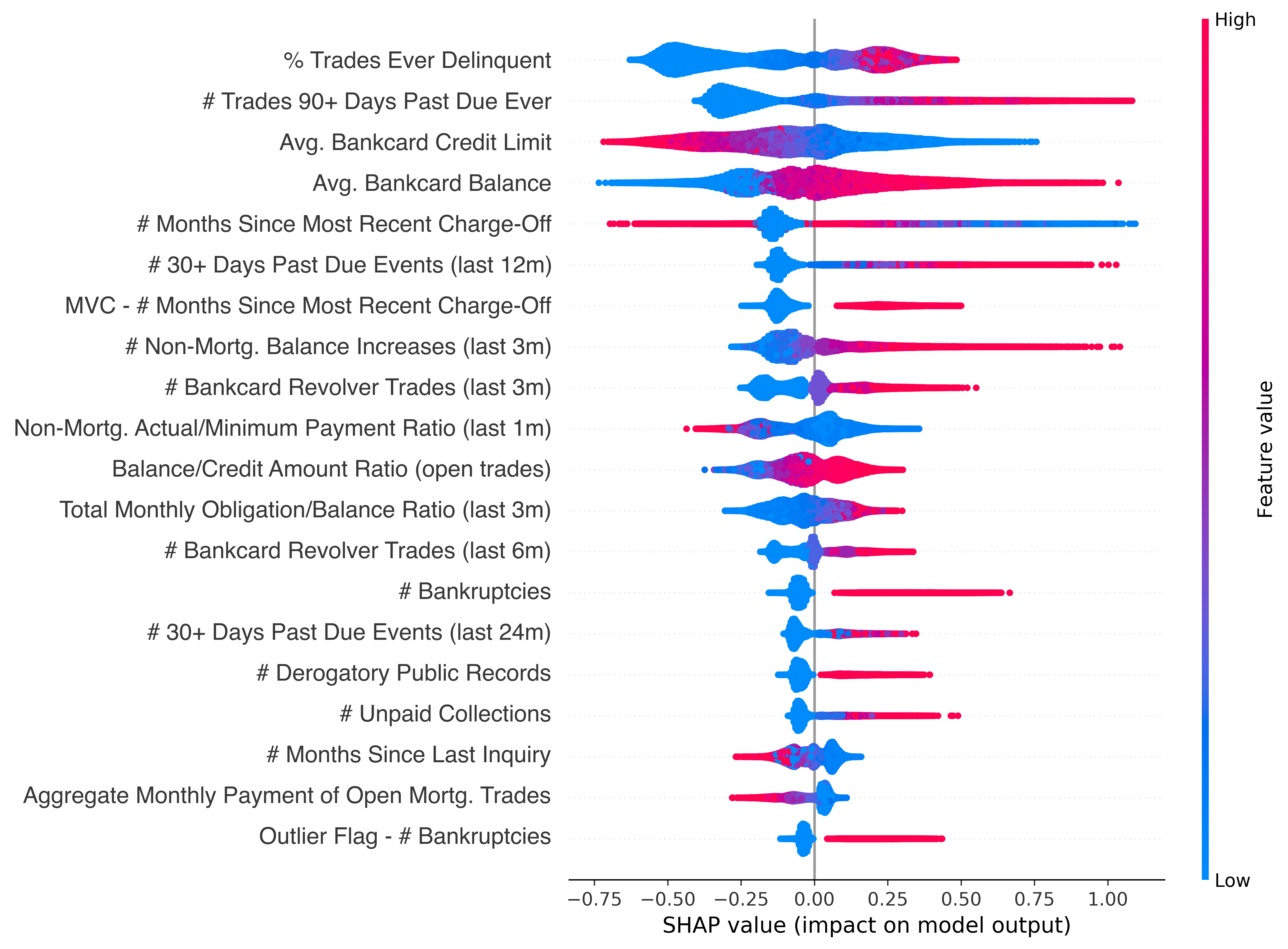}\caption{Non-minority}
	\end{subfigure}\\
	\begin{subfigure}[t]{0.45\textwidth}
		\includegraphics[width=1\textwidth] {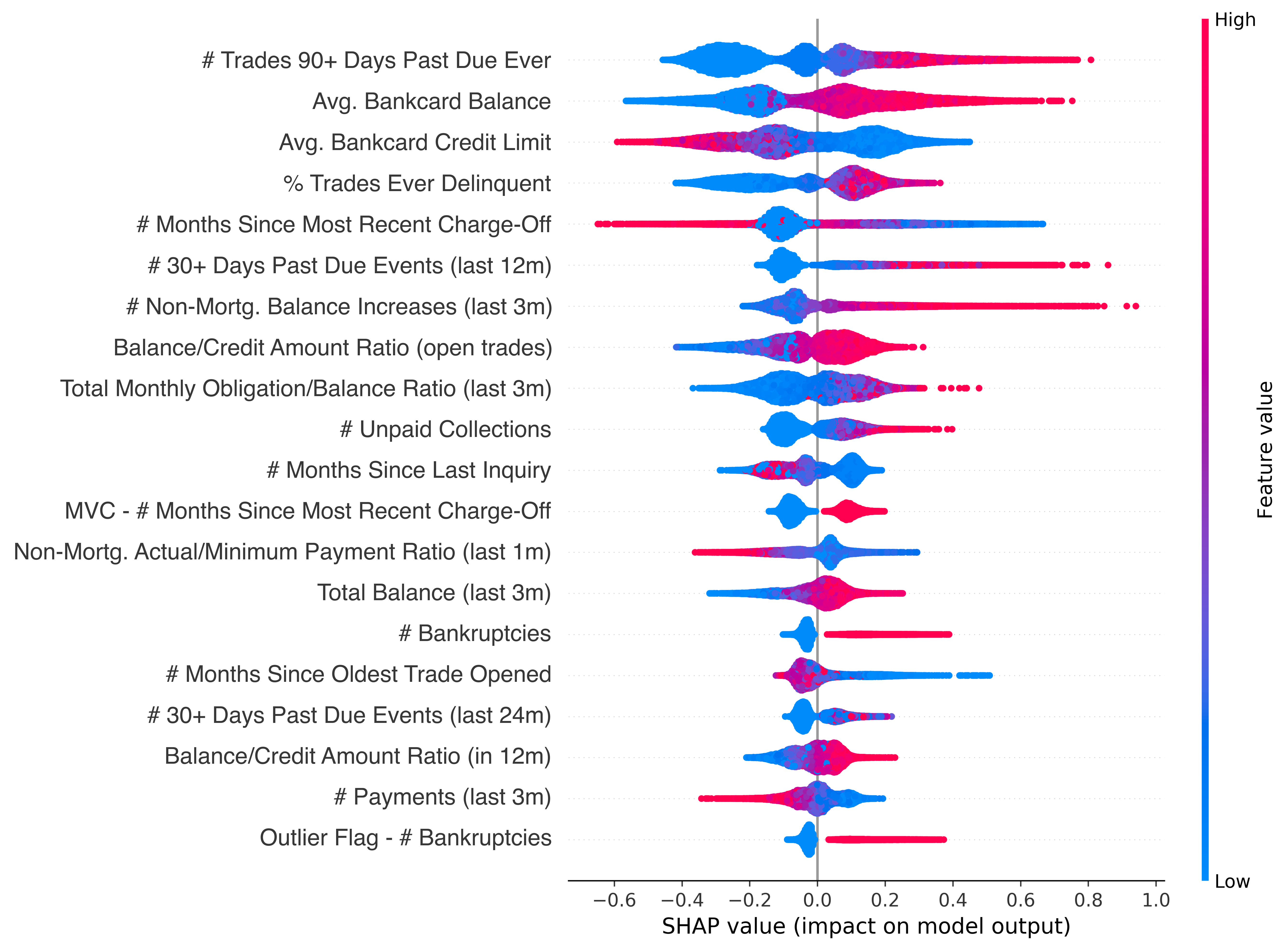}\caption{Low Income}
	\end{subfigure} 
	\begin{subfigure}[t]{0.45\textwidth}
		\includegraphics[width=\textwidth]{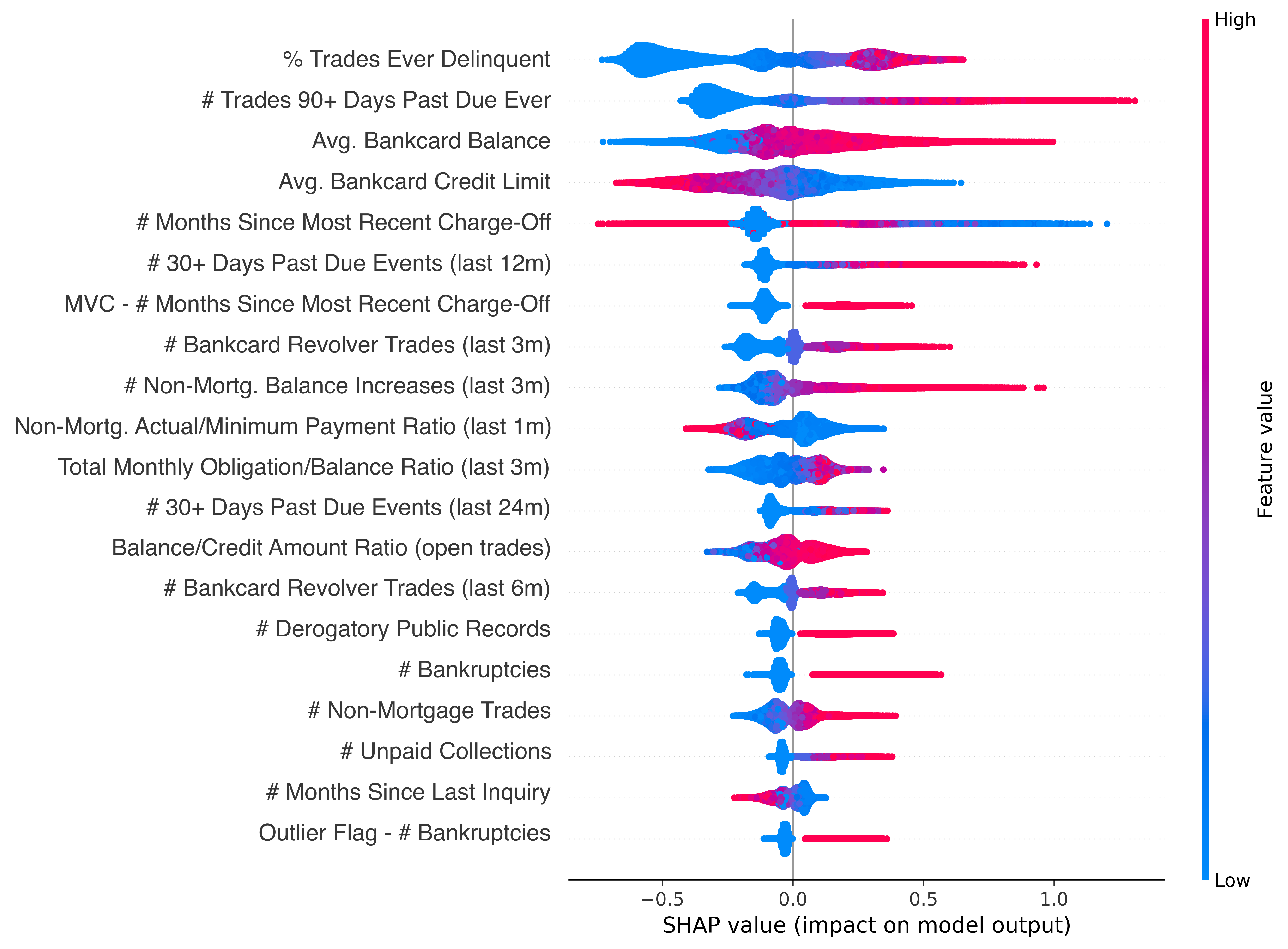}\caption{Higher Income}
	\end{subfigure}
	\begin{flushleft}
		\begin{spacing}{1.1}
			\small{Note: This figure shows variable importance measures (SHAP values) for XGBoost models trained separately by group \citep{lundberg2017unified}. In the figure, variables are sorted from top to bottom in order of average variable importance across applicants, while the distribution of (signed) importance measures (local SHAP values) across applicants is shown in each row. SHAP values (x-axis) are signed and sized such that negative values indicate lower predicted default risk, and larger values indicate greater impact on predicted default risk. Each panel corresponds to a model trained separately for that group.  SHAP values are calculated using the tree explainer implementation in  Python. Each point refers to an observation in the test data. The color indicates the value of the variable for that observation.  ``MVC" refers to categorical variables that preserve only missing value codes relative to the original underlying variable.}
		\end{spacing}
	\end{flushleft}
\end{figure}

Our findings are instead consistent with the second explanation: conditional expectation functions simply are similar across groups.\footnote{One potential caveat is that the improvement from solving aggregation bias is masked by an overfitting problem that is more acute for the group with the smaller sample size when training separate models (see \cite{wang_split_2020}). The gain in fitting a group-specific model could be erased by a reduction in performance due to the smaller sample size. However, this overfitting problem should also apply to the non-minority group when we fit separate models with identical samples sizes. As there should be little gain from eliminating aggregation bias, we would expect the overfitting problem to induce a large deterioration in performance for the non-minority group when limiting the sample size to the size of the minority group. The row labeled \emph{Different models: Same N} of  Table \ref{table_ml_applicant_mc} shows that we do not observe large drops in performance for the non-minority group suggesting that the overfitting problem is unlikely to be a problem given our sample sizes.} This interpretation is supported by Figure \ref{fig:shap_xgboost_groups_mc} which shows variable importance measures (SHAP values) for the models trained separately by group \citep{lundberg2017unified}. In the figure, variables are sorted from top to bottom in order of average absolute variable importance across consumers, while the distribution of (signed) importance measures across consumers, in the sense of having a positive or negative effect on the predicted probability of default, is shown in each row. The figure shows that the group-specific models rely on similar variables and that these variables play similar roles (in terms of sign and magnitude) in driving the predictive output of the respective models.\footnote{We obtain similar results when comparing coefficients of the logit models trained separately by group.}

\subsubsection{Data Bias\label{sec:databias}}

This section characterizes three forces that drive differences in the informativeness of credit report data: compositional differences; differences in how default is or is not reported to credit bureaus for different groups; and differences in outcome predictability. Of these, we show that compositional differences explain 46\%-53\% of the informativeness gap across groups. We then find suggestive evidence that reporting differences and differential outcome predictability, for example due to different exposure to future shocks, both help explain the remaining differences.

\paragraph{Compositional Differences}

\begin{figure}
	\caption{AUC and Compositional Differences in Credit Files \label{fig:auc_comp}}
	\centering
	
	\begin{subfigure}[t]{0.4\textwidth}
		\includegraphics[width=1\textwidth]{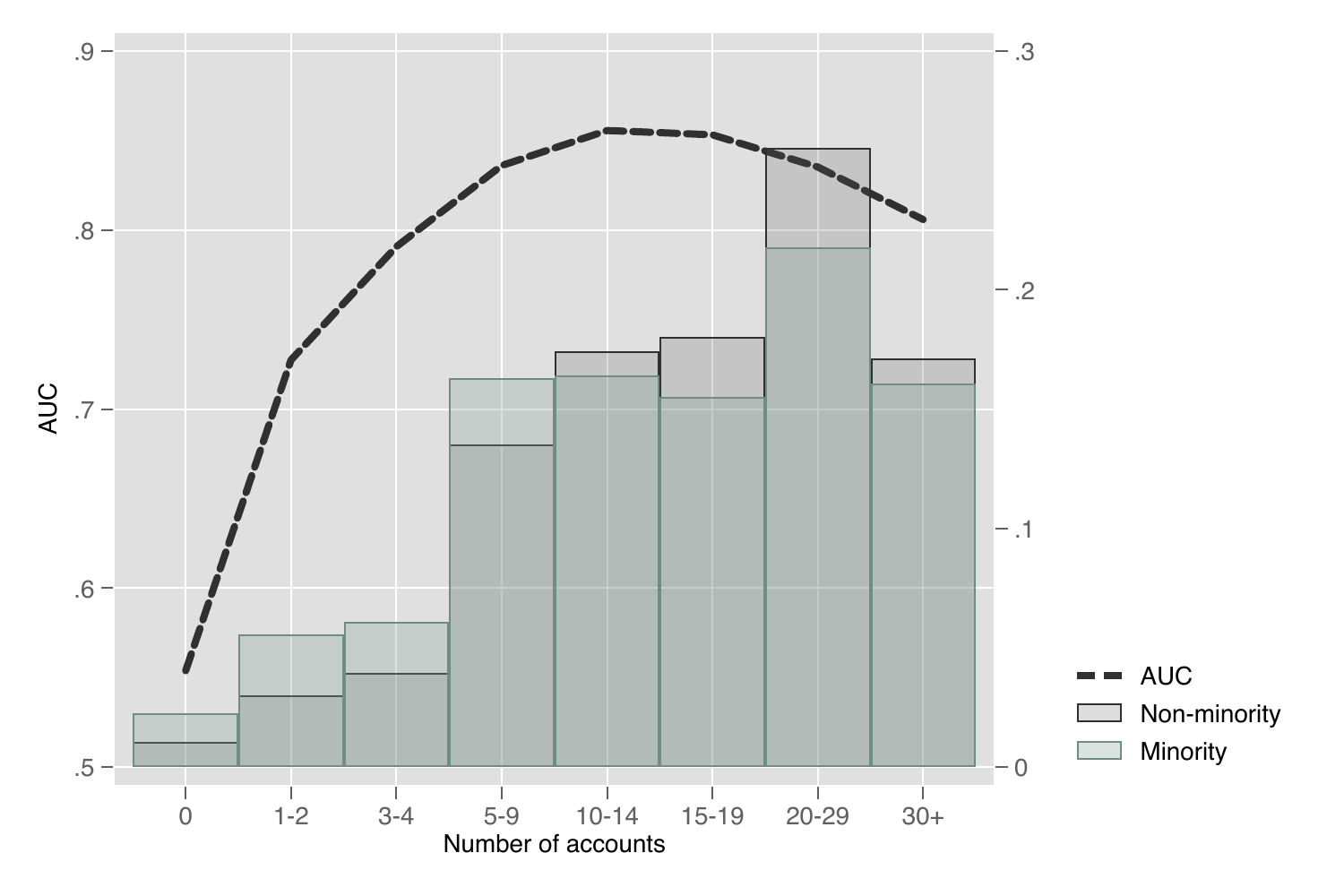}\caption{File Thickness by Minority Status}
	\end{subfigure} 
	\begin{subfigure}[t]{0.4\textwidth}
		\includegraphics[width=1\textwidth]{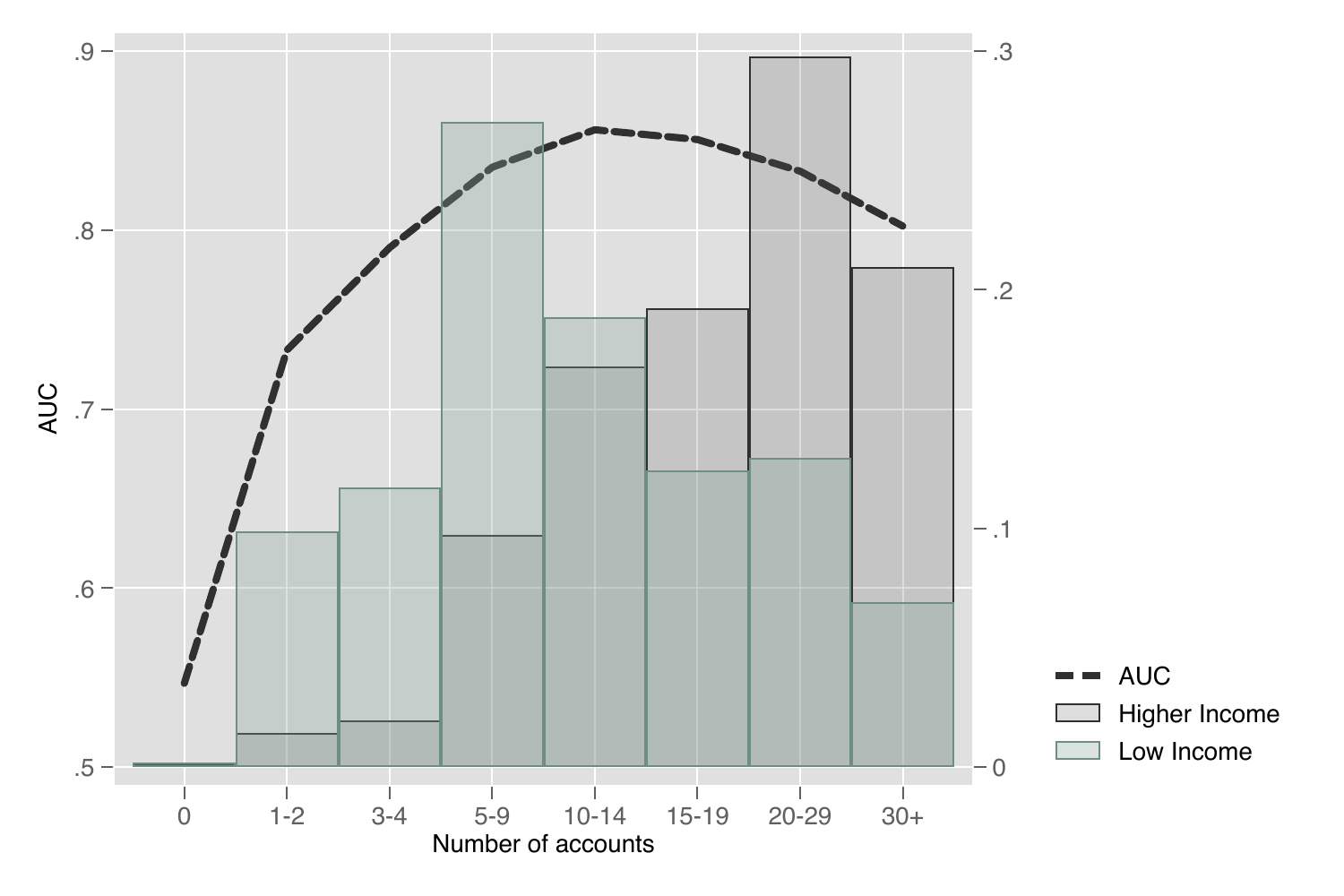}\caption{File Thickness by Income}
	\end{subfigure}\\
	
	\begin{subfigure}[t]{0.4\textwidth}
		\includegraphics[width=1\textwidth]{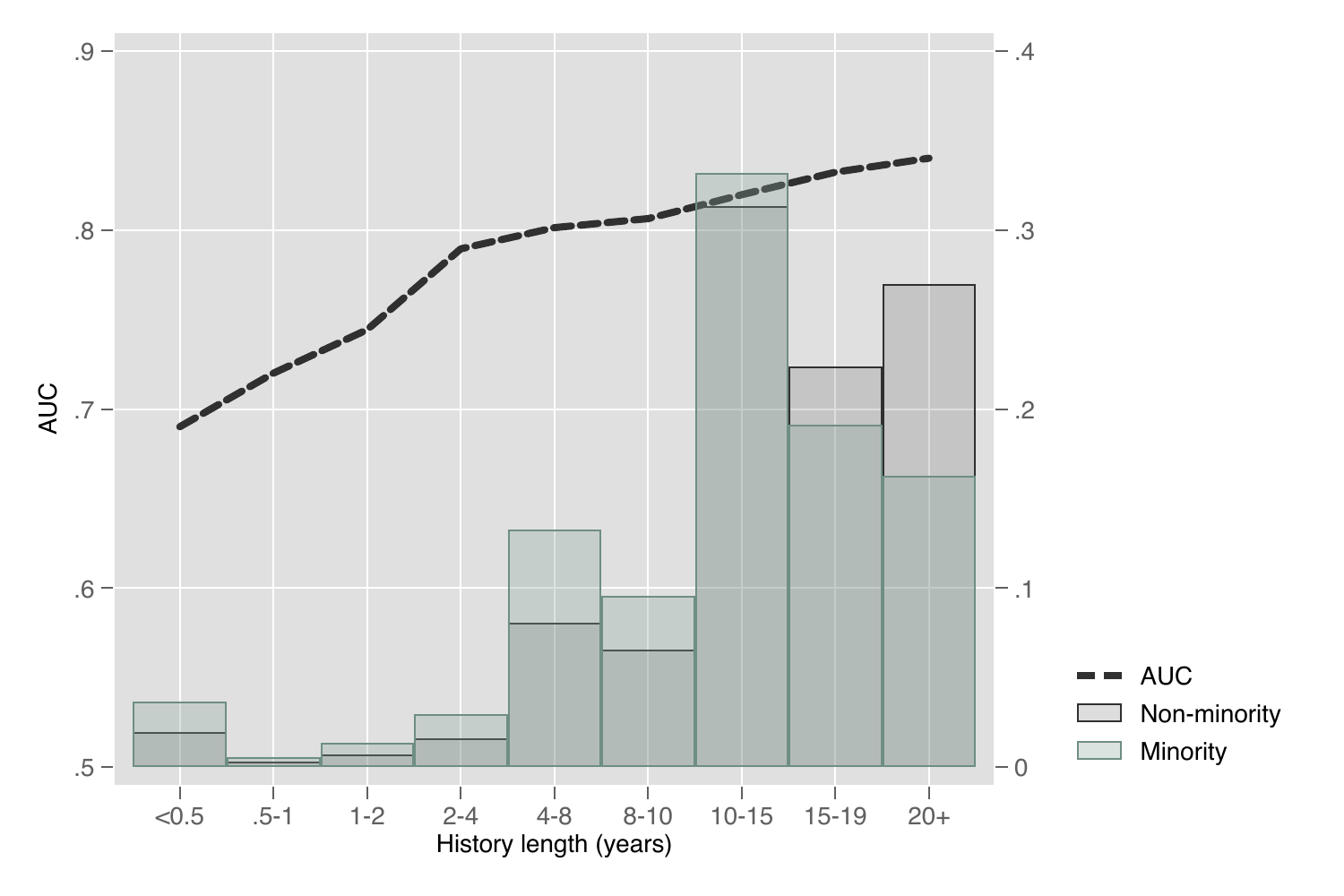}\caption{File History by Minority Status}
	\end{subfigure} 
	\begin{subfigure}[t]{0.4\textwidth}
		\includegraphics[width=1\textwidth]{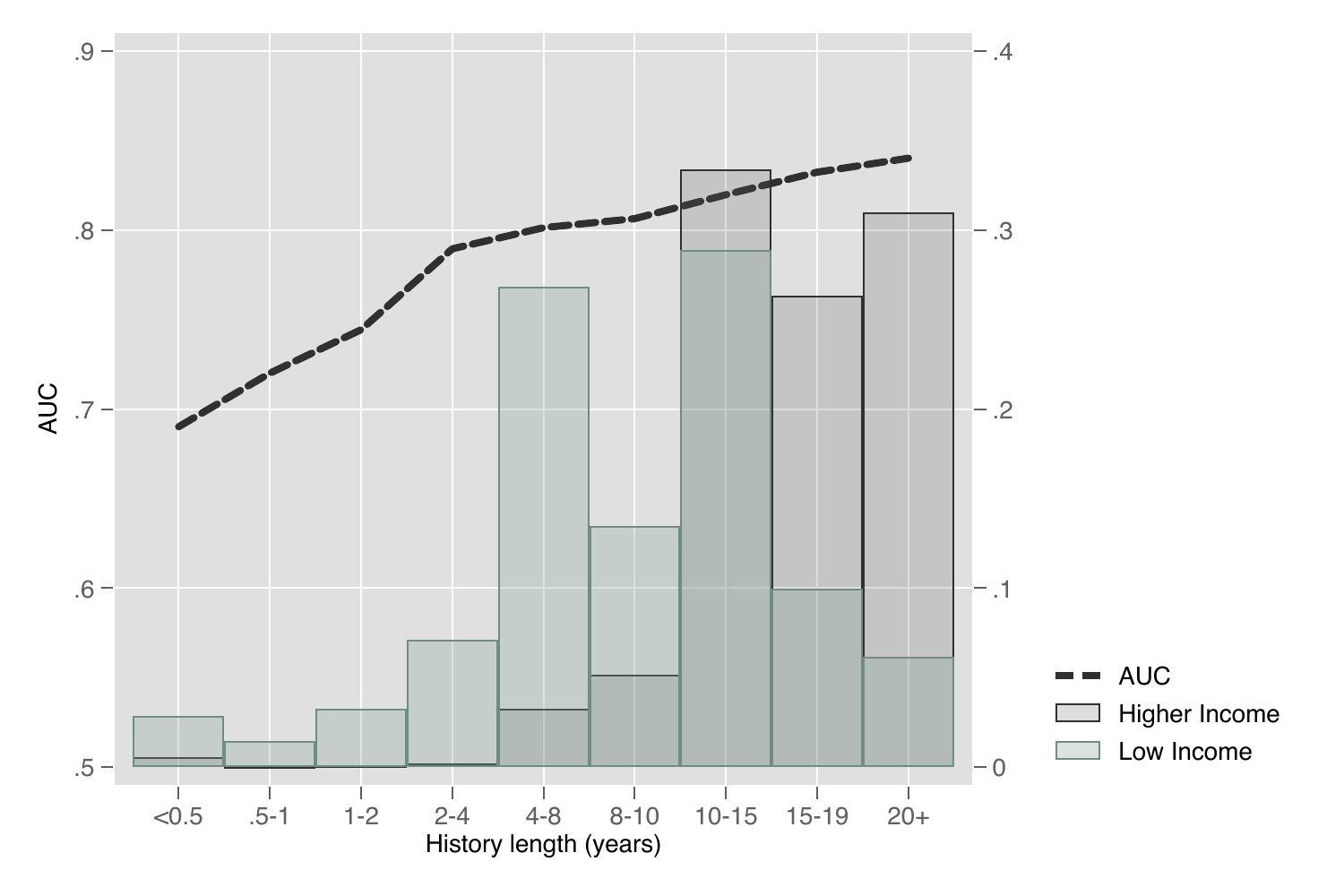}\caption{File History by Income}
	\end{subfigure}\\
	
	\begin{subfigure}[t]{0.4\textwidth}
		\includegraphics[width=1\textwidth]{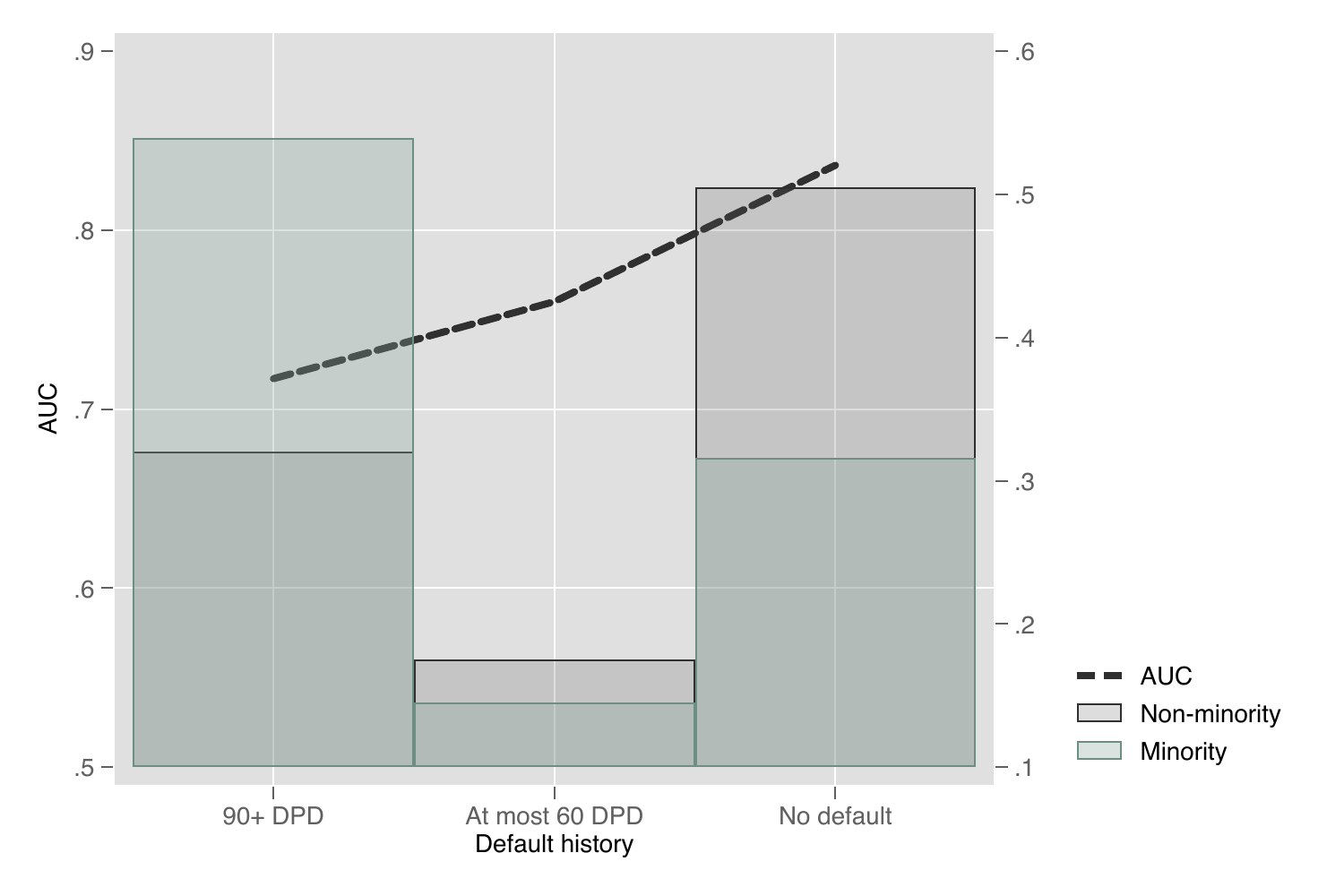}\caption{Default History by Minority Status}
	\end{subfigure} 
	\begin{subfigure}[t]{0.4\textwidth}
		\includegraphics[width=1\textwidth]{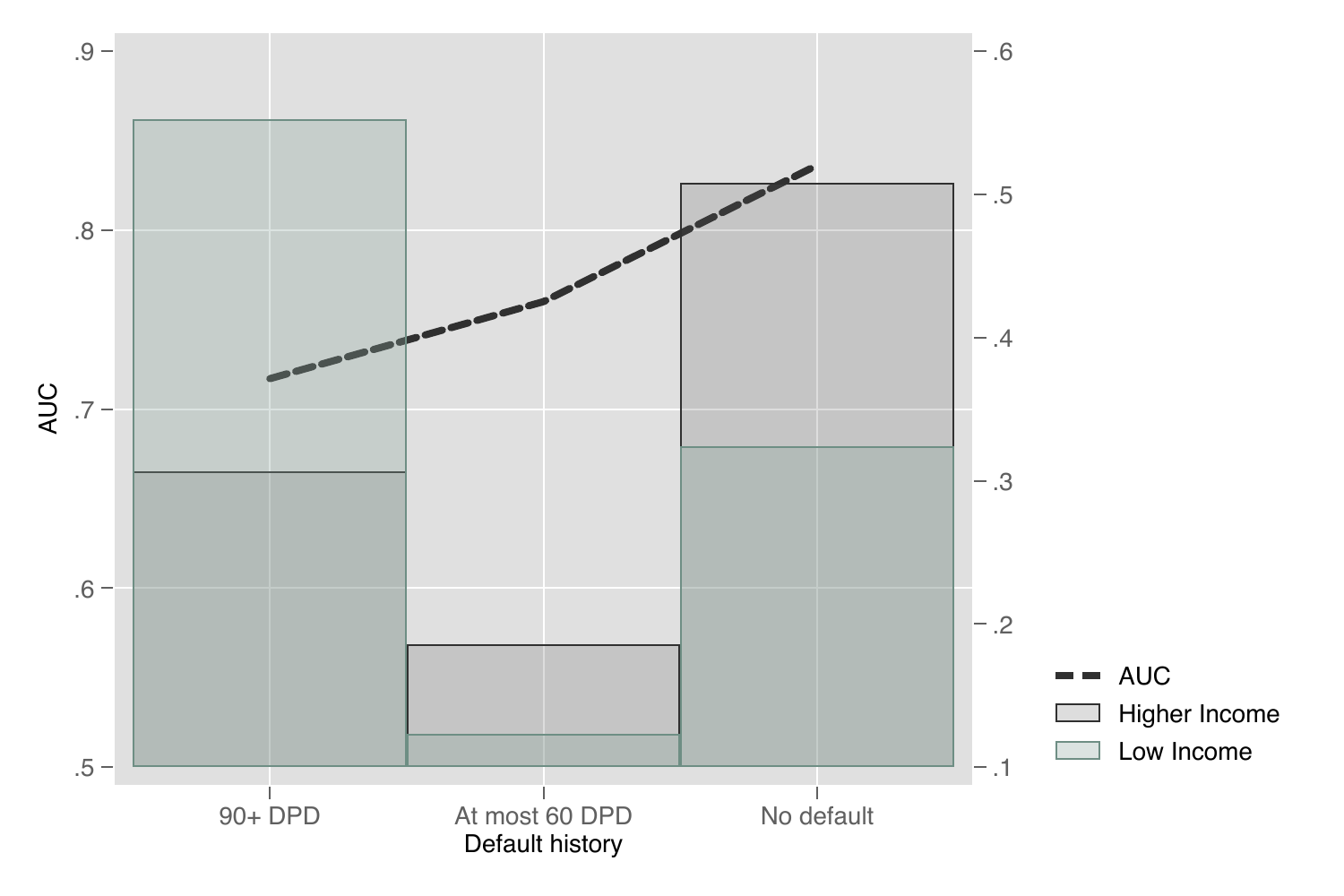}\caption{Default History by Income}
	\end{subfigure}\\
	
	\begin{subfigure}[t]{0.4\textwidth}
		\includegraphics[width=1\textwidth]{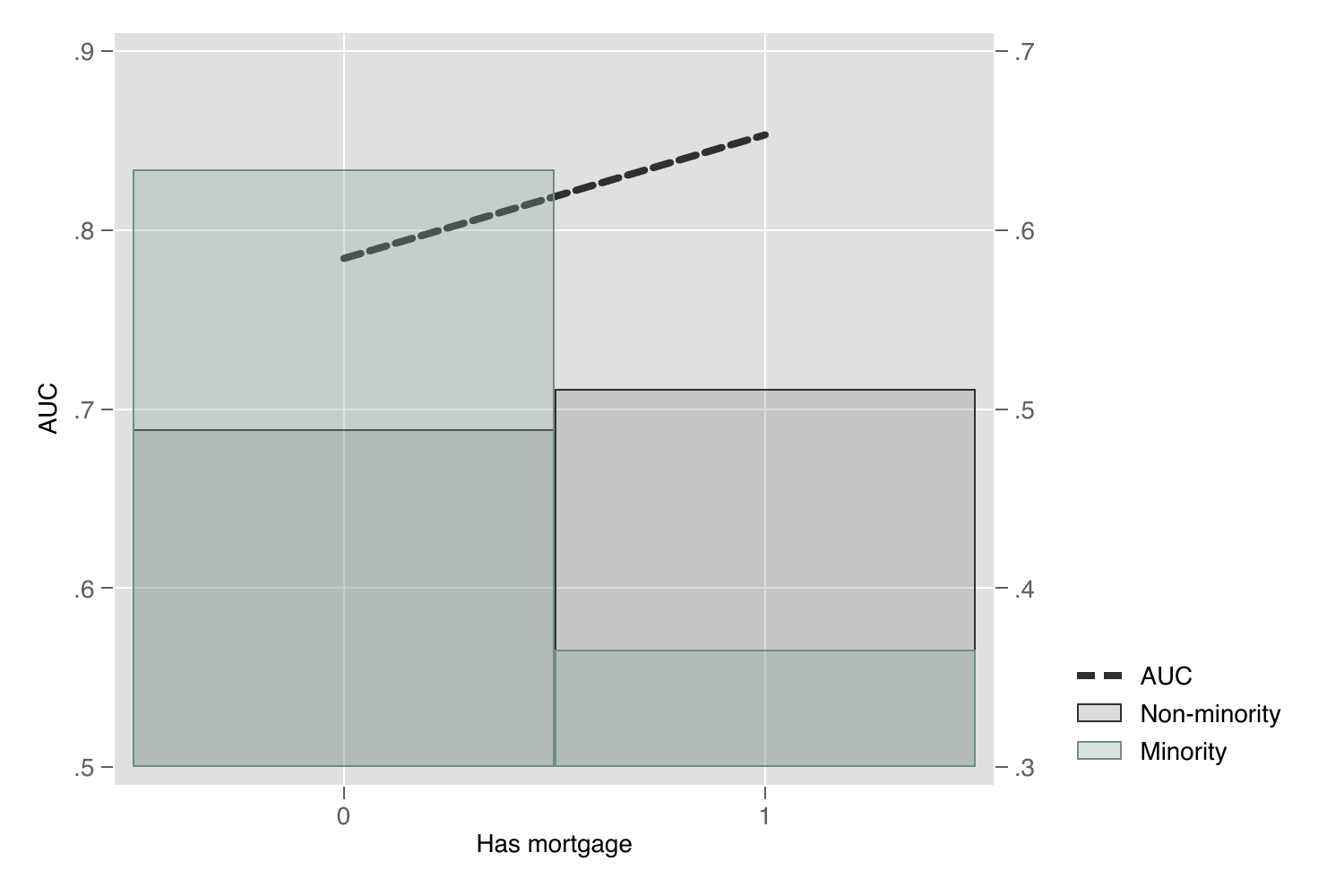}\caption{Mortgage by Minority Status}
	\end{subfigure} 
	\begin{subfigure}[t]{0.4\textwidth}
		\includegraphics[width=1\textwidth]{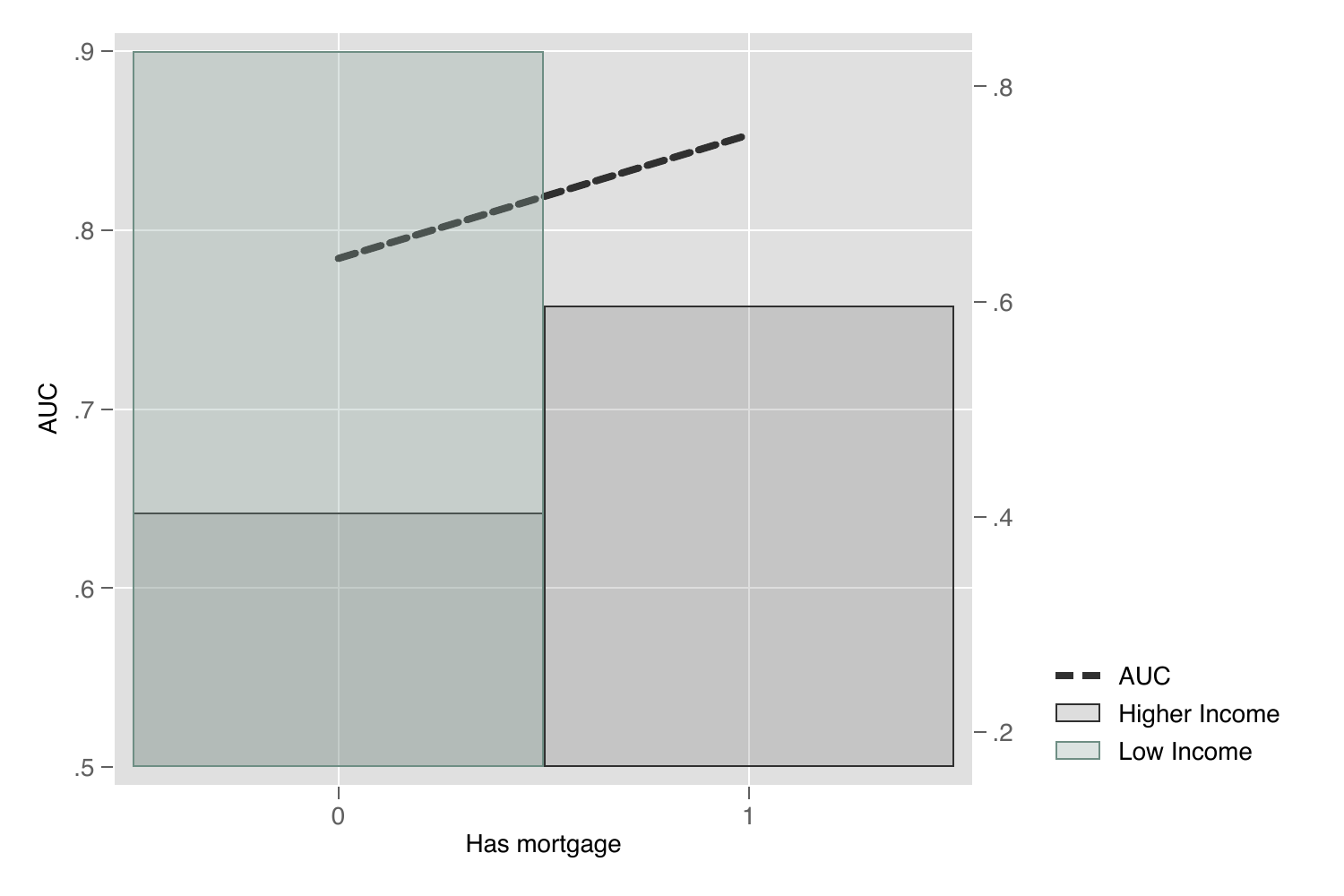}\caption{Mortgage by Income}
	\end{subfigure}\\
	
	\begin{flushleft}
		\begin{spacing}{1.1}
			\small{Note: Each panel of the figure highlights a different characteristic of credit report data (e.g., file thickness) for two complementary consumer groups (e.g. minority and non-minority applicants). Different subsamples defined by a given characteristic of credit report data are arrayed on the x-axis (e.g., different levels of file thickness). The y-axes of the plot then show subsample-specific Area Under the Curve (AUC) for the VantageScore 3.0 credit score (left axis), and group-specific distributions across these subsamples (y axis). All subsamples are drawn from the sample of mortgage applicants. DPD denotes days past due.}
		\end{spacing}
	\end{flushleft}
\end{figure}

We first investigate whether compositional differences in credit report data can explain the informativeness gap across consumer types. We center our analysis on the five key components of credit report data that are commonly understood to determine credit scores.\footnote{These components are: number of accounts, length of file history, past default history, usage (utilization and balances), and account mix (how many types of accounts a borrower has). 
}  Across these components, we show that credit report data are less informative about future default when the data are sparse (e.g., thin or young credit files), ``dirty" (e.g., showing past default), or non-diverse (e.g., only holding one type of loan). We then show disadvantaged groups' credit report data are more likely to be sparse, dirty, and non-diverse.  Figure \ref{fig:auc_comp} shows the AUC of the  VantageScore 3.0 credit score as we condition on various characteristics of credit report data, together with with estimated distributions across these characteristics for different groups. The top two panels of the figure show that ``thicker" credit files -- those with data from more past or current loans -- are associated with higher AUCs, while the distribution of minority (in panel (a)) and low-income (in panel (b)) consumers' credit report data is skewed toward ``thinner" credit files. For example, moving from 1-2 accounts to 10-14 accounts increases the AUC from 0.72 to 0.85, while the share of minority consumers with 1-2 accounts is roughly twice the share for non-minority consumers. Next, panels (c) and (d) show a similar result for history length, or how many years an individual has had a credit report: sparser data are associated with lower AUCs, and disadvantaged groups are more likely to have sparse data. 

Moving to the bottom half of Figure \ref{fig:auc_comp}, panels (e) and (f) show that credit files with no past default history (``clean files") have higher AUCs than those with past default (``dirty files") and that minority and low-income groups are more likely to have ``dirty" files. For example, the average AUC of clean file applicants is 0.82 compared to 0.72 of applicants with a past severe default (90+ days overdue), and minority applicants are roughly twice as likely as non-minority applicants to have a past severe default. Finally, panels (g) and (h) show credit files that include data from at least one mortgage account have higher AUCs than those that do not and that, again, minority and low-income groups are less likely to have a mortgage. Similar evidence for other credit report characteristics is shown in Appendix Figure \ref{fig:auc_comp1} for total balances, credit utilization, and the presence of a credit card. Overall, we consistently find that sparse, dirty, and non-diverse credit report data are associated with lower informativeness of credit scores and such data also are substantially more likely to belong to disadvantaged groups. 

To quantify the effect of these compositional differences, we present a simple back-of-the-envelope decomposition. We first divide applicants into nine (mutually exclusive) sub-samples that condition on similar default history, account mix, and file thickness -- the credit report characteristics associated in the previous figure with the largest AUC differences.\footnote{The nine sub-samples are defined by all pairwise interactions of two triples. The first triple is whether a file is thin (less than nine life-time accounts or less than 10 years of history), or thick (the complement of thin) with no mortgage, or thick with a mortgage. The second triple is whether the file has no default history, mild default history (at most 60 days past due (DPD)), or severe default history (90+ DPD).} We can then decompose the overall difference in AUC between groups into differences in each group's distribution \textit{across} subsamples, and a residual term that captures differences in AUC \textit{within} each sub-sample. While this back-of-the-envelope estimate should be interpreted carefully,\footnote{One important caveat to note is that this exercise is a linear approximation to the nonlinear relationship between population-level AUC and sub-population AUCs. A separate caveat is, of course, that some unobservables may be correlated with both credit report data composition and future default outcomes that determine AUCs. Nevertheless, this exercise provides a useful estimate in the spirit of a Kitagawa (Oaxaca-Blinder) decomposition of the importance of group-specific distributions of observables, similar to work that decomposes black-white earnings gaps into the composition of observables and other residual effects \citep{kitagawa1955components, oaxaca1973male, blinder1973wage}. As we emphasize when we discuss the potential for credit score hysteresis (see e.g. Section \ref{conclusion}), noting that these observables explain much of the observed gap should be seen as underscoring, rather than minimizing, the importance of understanding the history that gives rise to these different observables \citep{spriggs2020now}.} it provides a useful quantification of how much compositional differences in credit report data drive the gap in credit score informativeness. 

Table \ref{table_data_bias_main_lb} shows that 46\%-53\% of the overall AUC difference is accounted for by the fact that disadvantaged groups are over-represented in sub-samples where credit report data have less predictive power. Concretely, this is calculated by taking the average (not group-specific) AUC in each subsample, and averaging these across sub-samples using the group-specific distributions; for example, in the case of AUC gaps between minority and non-minority applicants considered in panel (a) of the table, this average is .025, which is 46\% of the .054 overall gap between groups. The residual term can then be calculated as the unexplained part of the overall gap between groups, or equivalently by taking group-specific AUCs within each subsample and averaging across them using a population-level (not group-specific) distribution across sub-samples. These residual terms capture the other near-half of AUC differences.

\paragraph{Residual Differences}

Table \ref{table_data_bias_main_lb} also provides important clues as to what drives these residual gaps in predictive power. The residual gap is not uniform across sub-samples but rather is concentrated in sub-samples that have clean files, indicating  residual credit score noise is driven by disparities in the data's ability to predict a consumer's \textit{first} observed default.\footnote{This finding -- that credit score informativeness is similar across minority and non-minority individuals after conditioning on those that have a history of default -- is interestingly consistent with results in \cite{bartik2020deleting}, where noise in credit report data is estimated to be similar across groups in a labor market context where employers largely focus on negative information such as past loan default. Credit report data's predictive power for future loan default and for worker match quality may also just differ due to other differences in these target outcomes or in these markets.
} Two possible explanations appear most consistent with this pattern: differential measurement error arising from how first defaults are reported to credit bureaus;
and second, differences in the inherent predictability of a first-ever default, for example due to differences in groups' exposure to future shocks.

While prior work has suggested various types of measurement error may be important in this setting,\footnote{For example, \cite{cherry_government_2021} find that forbearance rates are higher for low income borrowers and in areas with more minority representation. This finding suggests that default outcomes might be differentially under-reported across groups. See also \cite{FTCerrorsreport} for other evidence on the prevalence of measurement error.} we focus on a particular type of measurement error that our data are uniquely well-suited to study: disadvantaged groups may be more likely to borrow from lenders that do not report defaults to traditional credit bureaus, for example payday lenders.\footnote{For background and related research see e.g. \cite{morse_payday_2011}, \cite{bhutta_payday_2014}, and \cite{carrell_harms_2014}.} Using linked individual-level data from FactorTrust, a leading alternative credit bureau that covers payday borrowing, we investigate this possibility by including information on payday loan use and default in our predictive models of default, previously described in Section \ref{sec:modelbias}. However, we find that the inclusion of these additional variables does not reduce the across-group differences in credit score informativeness (see Table \ref{tab:ml_ft} in Appendix \ref{appendix_data_bias}).\footnote{This finding is consistent with prior research by \cite{bhutta2015payday} who find that consumers who use alternative credit such as payday loans also tend to be delinquent on traditional credit products; their Table 2 shows the average payday loan applicant is delinquent on more than half of their balances reported to a traditional credit bureau.}

Another explanation for the residual differences in informativeness is that available observables are less informative for predicting default for minority or low-income applicants due to differences in the inherent predictability of future default. A long-standing literature finds that minority individuals \citep{bils1985real, shin1994cyclicality, solon1994measuring} and low-income individuals \citep{guvenen2017worker, patterson2019matching} have earnings that are more exposed to business-cycle risk. In addition, uninsured idiosyncratic risk may also plausibly vary across groups. At the same time, there is evidence that liquidity constraints play an important role in driving default (\cite{indarte_moral_2020}, \cite{ganong_why_2020}). Together, these findings imply that more volatile incomes might drive differences in the volatility of default. We find some suggestive evidence consistent with such differences: transition probabilities from no default to default states are higher for low income and minority applicants for clean, but not dirty, credit files (see Table \ref{tab:default_persistence} in Appendix \ref{appendix_data_bias}).


\section{A Model of Information in Mortgage Lending} \label{sec:model}

Thus far we have shown two main results: quantifying the differential precision of credit report information across groups, and illustrating the role of data bias rather than modeling bias in generating these differences. We now develop and estimate a model of credit market information to assess the effects of these precision differences on economic efficiency, focusing on the context of the US mortgage market.  We then use the model to study counterfactual information structures relevant to potential or ongoing changes in consumer credit markets.

\subsection{Setup} \label{sec:model:setup}

We model a lender who receives signals $s \in \mathbb{R}^K$ about a borrower's unobservable type $\theta \in \mathbb{R}$. These signals may include, for example, a credit score, home equity, or soft information gathered through interactions or relationship lending.\footnote{For evidence on the role of soft information in US mortgage lending, see \cite{keys2012lender}, or \cite{RajanSeru2015} for the role of hard information not captured by credit scores.} We define noise in a standard way relative to the type $\theta$: for each signal $s_k$, the signal error $u_k$ is the gap between the signal value and the true type, $s_k = \theta + u_k$, such that the noise in a signal is $\sigma^2_k = \text{Var}(u_k)$.

We specify that approved loans default with probabilities $\delta(\theta)$ dependent on borrower types. Risk-neutral lenders make approval decisions based on their posterior belief $x(s)$ about a borrower's type given observed signals $s$ and approve loans for which their posterior falls above an approval threshold $\hat{x}$. This yields profits equal to,
\begin{align} \label{eqn:Pi_max_basic}
    \Pi(\hat{x}) = \int_\theta \left(\alpha + \beta R^\star -     \gamma\delta(\theta) - c\right) 
    \textbf{1}\{ x(s(\theta)) \geq \hat{x} \} d\theta
\end{align}
        
Here the parameters $\alpha$, $\beta$, and $\gamma$ respectively capture non-interest revenue, the present value of interest revenue retained by the lender given net interest rate $R$, and the lender's exposure to losses from loan default. Meanwhile the marginal cost $c$ captures other per-loan variable costs not driven by default. This profit function is relatively general in the mortgage-lending context because it describes both portfolio lending, where the lender fully internalizes interest revenue and default losses, as well as originate-to-distribute and GSE-funded business models, where lenders share revenue and risk externally. See also \cite{bhutta2020minorities} and \cite{woodward2012diagnosing} for discussions of similar mortgage lender profit functions.\footnote{This profit function also includes flexibility across groups in view of \cite{canay2020use}'s emphasis on how different groups studied in marginal outcome tests may also differ in unobservable dimensions of the signal receiver's payoff function.}

We suppose that posterior beliefs $x(s)$ are Bayesian posterior expectations of the applicant's type, and lenders observe the true distribution of signal noise for each applicant. Using this type of posterior is profit-maximizing for a risk-neutral lender in many settings, including the normal-normal parameterization we develop more below.

\subsection{Estimating the Model} \label{sec:model:estimation}

We use an empirical analog of this model in order to study a mortgage lender's problem of which applications to approve given noisy information about default.

To make empirical headway, we suppose that unobserved types for individuals $i$ from each applicant group $m$ are normally distributed with potentially group-specific mean and inverse variance,
\begin{equation}
    \theta_{im} \sim  \mathcal{N} \left( \mu_m, \frac{1}{h_{0m}} \right)
\end{equation}

Likewise signal noise is normally distributed with potentially group-specific variance,
\begin{align}
    u_{ikm} \sim  \mathcal{N} \left( 0, \frac{1}{h_{km}} \right)
\end{align}

Recalling $s_{ikm} = \theta_{im} + u_{ikm}$, this structure implies the lender's Bayesian posterior of each applicant's type is,
\begin{align}
    x_{im} = \frac{\sum_{k=1}^{K}h_{km} s_{ikm} + h_{0m}     \mu_m}{\sum_{k=1}^{K}h_{km} + h_{0m}}
\end{align}

We also map from unobserved types to observable default with default probabilities parameterized as,
\begin{equation}
    \delta(\theta) = \frac{\exp(\theta)}{1+\exp(\theta)}
\end{equation}

We follow the standard parameterization in the credit scoring industry \citep{thomas2009consumer} and suppose that observed credit scores are an affine transformation of a predicted log odds of default, 
 \begin{align}
        Score_{i} = a_1 + a_2 * log\Big(\frac{\delta(s_{ik})}{1-\delta(s_{ik})}\Big)
\end{align}

With these parameterizations in hand, we use a simulated method of moments strategy to estimate the key primitives of the environment: two parameters for each group describing the distribution of unobserved types for that group's loan applicants, $\mu_m$ and $h_{0m}$; signal noise for each group and for each signal, $h_{km}$; the scalar parameters of the affine mapping from credit report signals to credit scores, $a_1$ and $a_2$; and finally the approval threshold $\hat{x}_m$ for each group.\footnote{Note that the approval threshold can differ across groups even if the profit function is the same across groups, because of non-linearity in $\delta$ will make groups with noisier signals exhibit higher average default for the same posterior $x$.} Given the existence of an (unobserved) approval threshold  $\hat{x}$, we do not need to estimate the profit function parameters directly in order to recover these primitives on signal noise, so we postpone profit-function estimation until the counterfactual exercises described in Section \ref{sec:estimation_results:CFs}.

We estimate a model with two signals ($K=2$). We take one of of these two signals to be a credit score, and the other signal to be a composite of all other information available to the lender, which includes factors such as loan-to-value ratios, debt-to-income ratios, soft information, and other information sources potentially unobservable in our data. Combining all non-credit-score signals into a single composite is without loss of generality for i.i.d. signal noise.

To estimate the model primitives, we simulate the following moments and match them to corresponding empirical moments for each group: loan approval rates; the mean of credit scores among accepted applicants; the mean of credit scores among \emph{rejected} applicants; the slope of default outcomes with respect to credit scores among approved loans; the slope of the ``reverse regression" of scores on default outcomes; default rates among approved applicants; and the default rate of \emph{marginally} approved loans. This final moment corresponds to the expectation $\mathbb{E}[\delta(\theta_{im})        \hspace{1pt} | \hspace{1pt} x_{im} =        \hat{x}]$.

The model is transparently identified off of the target moments. First, the forward and ``reverse" regression slopes between credit scores and default realizations help identify the amount of signal noise in credit scores, similar to how reverse regressions can be used to test for measurement error \citep{black2000bounding}. Intuitively, the same attenuation in regression coefficients that we showed as evidence of credit score noise in Section \ref{sec:noiseresults} is used in model estimation to identify how much noise there is, quantified in terms of the variance of $u_{ikm}$ terms in a credit score signal.

\begin{figure}[h]\caption{Model Identification \label{fig:modelidentification}}
	\centering
	\includegraphics[width=.7\textwidth]{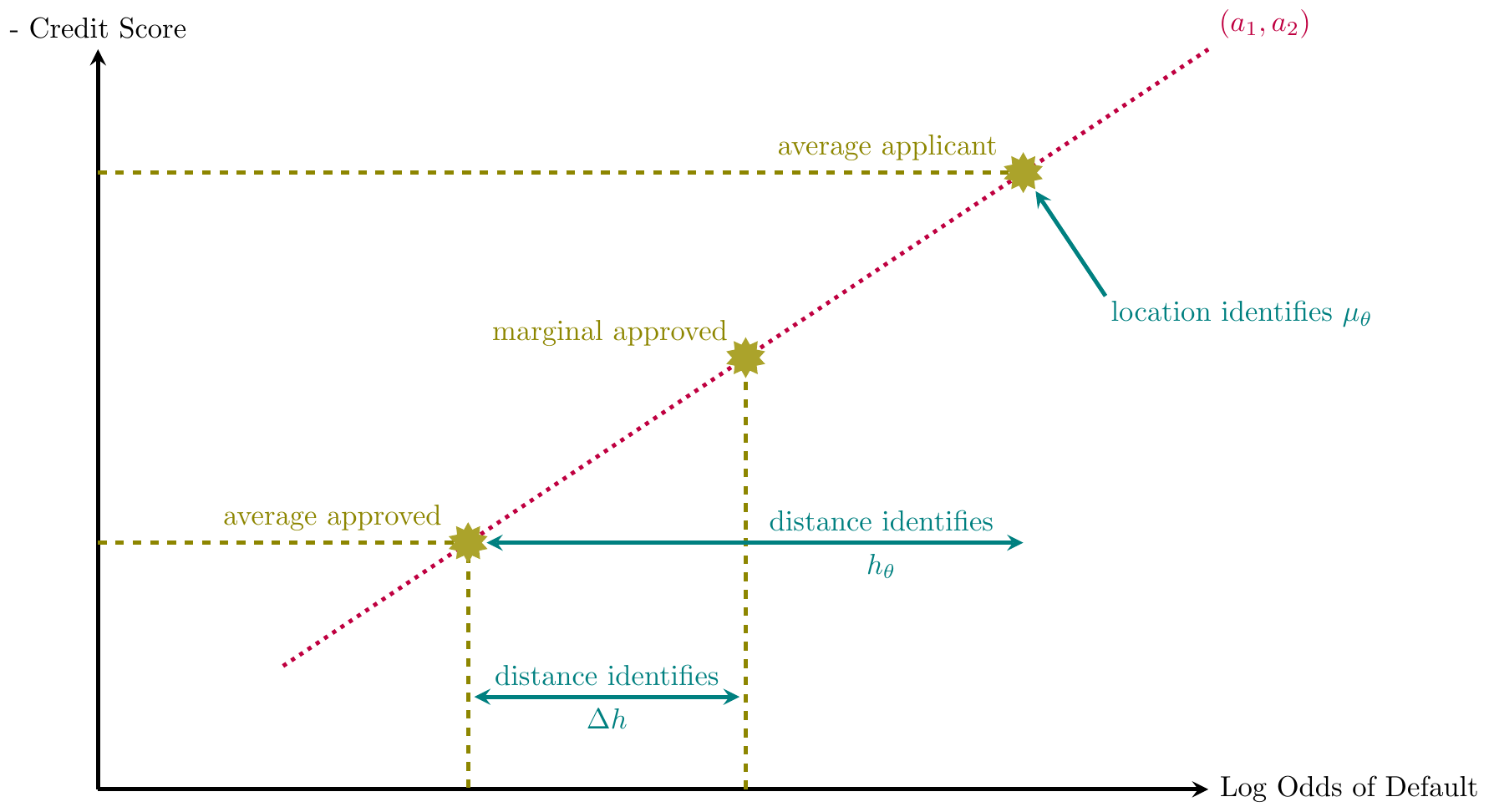}
	\begin{flushleft}
		\begin{spacing}{1.1}
			\small{Note: This figure illustrates how our target moments identify model parameters related to the shape of the unobserved risk distribution and to the combined precision $\Delta h$ of credit scores and other signals ($\Delta h \equiv h_1 + h_2$). The dotted red line denotes the unknown affine mapping between credit scores and log odds of default, which is described by the slope and intercept parameters $a_1$, $a_2$ and is identified by population-specific regression coefficients of credit score on realized default. Conditional on that mapping, the gold stars and dashed gold lines denote target moments that we can estimate. The blue labels and solid arrows then illustrate how these moments identify model parameters. Note that this figure illustrates the case where observed approval rates are below 50\%, because the average applicant is higher risk than the marginal applicant.}
		\end{spacing}
	\end{flushleft}
\end{figure}

To further illustrate how our empirical moments identify model parameters, Figure \ref{fig:modelidentification} shows visually how we identify other key parameters of interest: the mean and variance of the unobserved risk distribution, and the total precision of all signals available about a given applicant, summed across the credit score signal and other, non-score signals. This ``total precision'' measure, denoted $\Delta h$, is the sum of the model parameters that describe the precision of the credit score and non-score signals ($\Delta h_m \equiv h_{m1} + h_{m2}$).

In the figure, the gold stars and dashed gold lines denote moments that we can observe, corresponding to average applicants, marginally approved applicants, and average approved applicants. Moments for these subgroups of applicants can be observed either in terms of their credit score, or in terms of their default outcomes, or both, with the two axes of the figure corresponding to these two types of observables. The dashed gold lines in the figure then illustrate which types of observables are available for which subgroups. For example, default for \textit{non}-approved applicants is not observed, but credit score is. Connecting these two types of observables, the dotted red line illustrates the affine mapping to be estimated between the x- and y-axis outcomes.

Based on those empirical moments illustrated in the figure, the blue-colored labels then illustrate how these moments identify model parameters. It is straightforward that the location of the average applicant in this figure identifies the average applicant type, $\mu_\theta$. Conditional on other parameters, the distance between approved and rejected applicants then reveals how much dispersion there is in unobserved risk types. Finally, the difference between average and marginally approved applicants in terms of default odds reveals, conditional on an observed approval rate and the dispersion of risk types, how much information is conveyed by all available signals about default risk. Intuitively, if very little information were available to lenders about mortgage applicants, then lender posteriors would be minimally diffuse, and, holding fixed a given approval rate, marginally approved borrowers would be minimally different from average approved borrowers. Conversely (but again holding fixed the  approval rate), when marginally approved and average borrowers are more different from each other in terms of default outcomes, this reveals that the available signals are more informative about borrower types.


\section{Model Estimation and Results} \label{sec:estimation_results}

\subsection{Construction of the Instrument} \label{sec:estimation_results:instrumentconstruction}

Important in our model estimation is the ability to identify the characteristics of \emph{marginally} approved mortgage applicants; this is the moment $\mathbb{E}[\delta(\theta_{im})        \hspace{1pt} | \hspace{1pt} x_{im} =        \hat{x}]$ described in Section \ref{sec:model:estimation}. We use the idiosyncratic timing of Community Reinvestment Act (CRA) exams as an instrumental variable to obtain exogenous variation in the approval leniency of banks, such that the local average treatment effect in a two-stage least squares regression reveals the characteristics of the marginally approved.\footnote{Focus on marginal applicants in credit and labor markets has a long tradition dating back at least to \cite{Becker1957}. For a recent application of an IV strategy to identify characteristics of marginal loan applicants, see \cite{Dobbie2019}.}

The Community Reinvestment Act, passed in 1977, directs federal banking regulators ``to encourage insured depository institutions to help meet credit needs of all segments of their local communities" \citep{CRAdescription2018}. The Federal Reserve, FDIC, and the OCC conduct exams at regular intervals to evaluate whether lenders comply with the CRA lending requirements.  Unsatisfactory CRA exams can be grounds for the regulator to deny applications for mergers and acquisitions or the opening of new branches.

Compliance is measured in loan quantity (not in terms of the interest rates changed on the loans). Two types of loans satisfy the CRA requirements: first, any loans made in low- and moderate-income areas where the bank has a physical branch and takes deposits, or made in moderate-income areas deemed depressed or under-served by banking regulators (``CRA-eligible census tracts''); second, lending to low- and moderate-income individuals in any area regardless of area-average income.  

We construct the CRA instrument following \cite{Agarwal2012} and consider a twelve-month window prior to and during a lender's CRA exam. These exams occur at fixed intervals -- every five (resp. two) years for banks with assets under (resp. over) \$250 million -- and therefore, as argued by \cite{Agarwal2012}, the timing of these exams is likely unrelated to local economic conditions and, in particular for our setting, credit demand and credit risk.\footnote{\cite{foote2013government} argue that the regulator would also take into consideration loans originated long before the three-quarters prior to an exam. This makes it potentially challenging to evaluate the CRA program \citep{Agarwal2012}. Unlike \cite{Agarwal2012}, we just require that the CRA increased leniency in the time window relative to non-examined banks. Other work examining channels through which the CRA can incentivize mortgage lending includes, for example, \cite{avery2015subprime, bhutta2015assessing, bhutta2011community, ringo2017mortgage, avery2005assessing}.}

To establish that our instrument affects origination, we estimate the following first stage in a half-year ($t$) - treatment geography ($g$) - bank ($b$) panel, separately by observable borrower groups $m$: 
\begin{align}\label{eq:firststage}
\begin{split}
    Q_{bgmt}= & \hspace{4pt} \beta_m \text{CRA exam}_{bt} \times \text{CRA tract}_{gt} +\gamma_m \text{CRA exam}_{bt} \times \text{CRA ineligible tract}_{gt}  \\ &+  \alpha_{bgm} + \alpha_{gmt} + \epsilon_{bgmt} 
\end{split}
\end{align}
where $ Q_{bgmt}$ is the quantity of loans originated,  $\alpha_{bgm}$ is a bank-geography fixed effect for group $m$ and $\alpha_{gmt}$ is a geography-quarter fixed effect for group $m$. We weight by the bank's origination share in a given geography, and standard errors are clustered at the geography level. 

Geographies are defined by measuring CRA eligibility at the Census-tract level, and then pooling tracts within county that have the same CRA eligibility status across years. Hence our treatment geography is essentially a partition of a county that is defined using tract-level CRA eligibility. This results in a relatively fine level of geography fixed effects, including thousands of sub-county-specific partitions across the US. Given these fixed effects, our first stage regressions are identified off of \emph{within}-geography-time variation and within-lender variation by comparing lenders that are and are not subject to a CRA exam at a particular place and time.

We define a CRA-eligible tract using low- and moderate-income status relative to area median income as defined by the FFIEC.\footnote{We focus on CRA-eligible tracts that are eligible by virtue of their income level, rather than higher-income tracts deemed distressed or under-served in a given year; by doing so we focus on the vast majority of loans made in CRA-eligible tracts.} We find that the CRA instrument is strongest within CRA-eligible tracts, so we limit our estimation to treatment geographies that are CRA-eligible throughout our sample period. 

Because refinance loans constitute the bulk of all mortgage loans, we focus on refinance loans in our main results.\footnote{Results for purchase loans are available upon request. The institutional details of the refinancing market also make our instrument and estimates more straightforward to interpret for refinance loans than for purchase loans: the much smaller market share of FHA and VA loans among refinances means that there are fewer margins of adjustment for lenders in response to the instrument, and smaller effective choice sets of borrowers; we discuss this choice in the context of the standard IV monotonicity assumption more in the following section.}

\subsection{Instrument validity}  \label{sec:estimation_results:instrumentvalidity}

Our identification assumption is that a CRA exam affects origination only through the increased leniency of the lender.  The exclusion restriction would be violated if there are other time trends coinciding with the CRA examinations driving loan demand specific to the examined lenders. Including geography-time and bank-geography fixed effects helps to address these concerns. Additional loan-level controls are only required if we believe that the instrument is only valid conditional on these additional controls; we view this as unlikely, given how the bank-specific timing of CRA exams is set independently of local market conditions.

We see three reasons why a bank may choose to defer CRA-eligible lending until just before an exam. First, CRA-induced lending involves, almost by definition, loans that have lower-NPV than other loans for the bank: these are the marginal loans the bank is only induced to make because of the CRA restrictions. Making these lower NPV loans as late as possible -- just before an exam -- therefore maximizes overall value for the bank. Second, there may be option value in delaying CRA lending, as the bank potentially will issue sufficient CRA-eligible loans without needing to change its approval leniency. Third, there may be behavioral or agency frictions that make it difficult to incentivize loan officers to make CRA-eligible loans, which may result in under-production of CRA loans that the bank has to correct for just before its exam.

To illustrate the validity of the instrument, Table \ref{table_firststage_refi_mc} in Appendix \ref{appendix_more} presents the estimates of the first stage (equation \ref{eq:firststage}) for refinance loans in our data. We see this table as the most straightforward way of presenting the first stage; the instrument cannot be viewed as a classic staggered event study because the same bank can be treated repeatedly by the instrument over multiple CRA exams. We estimate the first stage separately for different income levels as well as minority status of the borrower. We observe a significant positive effect of the CRA exam instrument on the volume of originated refinance loans by examined banks across all subgroups, with first-stage F-statistics typically above 10, supporting the strength of the instrument \citep{stockyogo2005testing}.\footnote{The exception is the first stage for low-income applicants, where the F-statistic is 6.11. This may suggest greater caution in our model estimates for low-income applicants than for minority applicants.} For robustness, we run the same regressions without bank share weights. Table \ref{table_firststage_uw_robustness_refi_mc} shows these results, which are qualitatively similar to our weighted estimates.\footnote{To further validate the instrument, we examined CRA exam reports to find evidence of institutions responding to CRA incentives with expanded lending just before a CRA exam date. An apt example comes from Fifth Third Bank's 2016 CRA exam, describing the bank's increased lending in CRA-eligible areas one year prior to its exam: ``In 2015, Fifth Third launched the ``\textit{Community Reinvestment} Mortgage Special," whereby all lender fees are waived for borrowers purchasing properties located in a \textit{low-income tract}... Overall, Fifth Third made extensive use of flexible lending practices in serving \textit{low- and moderate-income} needs within its assessment areas.'' Both the use of ``community reinvestment" in the program name and the emphasis on low- and moderate-income tracts point to this expanded lending being directly in response to CRA incentives; the regulator's report also notes how the bank made ``extensive use" of such incentives more broadly.}

Instrument validity also requires a monotonicity condition that the CRA exam does not cause any applicants in CRA-eligible tracts to be denied credit when they otherwise would be approved (i.e., no ``defiers" in the language of \cite{Angrist2008}). Our focus on refinance loans helps support the validity of the monotonicity condition in our setting, because borrowers' smaller choice sets for refinance loans implies that lenders' response to the instrument is less likely to change the mix or characteristics of multiple products available to borrowers, and rather, lenders' response is likely to entail an across-the-board credit expansion.

\subsection{Empirical specification}  \label{sec:estimation_results:2sls}

We use the first-stage variation in loan volume from the previous section to estimate the default rate of marginal borrowers. We can recover these marginal default rates in a two-stage least squares framework where the second stage regression relates loan default volume on total (instrumented) origination volume. Specifically, the second stage at the quarter ($t$) - tract ($g$) - bank ($b$) level is, separately by observable groups $m$,
    \begin{align}\label{eq:2sls}
        Y_{bgmt} = \beta^{IV} Q_{bgmt} + \alpha_{bgm} + \alpha_{gmt} + \eta_{bgmt} 
   \end{align}
where  $Y_{bgct}$ is the volume of defaulted loans of group $g$ (e.g., income group), $Q_{bgct}$ is the (instrumented) volume of loans originated, and $\alpha_{bgm}$ and $\alpha_{gmt}$ are geography-time and bank-geography fixed effects respectively. Intuitively, this regression captures in the coefficient $\beta^{IV}$ the relationship $\partial Y / \partial Q$, or a marginal default rate, evaluated at equilibrium quantities in the absence of an exam. For example, if the instrument leads banks to originate 10 additional loans and we observe 3 additional defaults, then the 2SLS estimator would estimate a default rate of 30\% among the marginal loan applicants. As in the first stage, we weight by the bank's origination share in a given geography, and standard errors are clustered at the geography level.

We focus on default at a 24-month horizon after origination, where default is defined  as three consecutive monthly payments below the minimum due. This is the same default outcome that standard credit scores are designed to predict \citep{board2007report}.

\subsection{Marginal Default Rates}  \label{sec:estimation_results:results}

Table \ref{table_secondstage_refi_mc} shows our second stage estimates, separately for different consumer types. The coefficients can be interpreted as default rates for the marginal borrower. We present the 2SLS estimates of \emph{marginal} default rates alongside standard OLS estimates of \emph{average} default rates to show both moments of the default-probability distribution that we later target in model estimation.

Our baseline estimates suggest that marginally approved applicants within each group have between 1 and 6 p.p. higher default rates than average borrowers within that group.\footnote{Tables \ref{table_secondstage_uw_robustness_refi_mc} and \ref{table_secondstage_90plus_robustness_refi_mc} show that second stage results are qualitatively similar and overall robust to changes in weights (unweighted) and the definition of default used (including only cases of 90+ day delinquencies).} This is consistent with the screening process specified in our structural model, where approval decisions are driven by lenders' assessments of risk so that the riskiest borrowers are those closest to the approval/rejection margin.

The table also shows that marginal minority and marginal non-minority borrowers have approximately equal risk, defaulting with 6.9\% and 6.3\% probability respectively. This equality of marginal risk across groups is not the focus of our analysis but shows non-rejection of a \cite{Becker1957} test for taste-based discrimination in loan underwriting.\footnote{Other recent work suggesting a limited role for taste-based discrimination in the modern mortgage market includes \cite{gerardi2020mortgage}, \cite{Bartlett2019}, and \cite{bhutta2020minorities}.} This lack of bank bias is again consistent with the screening process specified in our structural model.

\subsection{Model Estimation Results} \label{sec:estimation_results:modelparams}

We use these coefficients and other moments of the data outlined in Section \ref{sec:model:estimation} to estimate our model of mortgage lenders screening applicants using noisy credit scores and other signals. The model's target moments are presented together in Table \ref{tab:moments}, separately by applicant income and by applicant minority status. Consistent with the analysis of low-income mortgage applicants in our reduced-form work, our low-income category here includes the bottom quartile of applicants by income; we then include all other quartiles in the higher-income category.

Table \ref{tab:modelestimates} presents a summary of model parameter estimates. We find that mortgage applicants differ substantially in how underlying default risk is distributed for higher- and low-income applicants as well as for minority and non-minority applicants; low-income and minority applicants both have higher average default risk (lower average $\theta$) and lower dispersion in default types. Meanwhile, we also find that low-income and minority applicants have substantially lower precision of credit score information: the estimated standard deviation of credit score noise is 2.2 times higher than that of minority applicants in particular. While these estimates of precision differences appear substantial, the model provides a framework for assessing the consequences of these differences in economic terms. These differences can be large or small depending on other differences across groups in terms of either the precision of other signals available to mortgage lenders or the underlying distribution of default risk. We use the model to turn to that quantification exercise next.

\subsection{The Costs of Noise} \label{sec:estimation_results:CFs}

This section uses the estimated model as a framework to assess the economic costs of credit score noise. To do so, we introduce counterfactual information structures in the model that change the level or role of such signal noise, and we then solve for counterfactual mortgage lending decisions. These exercises hold constant other model parameters that we estimated above, for example the distribution of applicants' risk types, in order to isolate the effects of signal precision differences in a precise \textit{ceteris paribus} sense.

We study two counterfactual information structures in particular. First, we study the removal of other, non-credit-score information from the market, in order to assess the role of non-credit-score information in either amplifying or reducing differences in information across groups of applicants. Second, we study an increase in the precision of the credit score signal for disadvantaged groups, in order to assess how mortgage market outcomes would change if credit score precision differences were reduced or eliminated.

For each counterfactual, we measure how changes in information affect approval rates and the efficiency of credit allocation. The measure of efficiency we focus on is an ex-ante full-information benchmark: which applicants would be approved for mortgages if lenders knew each applicant's probability of default perfectly (hence ``full information"), but did not know future realizations of this risk (hence ``ex-ante"). For example, applicants with a 2\% probability of mortgage default would be known to have a 2\% probability, but lenders would not foresee which of these applicants would in fact experience such a default. 

We define two types of errors that allow us to characterize deviations from the ex-ante efficiency benchmark. Type I errors refer to the case where an applicant's true risk type $\theta$ exceeds the lenders' threshold but the applicant is not approved (due to a low draw of signal noise). Type II errors refer to the case where the applicant's true risk type $\theta$ is below the lenders' approval threshold but the applicant is approved (due to a high draw of signal noise). 

Allocative efficiency of this form can be studied in our model in two steps. First, we recover profit function parameters in equation \ref{eqn:Pi_max_basic} that are consistent with a zero-profit condition or a perfectly competitive market.\footnote{Whereas a growing body of evidence highlights the importance of imperfect competition in the US mortgage market \citep{buchak2018beyond, buchak2021mortgage, jiang2020shadowbanking}, we focus on the perfectly competitive case in order to simplify exposition and focus on changes in information structure in a benchmark case.} To do so, we calibrate the loss-given-default parameter to $ \gamma=40 \% $, following \cite{an2017regime}, and we solve for other profit function parameters that yield zero expected profit given our other parameter estimates from Section \ref{sec:estimation_results:modelparams}. 

Then, while holding fixed these parameters and the applicant distribution, we change the signals available to lenders and calculate new break-even approval thresholds $\hat{x}$ given these counterfactual information structures. Intuitively, when more (resp. less) precise signals become available to lenders, lenders' posteriors about applicant risk become more (resp. less) spread out from their priors, and approval thresholds adjust in response to higher or lower expected risk among approved loans. Under these counterfactual thresholds $\hat{x}$, we then quantify what share of approved loans had types $\theta$ below the threshold and hence should have been rejected under a full-information ex-ante efficient benchmark (Type II error) -- and what share of denied loans had types $\theta$ above the threshold and hence should have been approved (Type I error). 

Table \ref{tab:CFs} presents results from the two counterfactuals described above and also characterizes efficiency in the baseline (no-counterfactual) case. The first panel of the table shows the baseline case. We find that misallocation is substantially higher at baseline for both low-income and minority mortgage applicants, especially in terms of Type I errors (inefficient rejections). Partly reflecting these Type I errors, approval rates are significantly lower for low-income and minority applicants: approximately 35\% of applicants in each disadvantaged group are approved, in contrast with approval rates over 50\% for higher income and non-minority applicants.

The bottom two panels of the table then show how approval rates and misallocation change under counterfactual information structures. First, in the middle panel we study a counterfactual where we remove non-credit-score information, so that approval decisions are made based on credit score alone. This change primarily increases disparities in Type I error rates (inefficient rejections) between groups. In other words, the use of non-credit-score information in the actual (``baseline") mortgage market helps reduce disadvantaged groups' inefficient exclusion from credit markets, relative to a counterfactual where underwriting uses credit scores alone.

In our second counterfactual (the third panel of the table) we study the effects of equalizing credit score precision across groups. We implement this counterfactual by increasing the precision of (i.e., decreasing the variance of noise in) the credit score signal for low-income and minority applicants, to be equal to the credit score signal precision of higher-income and non-minority applicants respectively. Despite baseline differences across  groups in applicant risk, this improvement in information substantially reduces disparities in both approval rates and allocative efficiency for minority mortgage applicants. Minority applicants' approval rates increase by 8 p.p., eliminating about 50\% of the baseline gap in approval rates relative to non-minority applicants. Nearly half of this gain comes from the reduction in minority applicants' Type I error rates (inefficient rejections). In contrast, for low-income applicants, the reduction in approval rate disparities is more modest -- 0.8 p.p. -- partly reflecting an efficiency effect where Type II error rates (inefficient approvals) fall as credit score precision improves.

It is worth emphasizing that this second counterfactual does not change other patterns of disadvantage for low-income and minority applicants that contribute to differences in underlying default risk -- labor market discrimination, for example -- so the results of this counterfactual highlight the potential benefits of reducing signal noise alone. These results motivate the importance of, and potential benefits from, changes such as expansion and innovation in credit report data, which can provide disadvantaged groups with greater ability to show their quality as borrowers.


\section{Conclusion}\label{conclusion}

Consumer credit scores are increasingly prevalent in markets with asymmetric information. Insurance, rental housing, labor, and lending markets rely on these scores when allocating insurance coverage, housing, employment, or credit. Moreover, recent advances in machine learning and artificial intelligence suggest that the economic roles of credit scores -- and of similar automated prediction tools -- will only grow. These facts make it crucial to understand how properties of credit scores affect allocative efficiency, and to understand how credit scores could thus contribute to disparities in outcomes across social groups.

In this paper, we use consumer lending and especially the US mortgage market as a setting to study these questions. We have three broad findings: (1) statistical noise in credit scores is greater for historically disadvantaged consumers, in particular minority and low-income individuals; (2) the resulting information disparity faced by lenders is a quantitatively important driver of disparities in credit access and credit misallocation across social groups; and (3) this information disparity is primarily due to features of the underlying credit report data, not due to features of scoring algorithms themselves. 

Together these findings point to the potential for credit score hysteresis in the tradition of \cite{blanchard1986hysteresis}: credit scores may exhibit path dependence whereby disparities in credit report data and score precision drive disparities in credit access and credit misallocation, which may then create persistent disparities in future credit report data. Precisely because we find the differential noise in credit scores across groups is due to features of the underlying credit report data, innovations in credit scoring alone may be insufficient to address these issues. Rather, changes to different groups' opportunities to generate credit report data may be necessary.

We show these results using complementary reduced-form and structural work. In our reduced-form work, we quantify the lower fit of widely used credit scores as predictors of default among minority and low-income consumers: a gap of 7-9\% in terms of AUC or 16-18\% in terms of R2.\footnote{As emphasized in footnote \ref{foot:vantagethroughout}, throughout the paper we study the VantageScore 3.0 credit score.} We replicate these patterns using advanced machine learning models and use these models to assess the roles of credit scoring (modeling bias) versus credit report data (data bias) in driving lower model fit for disadvantaged groups. We find the role of modeling bias is negligible. In contrast, we document that approximately half of the gap in credit scores' overall explanatory power, and at least half of data bias, can be explained by observable features of credit report data, such as data sparsity.

Our structural work helps us quantify the economic importance of these information disparities. We use a model of competitive lenders using heterogeneous information sources to screen loan applicants. Properties of information sources in the model are identified off of observed differences between marginal and average borrowers, which reveal the overall precision of lenders' information, and off of reverse-regression tests, which reveal the precision of credit scores in particular. The model suggests that the standard deviation of credit score noise is approximately twice as high for minority relative to non-minority consumers, and that this disparity in information drives roughly half of estimated disparities in credit access and credit misallocation. While this model focuses on particular forces in the US mortgage market, it helps contextualize the importance of our findings, and the importance of future work to understand how credit scores determine allocations in an increasing share of information-intensive markets.


\clearpage
\newpage
\bibliography{draft_tu_bib.bib}

\begin{thebibliography}{108}
\newcommand{\enquote}[1]{``#1''}
\expandafter\ifx\csname natexlab\endcsname\relax\def\natexlab#1{#1}\fi

\bibitem[\protect\citeauthoryear{Adams, Einav, and Levin}{Adams
  et~al.}{2009}]{adams2009liquidity}
\textsc{Adams, W., L.~Einav, and J.~Levin} (2009): \enquote{Liquidity
  constraints and imperfect information in subprime lending,} \emph{American
  Economic Review}, 99, 49--84.

\bibitem[\protect\citeauthoryear{Agan and Starr}{Agan and
  Starr}{2018}]{agan2018ban}
\textsc{Agan, A. and S.~Starr} (2018): \enquote{Ban the box, criminal records,
  and racial discrimination: A field experiment,} \emph{The Quarterly Journal
  of Economics}, 133, 191--235.

\bibitem[\protect\citeauthoryear{Agarwal, Benmelech, Bergman, and Seru}{Agarwal
  et~al.}{2012}]{Agarwal2012}
\textsc{Agarwal, S., E.~Benmelech, N.~Bergman, and A.~Seru} (2012):
  \enquote{Did the Community Reinvestment Act (CRA) Lead to Risky Lending?}
  \emph{NBER Working Paper No., 18609}.

\bibitem[\protect\citeauthoryear{Aigner and Cain}{Aigner and
  Cain}{1977}]{Aigner1977}
\textsc{Aigner, D.~J. and G.~G. Cain} (1977): \enquote{Statistical Theories of
  Discrimination in Labor Markets,} \emph{Industrial and Labor Relations
  Review}, 30, 175--187.

\bibitem[\protect\citeauthoryear{An and Cordell}{An and
  Cordell}{2017}]{an2017regime}
\textsc{An, X. and L.~Cordell} (2017): \enquote{Regime shift and the
  post-crisis world of mortgage loss severities,} .

\bibitem[\protect\citeauthoryear{Angrist and Pischke}{Angrist and
  Pischke}{2008}]{Angrist2008}
\textsc{Angrist, J.~D. and J.-S. Pischke} (2008): \emph{Mostly Harmless
  Econometrics: An Empiricistís Companion}, Princeton, NJ: Princeton Univ.
  Press.

\bibitem[\protect\citeauthoryear{Arnold, Dobbie, and Yang}{Arnold
  et~al.}{2018}]{arnold2018racial}
\textsc{Arnold, D., W.~Dobbie, and C.~S. Yang} (2018): \enquote{Racial bias in
  bail decisions,} \emph{The Quarterly Journal of Economics}, 133, 1885--1932.

\bibitem[\protect\citeauthoryear{Arnold, Dobbie, and Hull}{Arnold
  et~al.}{2020{\natexlab{a}}}]{arnold2020algodiscrim}
\textsc{Arnold, D., W.~S. Dobbie, and P.~Hull} (2020{\natexlab{a}}):
  \enquote{Measuring Racial Discrimination in Algorithms,} Tech. rep.

\bibitem[\protect\citeauthoryear{Arnold, Dobbie, and Hull}{Arnold
  et~al.}{2020{\natexlab{b}}}]{arnold2020measuring}
---\hspace{-.1pt}---\hspace{-.1pt}--- (2020{\natexlab{b}}): \enquote{Measuring
  racial discrimination in bail decisions,} Tech. rep., National Bureau of
  Economic Research.

\bibitem[\protect\citeauthoryear{Autor and Scarborough}{Autor and
  Scarborough}{2008}]{autor2008does}
\textsc{Autor, D.~H. and D.~Scarborough} (2008): \enquote{Does job testing harm
  minority workers? Evidence from retail establishments,} \emph{The Quarterly
  Journal of Economics}, 123, 219--277.

\bibitem[\protect\citeauthoryear{Avery, Bostic, Calem, and Canner}{Avery
  et~al.}{1996}]{avery1996}
\textsc{Avery, R., R.~Bostic, P.~Calem, and G.~Canner} (1996): \enquote{Credit
  Risk, Credit Scoring and the Performance of Home Mortgages,} \emph{Federal
  Reserve Bulletin}, 82, 621--648.

\bibitem[\protect\citeauthoryear{Avery, Bostic, and Canner}{Avery
  et~al.}{2005}]{avery2005assessing}
\textsc{Avery, R.~B., R.~W. Bostic, and G.~B. Canner} (2005):
  \enquote{Assessing the necessity and efficiency of the Community Reinvestment
  Act,} \emph{Housing Policy Debate}, 16, 143--172.

\bibitem[\protect\citeauthoryear{Avery and Brevoort}{Avery and
  Brevoort}{2015}]{avery2015subprime}
\textsc{Avery, R.~B. and K.~P. Brevoort} (2015): \enquote{The subprime crisis:
  Is government housing policy to blame?} \emph{Review of Economics and
  Statistics}, 97, 352--363.

\bibitem[\protect\citeauthoryear{Avery, Brevoort, and Canner}{Avery
  et~al.}{2012}]{avery2012does}
\textsc{Avery, R.~B., K.~P. Brevoort, and G.~Canner} (2012): \enquote{Does
  Credit Scoring Produce a Disparate Impact?} \emph{Real Estate Economics}, 40,
  S65--S114.

\bibitem[\protect\citeauthoryear{Avery, Brevoort, and Canner}{Avery
  et~al.}{2009}]{avery2009credit}
\textsc{Avery, R.~B., K.~P. Brevoort, and G.~B. Canner} (2009): \enquote{Credit
  scoring and its effects on the availability and affordability of credit,}
  \emph{Journal of Consumer Affairs}, 43, 516--537.

\bibitem[\protect\citeauthoryear{Avery, Calem, Canner, and Bostic}{Avery
  et~al.}{2003}]{avery2003overview}
\textsc{Avery, R.~B., P.~S. Calem, G.~B. Canner, and R.~W. Bostic} (2003):
  \enquote{An overview of consumer data and credit reporting,} \emph{Fed. Res.
  Bull.}, 89, 47.

\bibitem[\protect\citeauthoryear{Barocas, Hardt, and Narayanan}{Barocas
  et~al.}{2019}]{barocas2020}
\textsc{Barocas, S., M.~Hardt, and A.~Narayanan} (2019): \emph{Fairness and
  Machine Learning}, fairmlbook.org, \url{http://www.fairmlbook.org}.

\bibitem[\protect\citeauthoryear{Barocas and Selbst}{Barocas and
  Selbst}{2016}]{barocas2016}
\textsc{Barocas, S. and A.~D. Selbst} (2016): \enquote{Big Data's Disparate
  Impact,} \emph{California Law Review}, 104, 671--732.

\bibitem[\protect\citeauthoryear{Bartik and Nelson}{Bartik and
  Nelson}{2020}]{bartik2020deleting}
\textsc{Bartik, A.~W. and S.~Nelson} (2020): \enquote{Deleting a Signal:
  Evidence from Pre-Employment Credit Checks,} Working paper, Becker Friedman
  Institute.

\bibitem[\protect\citeauthoryear{Bartlett, Morse, Stanton, and
  Wallace}{Bartlett et~al.}{2019}]{Bartlett2019}
\textsc{Bartlett, R., A.~Morse, R.~Stanton, and N.~Wallace} (2019):
  \enquote{{Consumer-Lending Discrimination in the FinTech Era},} NBER Working
  Papers 25943, National Bureau of Economic Research, Inc.

\bibitem[\protect\citeauthoryear{Bartlett, Morse, Wallace, and
  Stanton}{Bartlett et~al.}{2020}]{bartlett_algorithmic_2020}
\textsc{Bartlett, R.~P., A.~Morse, N.~Wallace, and R.~Stanton} (2020):
  \enquote{Algorithmic {Discrimination} and {Input} {Accountability} under the
  {Civil} {Rights} {Acts},} {SSRN} {Scholarly} {Paper} ID 3674665, Social
  Science Research Network, Rochester, NY.

\bibitem[\protect\citeauthoryear{Bayer, Ferreira, and Ross}{Bayer
  et~al.}{2018}]{bayer2018drives}
\textsc{Bayer, P., F.~Ferreira, and S.~L. Ross} (2018): \enquote{What drives
  racial and ethnic differences in high-cost mortgages? The role of high-risk
  lenders,} \emph{The Review of Financial Studies}, 31, 175--205.

\bibitem[\protect\citeauthoryear{Becker}{Becker}{1957}]{Becker1957}
\textsc{Becker, G.~S.} (1957): \emph{The Economics of Discrimination}, Chicago:
  University of Chicago Press.

\bibitem[\protect\citeauthoryear{Bernstein, Diamond, McQuade, Pousada
  et~al.}{Bernstein et~al.}{2019}]{bernstein2019contribution}
\textsc{Bernstein, S., R.~Diamond, T.~McQuade, B.~Pousada, et~al.} (2019):
  \enquote{The contribution of high-skilled immigrants to innovation in the
  United States,} Tech. rep.

\bibitem[\protect\citeauthoryear{Bhutta}{Bhutta}{2011}]{bhutta2011community}
\textsc{Bhutta, N.} (2011): \enquote{The Community Reinvestment Act and
  mortgage lending to lower income borrowers and neighborhoods,} \emph{The
  Journal of Law and Economics}, 54, 953--983.

\bibitem[\protect\citeauthoryear{Bhutta}{Bhutta}{2014}]{bhutta_payday_2014}
---\hspace{-.1pt}---\hspace{-.1pt}--- (2014): \enquote{Payday loans and
  consumer financial health,} \emph{Journal of Banking \& Finance}, 47,
  230--242.

\bibitem[\protect\citeauthoryear{Bhutta and Hizmo}{Bhutta and
  Hizmo}{2020}]{bhutta2020minorities}
\textsc{Bhutta, N. and A.~Hizmo} (2020): \enquote{Do minorities pay more for
  mortgages?} .

\bibitem[\protect\citeauthoryear{Bhutta, Ringo et~al.}{Bhutta
  et~al.}{2015{\natexlab{a}}}]{bhutta2015assessing}
\textsc{Bhutta, N., D.~R. Ringo, et~al.} (2015{\natexlab{a}}):
  \enquote{Assessing the Community Reinvestment Act's Role in the Financial
  Crisis,} Tech. rep., Board of Governors of the Federal Reserve System (US).

\bibitem[\protect\citeauthoryear{Bhutta, Skiba, and Tobacman}{Bhutta
  et~al.}{2015{\natexlab{b}}}]{bhutta2015payday}
\textsc{Bhutta, N., P.~M. Skiba, and J.~Tobacman} (2015{\natexlab{b}}):
  \enquote{Payday loan choices and consequences,} \emph{Journal of Money,
  Credit and Banking}, 47, 223--260.

\bibitem[\protect\citeauthoryear{Bils}{Bils}{1985}]{bils1985real}
\textsc{Bils, M.~J.} (1985): \enquote{Real wages over the business cycle:
  evidence from panel data,} \emph{Journal of Political economy}, 93, 666--689.

\bibitem[\protect\citeauthoryear{Black, Berger, and Scott}{Black
  et~al.}{2000}]{black2000bounding}
\textsc{Black, D.~A., M.~C. Berger, and F.~A. Scott} (2000): \enquote{Bounding
  parameter estimates with nonclassical measurement error,} \emph{Journal of
  the American Statistical Association}, 95, 739--748.

\bibitem[\protect\citeauthoryear{Blanchard and Summers}{Blanchard and
  Summers}{1986}]{blanchard1986hysteresis}
\textsc{Blanchard, O.~J. and L.~H. Summers} (1986): \enquote{Hysteresis and the
  European unemployment problem,} \emph{NBER macroeconomics annual}, 1, 15--78.

\bibitem[\protect\citeauthoryear{Blinder}{Blinder}{1973}]{blinder1973wage}
\textsc{Blinder, A.~S.} (1973): \enquote{Wage discrimination: reduced form and
  structural estimates,} \emph{Journal of Human resources}, 436--455.

\bibitem[\protect\citeauthoryear{Blum and Stangl}{Blum and
  Stangl}{2020}]{BlumStangel2020}
\textsc{Blum, A. and K.~Stangl} (2020): \enquote{Recovering from {{Biased
  Data}}: {{Can Fairness Constraints Improve Accuracy}}?} \emph{Working paper}.

\bibitem[\protect\citeauthoryear{Buchak and J{\o}rring}{Buchak and
  J{\o}rring}{2021}]{buchak2021mortgage}
\textsc{Buchak, G. and A.~J{\o}rring} (2021): \enquote{Do Mortgage Lenders
  Compete Locally? Evidence Beyond Interest Rates,} \emph{Evidence Beyond
  Interest Rates (January 7, 2021)}.

\bibitem[\protect\citeauthoryear{Buchak, Matvos, Piskorski, and Seru}{Buchak
  et~al.}{2018}]{buchak2018beyond}
\textsc{Buchak, G., G.~Matvos, T.~Piskorski, and A.~Seru} (2018):
  \enquote{Beyond the balance sheet model of banking: Implications for bank
  regulation and monetary policy,} Tech. rep., National Bureau of Economic
  Research.

\bibitem[\protect\citeauthoryear{Butler, Mayer, and Weston}{Butler
  et~al.}{2020}]{butler2020racial}
\textsc{Butler, A.~W., E.~J. Mayer, and J.~Weston} (2020): \enquote{Racial
  Discrimination in the Auto Loan Market,} \emph{Available at SSRN 3301009}.

\bibitem[\protect\citeauthoryear{Canay, Mogstad, and Mountjoy}{Canay
  et~al.}{2020}]{canay2020use}
\textsc{Canay, I.~A., M.~Mogstad, and J.~Mountjoy} (2020): \enquote{On the use
  of outcome tests for detecting bias in decision making,} \emph{NBER working
  paper}.

\bibitem[\protect\citeauthoryear{Carrell and Zinman}{Carrell and
  Zinman}{2014}]{carrell_harms_2014}
\textsc{Carrell, S. and J.~Zinman} (2014): \enquote{In {Harm}'s {Way}? {Payday}
  {Loan} {Access} and {Military} {Personnel} {Performance},} \emph{The Review
  of Financial Studies}, 27, 2805--2840.

\bibitem[\protect\citeauthoryear{CFPB}{CFPB}{2014}]{CFPB2014ethn}
\textsc{CFPB} (2014): \enquote{Using publicly available information to proxy
  for unidentified race and ethnicity,} .

\bibitem[\protect\citeauthoryear{Chan and Gentzkow}{Chan and
  Gentzkow}{2020}]{chan2020selection}
\textsc{Chan, D.~C. and M.~Gentzkow} (2020): \enquote{Selection with variation
  in diagnostic skill: Evidence from radiologists,} Tech. rep., National Bureau
  of Economic Research.

\bibitem[\protect\citeauthoryear{Charles and Hurst}{Charles and
  Hurst}{2002}]{charleshurstReStat}
\textsc{Charles, K.~K. and E.~Hurst} (2002): \enquote{The transition to home
  ownership and the black-white wealth gap,} \emph{Review of Economics and
  Statistics}, 84, 281--297.

\bibitem[\protect\citeauthoryear{Charles and Hurst}{Charles and
  Hurst}{2003}]{charleshurstwealth}
---\hspace{-.1pt}---\hspace{-.1pt}--- (2003): \enquote{The correlation of
  wealth across generations,} \emph{Journal of political Economy}, 111,
  1155--1182.

\bibitem[\protect\citeauthoryear{Chatterjee, Corbae, Dempsey, and
  R{\'\i}os-Rull}{Chatterjee et~al.}{2020}]{CCDR}
\textsc{Chatterjee, S., D.~Corbae, K.~P. Dempsey, and J.-V. R{\'\i}os-Rull}
  (2020): \enquote{A quantitative theory of the credit score,} Tech. rep.,
  National Bureau of Economic Research.

\bibitem[\protect\citeauthoryear{Cherry, Jiang, Matvos, Piskorski, and
  Seru}{Cherry et~al.}{2021}]{cherry_government_2021}
\textsc{Cherry, S., E.~X. Jiang, G.~Matvos, T.~Piskorski, and A.~Seru} (2021):
  \enquote{Government and {Private} {Household} {Debt} {Relief} during
  {COVID}-19,} .

\bibitem[\protect\citeauthoryear{Corbae and Glover}{Corbae and
  Glover}{2018}]{corbae2018employer}
\textsc{Corbae, D. and A.~Glover} (2018): \enquote{Employer credit checks:
  Poverty traps versus matching efficiency,} Tech. rep., National Bureau of
  Economic Research.

\bibitem[\protect\citeauthoryear{Coston, Rambachan, and Chouldechova}{Coston
  et~al.}{2021}]{coston_characterizing_2021}
\textsc{Coston, A., A.~Rambachan, and A.~Chouldechova} (2021):
  \enquote{Characterizing {Fairness} {Over} the {Set} of {Good} {Models}
  {Under} {Selective} {Labels},} \emph{arXiv:2101.00352 [cs, stat]}, arXiv:
  2101.00352.

\bibitem[\protect\citeauthoryear{Cowgill and Tucker}{Cowgill and
  Tucker}{2020}]{cowgill2019economics}
\textsc{Cowgill, B. and C.~E. Tucker} (2020): \enquote{Algorithmic Fairness and
  Economics,} .

\bibitem[\protect\citeauthoryear{Diamond, Guren, and Tan}{Diamond
  et~al.}{2020}]{diamond2020effect}
\textsc{Diamond, R., A.~M. Guren, and R.~Tan} (2020): \enquote{The effect of
  foreclosures on homeowners, tenants, and landlords,} \emph{NBER Working
  Paper}.

\bibitem[\protect\citeauthoryear{Diamond, McQuade, and Qian}{Diamond
  et~al.}{2019}]{diamond2019effects}
\textsc{Diamond, R., T.~McQuade, and F.~Qian} (2019): \enquote{The effects of
  rent control expansion on tenants, landlords, and inequality: Evidence from
  San Francisco,} \emph{American Economic Review}, 109, 3365--94.

\bibitem[\protect\citeauthoryear{Dobbie, Liberman, Paravisini, and
  Pathania}{Dobbie et~al.}{2019}]{Dobbie2019}
\textsc{Dobbie, W., A.~Liberman, D.~Paravisini, and V.~Pathania} (2019):
  \enquote{Measuring Bias in Consumer Lending,} Working Paper 24953, National
  Bureau of Economic Research.

\bibitem[\protect\citeauthoryear{Doleac and Hansen}{Doleac and
  Hansen}{2020}]{Doleac2017}
\textsc{Doleac, J. and B.~Hansen} (2020): \enquote{{Does ``Ban-the-Box" help or
  hurt low-skilled workers? Statistical Discrimination and Employment Outcomes
  When Criminal Histories are Hidden},} \emph{Journal of Labor Economics}.

\bibitem[\protect\citeauthoryear{Einav, Jenkins, and Levin}{Einav
  et~al.}{2013}]{einav2013impact}
\textsc{Einav, L., M.~Jenkins, and J.~Levin} (2013): \enquote{The impact of
  credit scoring on consumer lending,} \emph{The RAND Journal of Economics},
  44, 249--274.

\bibitem[\protect\citeauthoryear{Elliott, Peter~Morrison, and Luri}{Elliott
  et~al.}{2009}]{census2009}
\textsc{Elliott, M., P.~P. Peter~Morrison, Allen~Fremont, and N.~Luri} (2009):
  \enquote{Using the Census Bureau’s surname list to improve estimates of
  race/ethnicity and associated disparities.} \emph{Health Serv Outcomes Res
  Methodn}, 9, 69--83.

\bibitem[\protect\citeauthoryear{{Federal Reserve Board}}{{Federal Reserve
  Board}}{2007}]{board2007report}
\textsc{{Federal Reserve Board}} (2007): \emph{Report to the Congress on Credit
  Scoring and its Effects on the Availability and Affordability of Credit},
  Federal Reserve Board of Governors.

\bibitem[\protect\citeauthoryear{{Federal Reserve Board}}{{Federal Reserve
  Board}}{2018{\natexlab{a}}}]{CRAdescription2018}
---\hspace{-.1pt}---\hspace{-.1pt}--- (2018{\natexlab{a}}): \enquote{The
  Community Reinvestment Act,} .

\bibitem[\protect\citeauthoryear{{Federal Reserve Board}}{{Federal Reserve
  Board}}{2018{\natexlab{b}}}]{FedCRA2018}
---\hspace{-.1pt}---\hspace{-.1pt}--- (2018{\natexlab{b}}): \enquote{Consumers
  \& Communities: Community Reinvestment Act (CRA),} Tech. rep.

\bibitem[\protect\citeauthoryear{{Federal Trade Commission}}{{Federal Trade
  Commission}}{2013}]{FTCerrorsreport}
\textsc{{Federal Trade Commission}} (2013): \enquote{Report to Congress Under
  Section 319 of the Fair and Accurate Credit Transactions Act of 2003,} .

\bibitem[\protect\citeauthoryear{Foote, Gerardi, and Willen}{Foote
  et~al.}{2013}]{foote2013government}
\textsc{Foote, C., K.~Gerardi, and P.~Willen} (2013): \enquote{Government
  policy and the crisis: The case of the Community Reinvestment Act,}
  \emph{Real Estate Research Blog, Federal Reserve Bank of Atlanta, August}, 1.

\bibitem[\protect\citeauthoryear{Fuster, Goldsmith-Pinkham, Ramadorai, and
  Walther}{Fuster et~al.}{2020}]{Fuster2020}
\textsc{Fuster, A., P.~Goldsmith-Pinkham, T.~Ramadorai, and A.~Walther} (2020):
  \enquote{Predictably Unequal? The Effects of Machine Learning on Credit
  Markets,} .

\bibitem[\protect\citeauthoryear{Ganong and Noel}{Ganong and
  Noel}{2020}]{ganong_why_2020}
\textsc{Ganong, P. and P.~Noel} (2020): \enquote{Why {Do} {Borrowers} {Default}
  on {Mortgages}? {A} {New} {Method} {For} {Causal} {Attribution},} .

\bibitem[\protect\citeauthoryear{Gerardi, Willen, Zhang et~al.}{Gerardi
  et~al.}{2020}]{gerardi2020mortgage}
\textsc{Gerardi, K., P.~Willen, D.~H. Zhang, et~al.} (2020): \enquote{Mortgage
  Prepayment, Race, and Monetary Policy,} Tech. rep.

\bibitem[\protect\citeauthoryear{Gillis and Spiess}{Gillis and
  Spiess}{2019}]{gillis_big_2019}
\textsc{Gillis, T. and J.~Spiess} (2019): \enquote{Big {Data} and
  {Discrimination},} \emph{University of Chicago Law Review}, 86, 00027.

\bibitem[\protect\citeauthoryear{Gillis}{Gillis}{2020}]{gillis_false_2020}
\textsc{Gillis, T.~B.} (2020): \enquote{False {Dreams} of {Algorithmic}
  {Fairness}: {The} {Case} of {Credit} {Pricing},} {SSRN} {Scholarly} {Paper}
  ID 3571266, Social Science Research Network, Rochester, NY.

\bibitem[\protect\citeauthoryear{Guvenen, Schulhofer-Wohl, Song, and
  Yogo}{Guvenen et~al.}{2017}]{guvenen2017worker}
\textsc{Guvenen, F., S.~Schulhofer-Wohl, J.~Song, and M.~Yogo} (2017):
  \enquote{Worker betas: Five facts about systematic earnings risk,}
  \emph{American Economic Review}, 107, 398--403.

\bibitem[\protect\citeauthoryear{Hanson, Hawley, Martin, and Liu}{Hanson
  et~al.}{2016}]{Hanson2016}
\textsc{Hanson, A., Z.~Hawley, H.~Martin, and B.~Liu} (2016):
  \enquote{Discrimination in mortgage lending: Evidence from a correspondence
  experiment,} \emph{Journal of Urban Economics}, 92, 48 -- 65.

\bibitem[\protect\citeauthoryear{Heimer and Yu}{Heimer and
  Yu}{2021}]{heimer2021using}
\textsc{Heimer, R. and E.~Yu} (2021): \enquote{Using High-Frequency Evaluations
  to Estimate Discrimination: Evidence from Mortgage Loan Officers,} Tech.
  rep., Federal Reserve Bank of Philadelphia.

\bibitem[\protect\citeauthoryear{Hertzberg, Liberti, and Paravisini}{Hertzberg
  et~al.}{2011}]{hertzberg2011public}
\textsc{Hertzberg, A., J.~M. Liberti, and D.~Paravisini} (2011):
  \enquote{Public information and coordination: evidence from a credit registry
  expansion,} \emph{The Journal of Finance}, 66, 379--412.

\bibitem[\protect\citeauthoryear{Hurst, Keys, Seru, and Vavra}{Hurst
  et~al.}{2016}]{KeysSeru2016}
\textsc{Hurst, E., B.~J. Keys, A.~Seru, and J.~Vavra} (2016): \enquote{Regional
  Redistribution through the US Mortgage Market,} \emph{American Economic
  Review}, 106, 2982--3028.

\bibitem[\protect\citeauthoryear{Indarte}{Indarte}{2020}]{indarte_moral_2020}
\textsc{Indarte, S.} (2020): \enquote{Moral {Hazard} versus {Liquidity} in
  {Household} {Bankruptcy},} .

\bibitem[\protect\citeauthoryear{Jiang}{Jiang}{2020}]{jiang2020shadowbanking}
\textsc{Jiang, E.} (2020): \enquote{Financing Competitors: Shadow Banks'
  Funding and Mortgage Market Competition,} \emph{Unpublished working paper}.

\bibitem[\protect\citeauthoryear{Kearns and Roth}{Kearns and
  Roth}{2019}]{KearnsRoth2019}
\textsc{Kearns, M. and A.~Roth} (2019): \emph{The Ethical Algorithm}, Oxford,
  UK: Oxford University Press.

\bibitem[\protect\citeauthoryear{Keys, Mukherjee, Vig, and Seru}{Keys
  et~al.}{2010}]{keys2010}
\textsc{Keys, B., T.~Mukherjee, V.~Vig, and A.~Seru} (2010): \enquote{Did
  Securitization Lead to Lax Screening? Evidence From Subprime Loans,}
  \emph{Quarterly Journal of Economics}, 125, 307--362.

\bibitem[\protect\citeauthoryear{Keys, Seru, and Vig}{Keys
  et~al.}{2012}]{keys2012lender}
\textsc{Keys, B.~J., A.~Seru, and V.~Vig} (2012): \enquote{Lender screening and
  the role of securitization: evidence from prime and subprime mortgage
  markets,} \emph{The Review of Financial Studies}, 25, 2071--2108.

\bibitem[\protect\citeauthoryear{Kitagawa}{Kitagawa}{1955}]{kitagawa1955components}
\textsc{Kitagawa, E.~M.} (1955): \enquote{Components of a difference between
  two rates,} \emph{Journal of the american statistical association}, 50,
  1168--1194.

\bibitem[\protect\citeauthoryear{Kleinberg, Ludwig, Mullainathan, and
  Rambachan}{Kleinberg et~al.}{2018}]{RambachanLudwig2018}
\textsc{Kleinberg, J., J.~Ludwig, S.~Mullainathan, and A.~Rambachan} (2018):
  \enquote{Algorithmic Fairness,} \emph{AEA Papers and Proceedings}, 108,
  22--27.

\bibitem[\protect\citeauthoryear{Kuhn, Schularick, and Steins}{Kuhn
  et~al.}{2020}]{kuhn2020income}
\textsc{Kuhn, M., M.~Schularick, and U.~I. Steins} (2020): \enquote{Income and
  wealth inequality in America, 1949--2016,} \emph{Journal of Political
  Economy}, 128, 3469--3519.

\bibitem[\protect\citeauthoryear{LaCour-Little and Fortowsky}{LaCour-Little and
  Fortowsky}{2004}]{Mays2004}
\textsc{LaCour-Little, M. and E.~Fortowsky} (2004): \enquote{Credit Scoring and
  the Fair Lending Issue of Disparate Impact,} in \emph{Credit Scoring for Risk
  Managers}, ed. by E.~Mays, Thomson Southwestern.

\bibitem[\protect\citeauthoryear{Lakkaraju, Kleinberg, Leskovec, Ludwig, and
  Mullainathan}{Lakkaraju et~al.}{2017}]{lakkaraju2017selective}
\textsc{Lakkaraju, H., J.~Kleinberg, J.~Leskovec, J.~Ludwig, and
  S.~Mullainathan} (2017): \enquote{The selective labels problem: Evaluating
  algorithmic predictions in the presence of unobservables,} in
  \emph{Proceedings of the 23rd ACM SIGKDD International Conference on
  Knowledge Discovery and Data Mining}, 275--284.

\bibitem[\protect\citeauthoryear{Liberman, Neilson, Opazo, and
  Zimmerman}{Liberman et~al.}{2018}]{liberman2018equilibrium}
\textsc{Liberman, A., C.~Neilson, L.~Opazo, and S.~Zimmerman} (2018):
  \enquote{The equilibrium effects of information deletion: Evidence from
  consumer credit markets,} Tech. rep., National Bureau of Economic Research.

\bibitem[\protect\citeauthoryear{Lundberg and Lee}{Lundberg and
  Lee}{2017}]{lundberg2017unified}
\textsc{Lundberg, S. and S.-I. Lee} (2017): \enquote{A unified approach to
  interpreting model predictions,} \emph{arXiv preprint arXiv:1705.07874}.

\bibitem[\protect\citeauthoryear{Mian and Sufi}{Mian and
  Sufi}{2018}]{mian2018credit}
\textsc{Mian, A. and A.~Sufi} (2018): \enquote{Credit supply and housing
  speculation,} Tech. rep., National Bureau of Economic Research.

\bibitem[\protect\citeauthoryear{Mian, Straub, and Sufi}{Mian
  et~al.}{2020}]{mian2020indebted}
\textsc{Mian, A.~R., L.~Straub, and A.~Sufi} (2020): \enquote{Indebted demand,}
  Tech. rep., National Bureau of Economic Research.

\bibitem[\protect\citeauthoryear{Morse}{Morse}{2011}]{morse_payday_2011}
\textsc{Morse, A.} (2011): \enquote{Payday lenders: {Heroes} or villains?}
  \emph{Journal of Financial Economics}, 102, 28--44.

\bibitem[\protect\citeauthoryear{Nelson}{Nelson}{2020}]{nelson2018private}
\textsc{Nelson, S.~T.} (2020): \enquote{Private Information and Price
  Regulation in the US Credit Card Market,} .

\bibitem[\protect\citeauthoryear{Oaxaca}{Oaxaca}{1973}]{oaxaca1973male}
\textsc{Oaxaca, R.} (1973): \enquote{Male-female wage differentials in urban
  labor markets,} \emph{International economic review}, 693--709.

\bibitem[\protect\citeauthoryear{Patterson et~al.}{Patterson
  et~al.}{2019}]{patterson2019matching}
\textsc{Patterson, C. et~al.} (2019): \enquote{The matching multiplier and the
  amplification of recessions,} \emph{Unpublished Manuscript, Northwestern
  University}.

\bibitem[\protect\citeauthoryear{Qian and Tan}{Qian and
  Tan}{2020}]{qian2020effects}
\textsc{Qian, F. and R.~Tan} (2020): \enquote{The Effects of High-skilled Firm
  Entry on Local Residents,} \emph{Working paper}.

\bibitem[\protect\citeauthoryear{Rajan, Seru, and Vig}{Rajan
  et~al.}{2015}]{RajanSeru2015}
\textsc{Rajan, U., A.~Seru, and V.~Vig} (2015): \enquote{The failure of models
  that predict failure: Distance, incentives, and defaults,} \emph{Journal of
  Financial Economics}, 115, 237--260.

\bibitem[\protect\citeauthoryear{Rambachan, Kleinberg, Mullainathan, and
  Ludwig}{Rambachan et~al.}{2020}]{rambachan_economic_2020}
\textsc{Rambachan, A., J.~Kleinberg, S.~Mullainathan, and J.~Ludwig} (2020):
  \enquote{An {Economic} {Approach} to {Regulating} {Algorithms},} Tech. Rep.
  w27111, National Bureau of Economic Research.

\bibitem[\protect\citeauthoryear{Rambachan and Roth}{Rambachan and
  Roth}{2019}]{rambachan2019}
\textsc{Rambachan, A. and J.~Roth} (2019): \enquote{Bias In, Bias Out?
  Evaluating the Folk Wisdom,} .

\bibitem[\protect\citeauthoryear{Ringo}{Ringo}{2017}]{ringo2017mortgage}
\textsc{Ringo, D.} (2017): \enquote{Mortgage lending, default and the Community
  Reinvestment Act,} \emph{Default and the Community Reinvestment Act (December
  14, 2017)}.

\bibitem[\protect\citeauthoryear{Ross, Turner, Godfrey, and Smith}{Ross
  et~al.}{2008}]{Ross2008}
\textsc{Ross, S.~L., M.~A. Turner, E.~Godfrey, and R.~R. Smith} (2008):
  \enquote{Mortgage lending in Chicago and Los Angeles: A paired testing study
  of the pre-application process,} \emph{Journal of Urban Economics}, 63, 902
  -- 919.

\bibitem[\protect\citeauthoryear{Shin}{Shin}{1994}]{shin1994cyclicality}
\textsc{Shin, D.} (1994): \enquote{Cyclicality of real wages among young men,}
  \emph{Economics Letters}, 46, 137--142.

\bibitem[\protect\citeauthoryear{Siddiqi}{Siddiqi}{2017}]{siddiqi2017intelligent}
\textsc{Siddiqi, N.} (2017): \emph{Intelligent credit scoring: Building and
  implementing better credit risk scorecards}, John Wiley \& Sons.

\bibitem[\protect\citeauthoryear{Solon, Barsky, and Parker}{Solon
  et~al.}{1994}]{solon1994measuring}
\textsc{Solon, G., R.~Barsky, and J.~A. Parker} (1994): \enquote{Measuring the
  cyclicality of real wages: how important is composition bias?} \emph{The
  quarterly journal of economics}, 109, 1--25.

\bibitem[\protect\citeauthoryear{Sood and Laohaprapanon}{Sood and
  Laohaprapanon}{2018}]{sood2018}
\textsc{Sood, G. and S.~Laohaprapanon} (2018): \enquote{Predicting Race and
  Ethnicity From the Sequence of Characters in a Nam,} .

\bibitem[\protect\citeauthoryear{Spriggs}{Spriggs}{2020}]{spriggs2020now}
\textsc{Spriggs, W.} (2020): \enquote{Is Now a Teachable Moment for
  Economists?} \emph{Federal Reserve Bank of Minneapolis, Minnesota}.

\bibitem[\protect\citeauthoryear{Stock and Yogo}{Stock and
  Yogo}{2005}]{stockyogo2005testing}
\textsc{Stock, J.~H. and M.~Yogo} (2005): \enquote{Testing for weak instruments
  in linear IV regression,} \emph{Identification and inference for econometric
  models: Essays in honor of Thomas Rothenberg}, 80, 1.

\bibitem[\protect\citeauthoryear{Suresh and Guttag}{Suresh and
  Guttag}{2020}]{sureshgattag2020}
\textsc{Suresh, H. and J.~V. Guttag} (2020): \enquote{A {Framework} for
  {Understanding} {Unintended} {Consequences} of {Machine} {Learning},}
  \emph{arXiv:1901.10002 [cs, stat]}, 00061 arXiv: 1901.10002.

\bibitem[\protect\citeauthoryear{Thomas}{Thomas}{2009}]{thomas2009consumer}
\textsc{Thomas, L.~C.} (2009): \emph{Consumer credit models: pricing, profit
  and portfolios: pricing, profit and portfolios}, OUP Oxford.

\bibitem[\protect\citeauthoryear{Traub}{Traub}{2013}]{traub2013discredited}
\textsc{Traub, A.} (2013): \enquote{DiscreDiteD,} \emph{How employment credit
  checks keep qualified workers out of a job}, 7.

\bibitem[\protect\citeauthoryear{Wang, Hsu, Diaz, and Calmon}{Wang
  et~al.}{2020}]{wang_split_2020}
\textsc{Wang, H., H.~Hsu, M.~Diaz, and F.~P. Calmon} (2020): \enquote{To
  {Split} or {Not} to {Split}: {The} {Impact} of {Disparate} {Treatment} in
  {Classification},} .

\bibitem[\protect\citeauthoryear{Woodward and Hall}{Woodward and
  Hall}{2012}]{woodward2012diagnosing}
\textsc{Woodward, S.~E. and R.~E. Hall} (2012): \enquote{Diagnosing consumer
  confusion and sub-optimal shopping effort: Theory and mortgage-market
  evidence,} \emph{American Economic Review}, 102, 3249--76.

\bibitem[\protect\citeauthoryear{Wozniak}{Wozniak}{2015}]{Wozniak2015}
\textsc{Wozniak, A.~K.} (2015): \enquote{{Discrimination and the Effects of
  Drug Testing on Black Employment},} \emph{Review of Economics and
  Statistics}, 95, 548--566.

\bibitem[\protect\citeauthoryear{Yang and Dobbie}{Yang and
  Dobbie}{2019}]{yang_equal_2019}
\textsc{Yang, C. and W.~Dobbie} (2019): \enquote{Equal {Protection} {Under}
  {Algorithms}: {A} {New} {Statistical} and {Legal} {Framework},} \emph{SSRN
  Electronic Journal}.

\bibitem[\protect\citeauthoryear{Zafar, Valera, Gomez~Rodriguez, and
  Gummadi}{Zafar et~al.}{2017}]{Zafar2017}
\textsc{Zafar, M.~B., I.~Valera, M.~Gomez~Rodriguez, and K.~P. Gummadi} (2017):
  \enquote{Fairness Beyond Disparate Treatment \& Disparate Impact,}
  \emph{Proceedings of the 26th International Conference on World Wide
  Web’17}.

\bibitem[\protect\citeauthoryear{Zhang and Willen}{Zhang and
  Willen}{2020}]{zhang2020lenders}
\textsc{Zhang, D.~H. and P.~Willen} (2020): \enquote{Do Lenders Still
  Discriminate? A Robust Approach for Assessing Differences in Menus,} .

\end{thebibliography}


\pagebreak \newpage
\FloatBarrier


\centering


\centering


\pagebreak \newpage
\FloatBarrier

\centering
\captionsetup{justification=centering}

\begin{landscape}
	\begin{table}[]
		\caption{Summary Statistics: Originated Mortgages}
		\label{table_summarystats}
		
		\centering
		
		\resizebox{1.25\textwidth}{!}{

\begin{tabular}{lccccc} \toprule
\multicolumn{1}{c}{}               & 1                         & 2                                  & 3                                & 4                                    & 5                               \\ 
                                   & TransUnion + Infutor & Matched CoreLogic address & Matched CL transactions & $<$ 20\% amount difference & Has lender information \\ \midrule
                                   &                  &                           &                         &                             &                        \\
                                   &                           &                                    &                                  &                                      &                                 \\
Minority share                     & 0.12                      & 0.11                               & 0.11                             & 0.11                                 & 0.10                            \\
                                   &                           &                                    &                                  &                                      &                                 \\
Low income share                   & 0.14                      & {0.12}           & 0.13                             & 0.13                                 & 0.13                            \\
                                   &                           &                                    &                                  &                                      &                                 \\
Credit Score                          & 741.00                    & 744.00                             & 746.50                           & 750.25                               & 751.47                          \\
                                   & (65.17)                   & (64.13)                            & (65.83)                          & (62.63)                              & (61.89)                         \\
                                   &                           &                                    &                                  &                                      &                                 \\
Mortgage amount (000s)                   & 219.57                & 219.96                         & 216.95                       & 213.55                           & 212.38                      \\
                                   & (184.40)              & (178.98)                       & (169.18)                     & (161.36)                         & (167.06)                    \\
                                   &                           &                                    &                                  &                                      &                                 \\
Default rate                       & 0.05                      & 0.05                               & 0.05                             & 0.05                                 & 0.04                            \\
                                   &                           &                                    &                                  &                                      &                                 \\
Refinance share                    & 0.52                      & 0.52                               & 0.49                             & 0.52                                 & 0.54                            \\
                                   &                           &                                    &                                  &                                      &                                 \\
Loan type (share)                  &                           &                                    &                                  &                                      &                                 \\
\multicolumn{1}{r}{Conventional}                     & 0.73                      & 0.74                               & 0.69                             & 0.68                                 & 0.72                            \\
\multicolumn{1}{r}{FHA}                                & 0.16                      & 0.16                               & 0.23                             & 0.24                                 & 0.20                            \\
\multicolumn{1}{r}{VA}                                 & 0.07                      & 0.07                               & 0.05                             & 0.05                                 & 0.05                            \\
                                   &                           &                                    &                                  &                                      &                                 \\
Monthly mortgage payment           & 1,551.53                  & 1,542.87                           & 1,521.91                         & 1,496.81                             & 1,493.22                        \\
                                   & (7,046.41)                & (7,053.62)                         & (3,243.93)                       & (2,894.80)                           & (2,688.71)                      \\
                                   &                           &                                    &                                  &                                      &                                 \\
Income (000s, estimated)                 & 114.00                    & 116.00                             & 112.41                           & 110.02                               & 111.64                          \\
                                   & (63.22)                   & (62.65)                            & (59.24)                          & (57.26)                              & (58.07)                         \\
                                   &                           &                                    &                                  &                                      &                                 \\
                                   &                           &                                    &                                  &                                      &    \\
N                                  & 15,311,720                & 10,234,555                         & 6,745,216                        & 5,553,044                            & 3,917,208                       \\
Retention rate                     &                           & 0.67                               & 0.66                             & 0.82                                 & 0.71                            \\

                                   \bottomrule                            
\end{tabular}
		} 
		\bigskip
		
		\begin{flushleft}
			\small{Note: The table shows descriptive statistics for five (increasingly restrictive) loan-level samples of originated mortgages. The initial sample frame is newly originated mortgages between 2009-2016 in credit bureau records that are matched to a sample of individuals in Infutor. Column 2 describes mortgages where we successfully matched an individual's Infutor address to an address in CoreLogic (CL) deed records. Column 3 describes mortgages where the bureau origination date lines up with the deed origination date. Column 4 describes the sample where in addition to the date accordance, we also see only small discrepancies in the mortgage amount. Column 5 describes the subset of mortgages where we successfully matched information on the mortgage lender. Retention rates refer to the share of observations that are matched relative to the previous column. We include match rates excluding northeast states, Puerto Rico, and South Dakota, which have no or very low coverage in our CoreLogic data. The estimates in the credit score row are based on the Vantage 3.0 score.}
			
		\end{flushleft}
		
	\end{table}
\end{landscape}


\centering
\captionsetup{justification=centering}

\begin{table}
	\centering
	\small
	\caption{Summary Statistics: Default Measures} 
	\label{table_default_summary}
	
	\resizebox{0.85\textwidth}{!}{

\begin{tabular}{lcccccc} \toprule
&  \multicolumn{2}{c}{ All applicants}  & \multicolumn{2}{c}{ Purchase applicants} & \multicolumn{2}{c}{ Refinance applicants } \\
& Accepts & Rejects & Accepts & Rejects & Accepts & Rejects \\\cline{2-7}
\textit{Any credit product}& & & & & & \\
Has any credit product & 0.997 & 0.910 & 0.997 & 0.871 & 0.998 & 0.978 \\
Has delinquency &  &  & & &  & \\
\multicolumn{1}{r}{At least 30 days} & 0.203 & 0.482 & 0.231 & 0.512 & 0.169 & 0.421 \\
\multicolumn{1}{r}{At least 60 days} & 0.113 & 0.390 & 0.139 & 0.430 & 0.082 & 0.313 \\
\multicolumn{1}{r}{At least 90 days} & 0.095 & 0.362 & 0.119 & 0.405 & 0.066 & 0.282 \\
& & & & & & \\
\textit{Auto loan}& & & & & & \\
Has auto loan & 0.510 & 0.473 & 0.519 & 0.452 & 0.495 & 0.503 \\
Has delinquency &  &  & & &  & \\
\multicolumn{1}{r}{At least 30 days} & 0.063 & 0.214 & 0.078 & 0.250 & 0.078 & 0.156 \\
\multicolumn{1}{r}{At least 90 days}  & 0.016 & 0.086 & 0.022 & 0.108 & 0.977 & 0.051 \\
& & & & & & \\
\textit{Credit card}& & & & & & \\

Has credit card & 0.968 & 0.842 & 0.961 & 0.638 & 0.977 & 0.928 \\
Has delinquency &  &  & & &  & \\
\multicolumn{1}{r}{At least 30 days} & 0.106 & 0.315 & 0.126 & 0.247 & 0.084 & 0.247 \\
\multicolumn{1}{r}{At least 90 days}   & 0.052 & 0.231 & 0.064 & 0.166 & 0.033 & 0.166 \\
& & & & & & \\
\textit{HELOC}& & & & & & \\

Has HELOC & 0.140 & 0.099 & 0.101 & 0.058 & 0.185 & 0.180 \\
Has delinquency &  &  & & &  & \\
\multicolumn{1}{r}{At least 30 days} & 0.021 & 0.084 & 0.025 & 0.082 & 0.019 & 0.083 \\
\multicolumn{1}{r}{At least 90 days}  & 0.006 & 0.039 & 0.009 & 0.043 & 0.005 & 0.036 \\
& & & & & & \\
\textit{Mortgage}& & & & & & \\

Has mortgage account & 0.939 & 0.447 & 0.933 & 0.229 & 0.947 & 0.817 \\
Has delinquency &  &  & & &  & \\

\multicolumn{1}{r}{At least 30 days} & 0.058 & 0.198 & 0.071 & 0.160 & 0.071 & 0.213 \\
\multicolumn{1}{r}{At least 90 days}   & 0.022 & 0.115 & 0.028 & 0.096 & 0.028 & 0.123 \\
& & & & & & \\
N & 20,345,046 & 27,416,619 & 8,276,748 & 13,167,856 & 9,107,935 & 7,403,992 \\ \bottomrule
\end{tabular}
	}
	\bigskip
	
	\begin{flushleft}
		\small{Note: The table shows different measures of default in the sample of mortgage applicants. All default outcomes are measured in the 24 months after the application. The first line in each sub-panel refers to the percentage of applicants who have an open account (of this type) in the 24 months after their mortgage application. All other numbers are default rates. Default is only defined for individuals who have had an open account (of that type) in the past 24 months.}
	\end{flushleft}
\end{table}


\begin{table}[]
	\caption{Confusion Matrices for Default Measures\label{tab:default_confusionmatrix}}
	
	\centering
	\resizebox{\textwidth}{!}{

\begin{tabular}{ccccccc}\toprule

 \multicolumn{3}{c}{ 30+ Days Past Due } & & \multicolumn{3}{c}{ 90+ Days Past Due } \\
\cmidrule(lr){1-3}\cmidrule(lr){5-7} 
& \multicolumn{2}{c}{ Mortgage delinquency } & & & \multicolumn{2}{c}{ Mortgage delinquency } \\

 Non-mortgage delinquency & 0 & 1 & & Non-mortgage delinquency & 0 & 1 \\

0 & 16,123,407 & 461,512 & & 0 &  17,244,537 &  685,306  \\
1 & 1,641,278 & 595,377 & & 1 &  520,148 & 371,583 \\
& & & & & & \\

\multicolumn{1}{r}{Precision} & 0.29 & & & \multicolumn{1}{r}{Precision} & 0.42 & \\
\multicolumn{1}{r}{Recall} & 0.56 & & & \multicolumn{1}{r}{Recall} & 0.35 & \\
\multicolumn{1}{r}{Accuracy} & 0.89 & & & \multicolumn{1}{r}{Accuracy} & 0.94 & \\

\bottomrule

\end{tabular}

	}
	\bigskip
	
	\begin{flushleft}
		\small{Note: The table shows confusion matrices in the sample of approved mortgage applicants. Mortgage delinquency refers to  whether an approved applicant is delinquent on their mortgage 24 months after origination. The matrix shows how well this is predicted by various measures of non-mortgage delinquency, ranging from at least 30 days to at least 90 days. For each confusion matrix, we show three performance metrics. Precision is defined as TP/(TP+FP). Recall is defined as TP/(TP+FN). Accuracy is defined as (TP+TN)/N. TP denotes true positives (defaults that were correctly predicted), and TN denotes true negatives (non-defaults that were correctly predicted as non-defaults). For brevity, we exclude panels that compare to two other different measures: results for  120+ days delinquent are similar to those of 90+; results for 60+ are approximately halfway between those for 30+ and 90+ days delinquent.}
		
	\end{flushleft}
	
\end{table}


\centering
\captionsetup{justification=centering}

\begin{table}\small \centering 
	\caption{Credit Score Precision by Group}
	\label{table_auc_summary_mc}
	
	\resizebox{\textwidth}{!}{

\begin{tabular}{lcccccccc}\toprule
Applicant sample & Higher Income & Low Income & Diff(pp) & Diff(\%) & Non-minority & Minority & Diff(pp) & Diff(\%) \\ \midrule

\multicolumn{9}{l}{Delinquency of 90+ days (baseline)} \\
\multicolumn{1}{r}{AUC}  & 0.868 & 0.787 & 0.081 & 0.093 & 0.863 & 0.805 & 0.058 & 0.067 \\
\multicolumn{1}{r}{R2} & 0.951 & 0.803 & 0.148 & 0.156 & 0.790 & 0.652 & 0.138 & 0.175 \\
& & & & & & & & \\

\multicolumn{9}{l}{Delinquency of 120+ days} \\
\multicolumn{1}{r}{AUC} & 0.871 & 0.787 & 0.084 & 0.096 & 0.863 & 0.805 & 0.058 & 0.067 \\
\multicolumn{1}{r}{R2} & 0.946 & 0.807 & 0.139 & 0.147 & 0.790 & 0.652 & 0.138 & 0.175 \\
& & & & & & & & \\

\multicolumn{9}{l}{Delinquency of 60+ days} \\
\multicolumn{1}{r}{AUC} & 0.860 & 0.788 & 0.072 & 0.084 & 0.857 & 0.804 & 0.053 & 0.062 \\
\multicolumn{1}{r}{R2} & 0.959 & 0.747 & 0.212 & 0.221 & 0.794 & 0.629 & 0.165 & 0.208 \\ \bottomrule
\end{tabular}
	}
	\bigskip
	
	\begin{flushleft}
		\small{Note: The table shows measures of predictive accuracy of credit scores by different consumer types in our sample of mortgage applicants. Each sub-panel uses a delinquency variable that differs in the severity of default. Credit scores are scores at times of the mortgage application. R2 refers to the R2 of a regression of the log odds of default on the score after aggregating the log-odds for a given credit score bin.  AUC refers to the Area Under the Curve measure. }
	\end{flushleft}
\end{table}


\begin{table}[]
	\caption{Testing Modeling Bias}
	\label{table_ml_applicant_mc}
	
	\centering
	\begin{subtable}[t]{\linewidth}
		\caption{\large{Panel A: ROC AUC}}

\begin{tabular}{lcccccccc}\toprule
& & Non-minority & Minority &$\Delta$ & & Higher Income & Low Income & $\Delta$  \\ \midrule
VantageScore & & 0.829 & 0.774 & 0.055 & & 0.830 & 0.744 & 0.086 \\
\emph{Baseline (pooled sample)} & & & & & & & & \\
\multicolumn{1}{r}{Random Forest} & & 0.879 & 0.833 & 0.046 & & 0.882 & 0.809 & 0.073 \\
\multicolumn{1}{r}{XGBoost} & & 0.887 & 0.840 & 0.047 & & 0.891 & 0.814 & 0.077 \\
\multicolumn{1}{r}{Logit} & & 0.835 & 0.786 & 0.049 & & 0.839 & 0.756 & 0.082 \\
\emph{Different models}& & & & & & & & \\
\multicolumn{1}{r}{Random Forest} & & 0.878 & 0.831 & 0.047 & & 0.881 & 0.808 & 0.073 \\
\multicolumn{1}{r}{XGBoost} & & 0.887 & 0.839 & 0.048 & & 0.888 & 0.814 & 0.074 \\
\multicolumn{1}{r}{Logit} & & 0.835 & 0.787 & 0.047 & & 0.840 & 0.759 & 0.082 \\
\emph{Different models: Same N}& & & & & & & & \\
\multicolumn{1}{r}{Random Forest} & & 0.875 & 0.831 & 0.044 & & 0.879 & 0.808 & 0.071 \\
\multicolumn{1}{r}{XGBoost} & & 0.884 & 0.839 & 0.046 & & 0.889 & 0.814 & 0.075 \\
\multicolumn{1}{r}{Logit} & & 0.835 & 0.787 & 0.047 & & 0.840 & 0.759 & 0.082 \\
\emph{Re-weight training data} & & & & & & & & \\
\multicolumn{1}{r}{Random Forest} & & 0.876 & 0.832 & 0.044 & & 0.881 & 0.809 & 0.072 \\
\multicolumn{1}{r}{XGBoost} & & 0.886 & 0.840 & 0.046 & & 0.888 & 0.812 & 0.076 \\
\multicolumn{1}{r}{Logit} & & 0.835 & 0.786 & 0.048 & & 0.838 & 0.757 & 0.080 \\ \bottomrule
\end{tabular}

	\end{subtable}
	\newline
	\vspace*{.35 cm}
	\begin{subtable}[t]{\linewidth}
		\caption{\large{Panel B: MSE}}
\begin{tabular}{lcccccccc}\toprule
& & Non-minority & Minority & $\Delta$ & & Higher Income & Low Income &  $\Delta$ \\\midrule
VantageScore& & 0.092 & 0.146 & -0.054 & & 0.085 & 0.166 & -0.081 \\

\emph{Baseline (pooled sample)} & & & & & & & & \\
\multicolumn{1}{r}{Random Forest} & & 0.080 & 0.127 & -0.047 & & 0.073 & 0.145 & -0.072 \\
\multicolumn{1}{r}{XGBoost} & & 0.077 & 0.124 & -0.047 & & 0.070 & 0.143 & -0.073 \\
\multicolumn{1}{r}{Logit} & & 0.088 & 0.142 & -0.054 & & 0.080 & 0.164 & -0.084 \\
\emph{Different models} & & & & & & & & \\
\multicolumn{1}{r}{Random Forest} & & 0.092 & 0.146 & -0.054 & & 0.086 & 0.166 & -0.080 \\
\multicolumn{1}{r}{XGBoost} & & 0.077 & 0.125 & -0.048 & & 0.070 & 0.143 & -0.073 \\
\multicolumn{1}{r}{Logit} & & 0.088 & 0.140 & -0.052 & & 0.080 & 0.160 & -0.080 \\

\emph{Different models: Same N} & & & & & & & & \\
\multicolumn{1}{r}{Random Forest} & & 0.093 & 0.146 & -0.053 & & 0.085 & 0.166 & -0.081 \\
\multicolumn{1}{r}{XGBoost} & & 0.078 & 0.125 & -0.047 & & 0.071 & 0.143 & -0.072 \\
\multicolumn{1}{r}{Logit} & & 0.088 & 0.140 & -0.052 & & 0.080 & 0.160 & -0.080 \\

\emph{Re-weight training data} & & & & & & & & \\
\multicolumn{1}{r}{Random Forest} & & 0.093 & 0.146 & -0.053 & & 0.085 & 0.166 & -0.081 \\
\multicolumn{1}{r}{XGBoost} & & 0.077 & 0.124 & -0.047 & & 0.071 & 0.143 & -0.072 \\
\multicolumn{1}{r}{Logit} & & 0.088 & 0.141 & -0.053 & & 0.081 & 0.162 & -0.081 \\ \bottomrule
\end{tabular}

	\end{subtable}
	\bigskip
	
	\begin{flushleft}
		\small{Note: The table shows results from testing aggregation and majority bias. All performance metrics are calculated on the test set. Vantage refers to the performance of either a logit or random forest model with the VantageScore 3.0 credit score as the only input. Baseline refers to models trained on a random sub-sample of our applicant sample, pooling both groups. The first exercise (different models) allows machine learning models to be fitted separately for each group, where we either allow the sample size to reflect group proportions (different N), or force sample sizes to be equal (same N). The second exercises runs a pooled model but re-weights the training data such that both groups are represented in equal proportions.}
	\end{flushleft}
\end{table}


\begin{table}[]
	\caption{Testing Data Bias}
	\label{table_data_bias_main_lb}
	
	\centering
	\resizebox{\textwidth}{!}{
		\begin{subtable}[t]{\linewidth}
			\caption{\large{Panel A: By Minority Status}}

\begin{tabular}{rcccccc} 

\toprule
& \multicolumn{3}{c}{AUC} & \multicolumn{3}{c}{Share} \\
& \multicolumn{1}{c}{Non-minority\ \ \ } &  \multicolumn{1}{c}{\ \ Minority\ \ \ \ \ \ } &  \multicolumn{1}{c}{ \ \ Diff  \ } &  \multicolumn{1}{c}{Non-minority\ \ \ } &  \multicolumn{1}{c}{\ \ Minority\ \ \ \ \ } &  \multicolumn{1}{c}{ \ \ Diff  \ } \\
\cline{2-7}
 \multicolumn{1}{l}{Thin file} & & & & & & \\
Clean & 0.827 & 0.752 & 0.075 & 0.151 & 0.143 & 0.008 \\
$<90$ DPD & 0.752 & 0.726 & 0.026 & 0.037 & 0.045 & -0.009 \\
90$+$ DPD & 0.688 & 0.665 & 0.023 & 0.109 & 0.238 & -0.126 \\
 \multicolumn{1}{l}{Thick file: no mortgage} & & & & & & \\
Clean & 0.837 & 0.788 & 0.049 & 0.101 & 0.060 & 0.040 \\
$<$90 DPD & 0.744 & 0.715 & 0.029 & 0.041 & 0.035 & 0.006 \\
90$+$ DPD & 0.709 & 0.705 & 0.005 & 0.108 & 0.169 & -0.058 \\
 \multicolumn{1}{l}{Thick file: with mortgage} & & & & & & \\
Clean & 0.820 & 0.792 & 0.028 & 0.253 & 0.111 & 0.141 \\
$<$90 DPD & 0.764 & 0.727 & 0.037 & 0.097 & 0.063 & 0.033 \\
90$+$ DPD & 0.756 & 0.738 & 0.018 & 0.098 & 0.124 & -0.028 \\
& & & & & & \\
Average & 0.779 & 0.725 & 0.054 & & & \\
 \multicolumn{1}{l}{\emph{Decomposition}} & & & & & & \\
Between group &0.025 & &  & & & \\
Residual & 0.029 & & & & & \\ \bottomrule 
\end{tabular}


		\end{subtable}
	}
	\vspace*{.35 cm}
	\resizebox{\textwidth}{!}{
		\begin{subtable}[t]{\linewidth}
			\caption{\large{Panel B: By Income Status}}

\begin{tabular}{rcccccc} 
\toprule
& \multicolumn{3}{c}{AUC} & \multicolumn{3}{c}{Share} \\
  &\multicolumn{1}{c}{Higher income} & \multicolumn{1}{c}{\ \ Low income}& \multicolumn{1}{c}{ \ \  Diff  \ \ } & \multicolumn{1}{c}{Higher income} &\multicolumn{1}{c}{\ \ Low income} & \multicolumn{1}{c}{ \ \ Diff  \ \ } \\
\cline{2-7}
& & & & & & \\
\multicolumn{1}{l}{Thin file} & & & & & & \\
Clean & 0.845 & 0.764 & 0.081 & 0.151 & 0.252 & -0.102 \\
$<$90 DPD & 0.752 & 0.733 & 0.019 & 0.039 & 0.079 & -0.040 \\
90$+$ DPD & 0.688 & 0.669 & 0.020 & 0.135 & 0.357 & -0.222 \\
\multicolumn{1}{l}{Thick file: no mortgage} & & & & & & \\
Clean & 0.830 & 0.725 & 0.105 & 0.093 & 0.037 & 0.056 \\
$<$90 DPD & 0.743 & 0.711 & 0.031 & 0.039 & 0.023 & 0.016 \\
90$+$ DPD & 0.711 & 0.699 & 0.011 & 0.120 & 0.134 & -0.014 \\
\multicolumn{1}{l}{Thick file: with mortgage} & & & & & & \\
Clean & 0.813 & 0.815 & -0.001 & 0.224 & 0.034 & 0.190 \\
$<$90 DPD & 0.760 & 0.743 & 0.017 & 0.089 & 0.020 & 0.070 \\
90$+$ DPD & 0.749 & 0.730 & 0.019 & 0.104 & 0.046 & 0.058 \\
& & & & & & \\
Average & 0.780 & 0.714 & 0.066 & & & \\
\multicolumn{1}{l}{\emph{Decomposition}} & & & & & & \\
Between group & 0.035& & & & & \\
Residual & 0.035 & & & & & \\ \bottomrule 
\end{tabular}
		\end{subtable}
	}
	\bigskip
	
	\begin{flushleft}
		\small{Note: The table shows results from testing data bias. Each row corresponds to a sub-sample of observations. Categories are mutually exclusive.  Share refers to the share of individuals in that sub-sample. AUC refers to the Area Under the Curve measure of the VantageScore 3.0 credit score. Average refers to the average AUC obtained by taking a weighted average across sub-samples by group. The between group number is obtained by summing the product of the difference in shares (the final column of the table) and the average AUC (not shown) across sub-samples (rows). The residual is obtained by summing the product of the AUC difference (the third column of the table) and the average share (not shown) across sub-samples (rows). The data sample is the same as in Table \ref{table_ml_applicant_mc} and Figure \ref{fig:auc_comp}. Note that the overall AUC difference between groups diverges from Figure \ref{rocgraph} since we linearly aggregate over sub-sample AUCs.}
	\end{flushleft}
\end{table}


\begin{table}[]
	\caption{Second Stage Regressions (Refinance Loans)}
	\label{table_secondstage_refi_mc}
	
	\centering
	
	\resizebox{\textwidth}{!}{

\begin{tabular}{lcccccccc} \toprule
                                                                  & \multicolumn{2}{c}{Minority}   & \multicolumn{2}{c}{Non-minority}    & \multicolumn{2}{c}{Low Income}   & \multicolumn{2}{c}{Higher Income}   \\
                                        \cmidrule(lr){2-5}\cmidrule(lr){6-9}
             Number of loans in default                                                     & Marginal  & Average     & Marginal    & Average   & Marginal    & Average & Marginal    & Average   \\ \midrule
            & & & & & & & & \\
    Originated Loans                                              &0.069***              &  0.055***            &        0.063***     & 0.042*** & 0.138*        & 0.067***          & 0.043***            & 0.038***        \\
                                                                  & (0.026)          &     (0.004)    &     (0.018)            & (0.005)  & (0.079)            & (0.009)           & (0.015)             & (0.003)            \\
                                                                  &                      &                      &                      &   &                      &                      &                      &                   \\
    \textit{Controls}                                                      &                      &                      &                      & &                      &                      &                      &                      \\
    Bank x Tract FE                                               & X                    & X                     & X                    &  X & X                    & X                     & X                    &  X                   \\
    Tract x Time FE                                               & X                    & X                     & X                    &  X    & X                    & X                     & X                    &  X                \\
                                                                  &                      &                      &                      &   &                      &                      &                      &                    \\ \midrule
     First-stage F-stat  &  10.37 & -  & 32.83  &  - & 6.111  &  - & 26.45  & - \\
    N                                                             & 64,294              & 64,294                &       130,871         &130,871   & 67,170               & 67,170               & 139,441              & 139,441                  \\
    R-sq                                                          & 0.258               & 0.773               & 0.119               & 639     & -0.014               & 0.646              & 0.243              & 0.751           \\ \bottomrule
    \end{tabular}
		
	} 
	
	\bigskip
	
	\begin{flushleft}
		\small{Note: The table shows results from estimating the second-stage regression of loan default volumes (any delinquency in 24 months post-origination) on loan origination volumes instrumented by variation in CRA exam activity, as described in equation \ref{eq:2sls}. Averages come from OLS regressions including the same fixed effects as the second stage. Group membership is described in the text. Standard errors (in parentheses) are clustered at the tract level.}
	\end{flushleft}
	
\end{table}


\captionsetup{justification=centering}

\begin{table}[]
	\caption{Model Estimation: Target Moments}
	\label{tab:moments}
	
	\centering
	
	\resizebox{0.68\textwidth}{!}{
		
    \begin{tabular}{lcccc} \toprule
    & Higher Income & Low Income  & Non-minority & Minority \\ \cline{2-5}
      & & & & \\
    
    Approval Rate 	            &	0.534     &	 0.357      &	0.545     &	 0.374     \\
    Average Default Rate 	    &	0.038     &	 0.067      &	0.042     &	 0.055     \\
    Marginal Default Rate 	    &	0.043     &	 0.138      &	0.063     &	 0.069     \\
    Avg. Approved Score 	    &	727       &	 685        &	729       &	 694       \\
    Avg. Rejected Scores 	    &	708       &	 652        &	697       &	 655       \\
    Default vs. Score Slope 	&	-0.00044  &	 -0.00095 	&	-0.00043  &	 -0.00086  \\
    Score vs. Default Slope 	&	-40.78    &	 -68.03 	&	-41.28    &	 -64.07    \\
    \bottomrule

    \end{tabular}

	} 
	
	\bigskip
	
	\begin{flushleft}
		\small{Note: This table reports the empirical moments we target in the full model estimation. Approval and rejection credit score statistics are taken from the credit report data; the regression slopes between default and credit score are estimated by regressing the log odds of default on score, and vice versa; average and marginal default rates are estimated via OLS and 2SLS and are taken from the results reported in Table \ref{table_secondstage_refi_mc}. For sake of comparison with the means in this table, we also note that the median credit scores of approved applicants are, respectively across the four columns of the table, 747, 710, 746, and 725, whereas the median scores for rejected applicants correspondingly are 721, 638, 720, and 664. Regression slopes in this table differ from those in Table \ref{attenuationtable} due to the tail-trimming strategy discussed in Appendix \ref{appendix_model}. Group membership is described in the text.}
		
	\end{flushleft}
	
\end{table}


\begin{table}[]
	\caption{Model Parameter Estimates}
	\label{tab:modelestimates}
	\centering
	\resizebox{0.75\textwidth}{!}{
\begin{tabular}{lcccc} 
        \toprule
     & Higher Income & Low Income  & Non-minority & Minority \\ \cline{2-5}
        & & & & \\
        $\mu_0$: mean($\theta$)	    &	2.89	&	0.96	&	2.89	&	1.13   \\
        $1/h_0$:  var($\theta$)	    &	21.10	&	4.48	&	28.73	&	15.96   \\
        $1/h_1$:  var(score noise)	&	0.32	&	0.48	&	0.50	&	2.46    \\
        \bottomrule
\end{tabular}

	}
	\bigskip
	
	\begin{flushleft}
		\small{Note: This table reports estimates of key model parameters in the model of Section \ref{sec:model:estimation}. $\theta$ denotes borrowers' unobserved risk types and score noise refers to the signal noise in credit scores. Group membership is described in the text.}
	\end{flushleft}
\end{table}


\begin{table}[]
	\caption{Allocative Efficiency with Counterfactual Information}
	\label{tab:CFs}
	
	\centering
	
	\resizebox{0.8\textwidth}{!}{
		
\begin{tabular}{lcccccc} \toprule
& \multicolumn{2}{c}{{Income}}   & & \multicolumn{2}{c}{{Minority}} & \\
& Higher & Low  & $\Delta$ & No & Yes & $\Delta$ \\ \cline{2-7}

  \multicolumn{7}{l}{\textit{Baseline:}} \\
Approval Rate	                            & 	0.535  &	0.357  &	-0.178  &   0.545 & 	0.374 & 	-0.171	 \\
Type I Error Rate	                        & 	0.032  &	0.062  &	0.030   &   0.032 & 	0.097 & 	0.065	 \\
( $\theta_i >= \hat{x}$ but not approved)   &            &           &           &           &           &           \\
Type II Error Rate	                        & 	0.028  &	0.099  &	0.071   &   0.027 & 	0.141 & 	0.113	 \\
($\theta_i < \hat{x}$ but approved)         &            &           &           &           &           &           \\ 
  
\midrule 

 \multicolumn{7}{l}{\textit{Remove non-credit-score information:}} \\
Approval Rate	     & 	0.529  &	0.337  &	-0.192  & 	0.536 & 	0.374 & 	-0.162	 \\
Type I Error Rate	 & 	0.037  &	0.126  &	0.089   & 	0.039 & 	0.140 & 	0.102	 \\
Type II Error Rate	 & 	0.041  &	0.076  &	0.035   & 	0.044 & 	0.098 & 	0.053	 \\

 \midrule
 
 \multicolumn{7}{l}{\textit{Equalize credit score precision:}} \\
Approval Rate	     & 	0.535  &	0.365  &	-0.170  & 	0.545 & 	0.454 & 	-0.091	 \\
Type I Error Rate	 & 	0.028  &	0.086  &	0.058   & 	0.027 & 	0.060 & 	0.033	 \\
Type II Error Rate	 & 	0.032  &	0.055  &	0.023   & 	0.032 & 	0.051 & 	0.019	 \\

 \bottomrule
    \end{tabular}

	} 
	
	\bigskip
	
	\begin{flushleft}
		\small{Note: The table shows estimates of approval rates, inefficient rejection rates (Type I errors), and inefficient approval rates (Type II errors) under the two counterfactual information structures described in the text: removing non-credit-score information from the market, and equalizing the precision of credit score information across groups. See Section \ref{sec:estimation_results:CFs} for the definition of inefficient rejections and approvals. Group membership is described in the text. The third column shows the difference between the first two columns for emphasis.}
		
	\end{flushleft}
\end{table}

\FloatBarrier

\appendix 
\flushleft

\section{Appendix: Additional Figures and Tables}\label{appendix_more}

\setcounter{figure}{0} \renewcommand{\thefigure}{A.\arabic{figure}} 
\setcounter{table}{0} \renewcommand{\thetable}{A.\arabic{table}} 


\begin{figure}[H]
\caption{Benefits from Model-Splitting}
\label{model_bias_intuition}
\centering

    \includegraphics[width=0.8\textwidth]{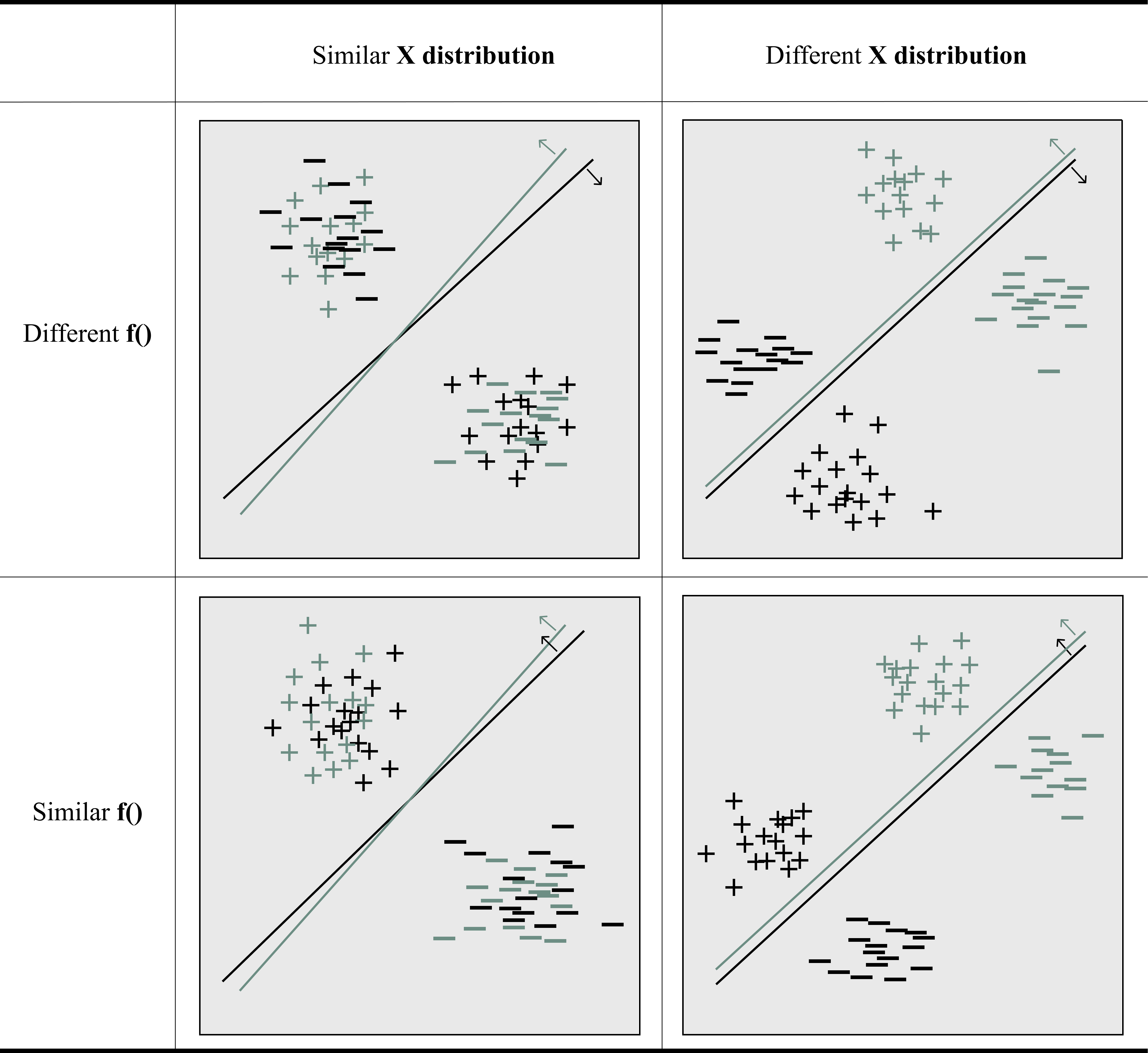}

    \begin{flushleft}
    \begin{spacing}{1.1}
    \small{Note: This figure is an illustration based on \cite{wang_split_2020} to show when model-splitting, or training separate models by group, will reduce precision differentials. The rows indicate whether the conditional expectation functions differ across groups. The columns indicate whether the distribution of observables differ across groups. Different colors indicate different groups. Realized default outcomes are represented by plus or minus.  Arrows indicate the regions where the predicted default outcomes is plus. }
    \end{spacing}
    \end{flushleft}
\end{figure}

\begin{figure}[H]
\caption{Relationship between Probability of Default and Credit Score}
\label{attenuation}
\centering
\begin{subfigure}[t]{0.45\textwidth}
    \includegraphics[width=1\textwidth]{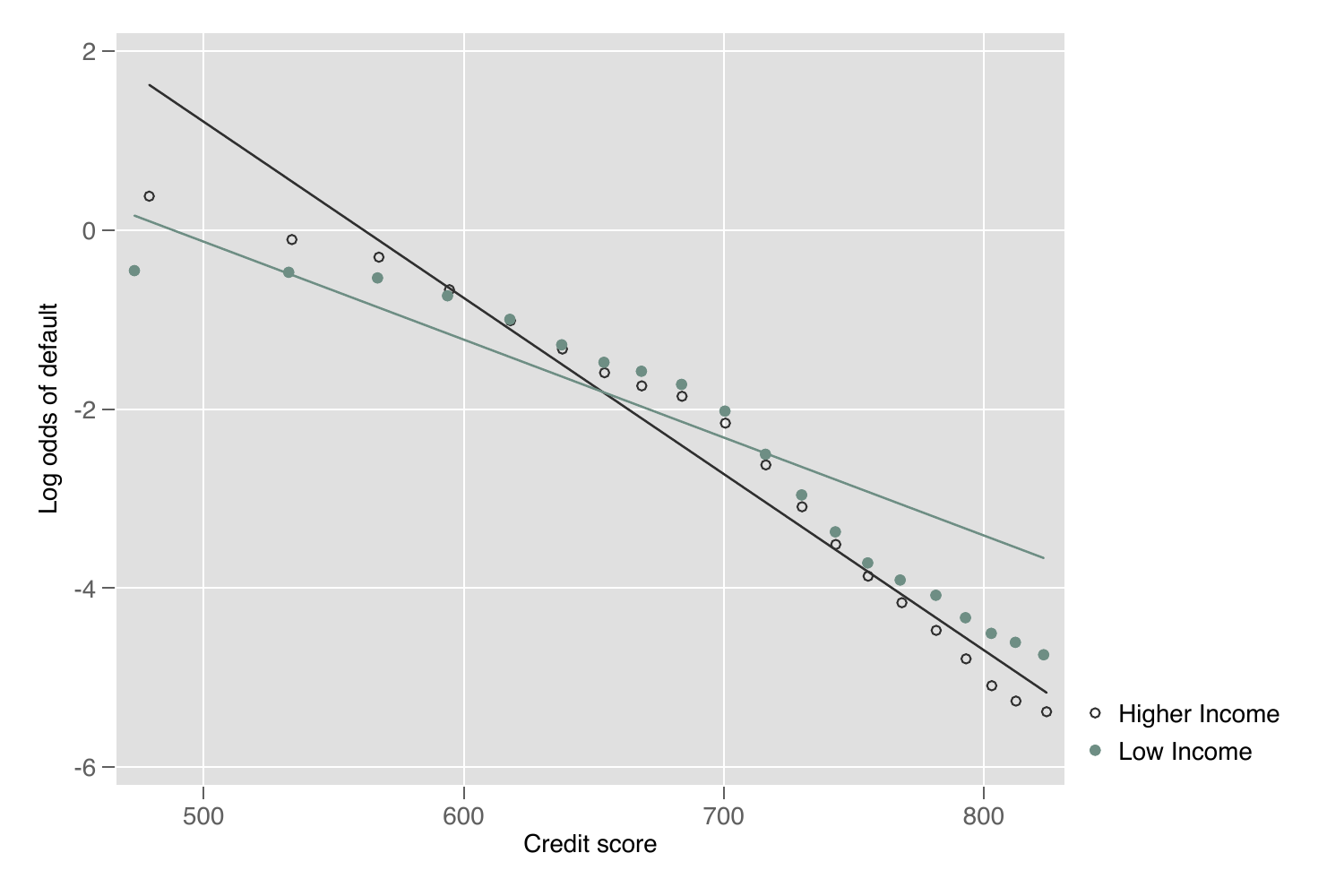}\caption{Applicant Sample by Income Status}
     \end{subfigure} 
    \begin{subfigure}[t]{0.45\textwidth}
    \includegraphics[width=1\textwidth]{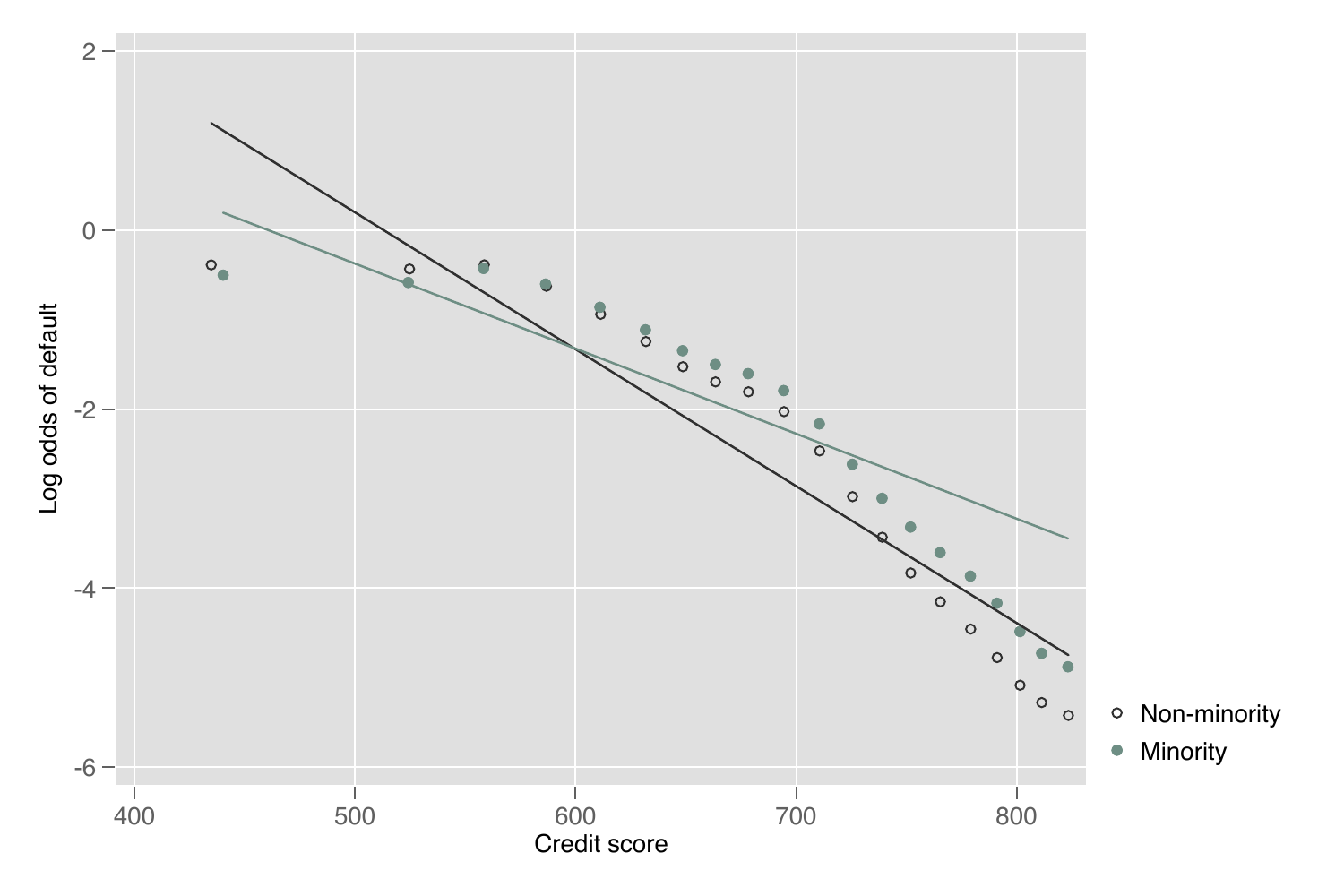}\caption{Applicant Sample by Minority Status}
    \end{subfigure}
       \begin{subfigure}[t]{0.45\textwidth}
    \includegraphics[width=1\textwidth]{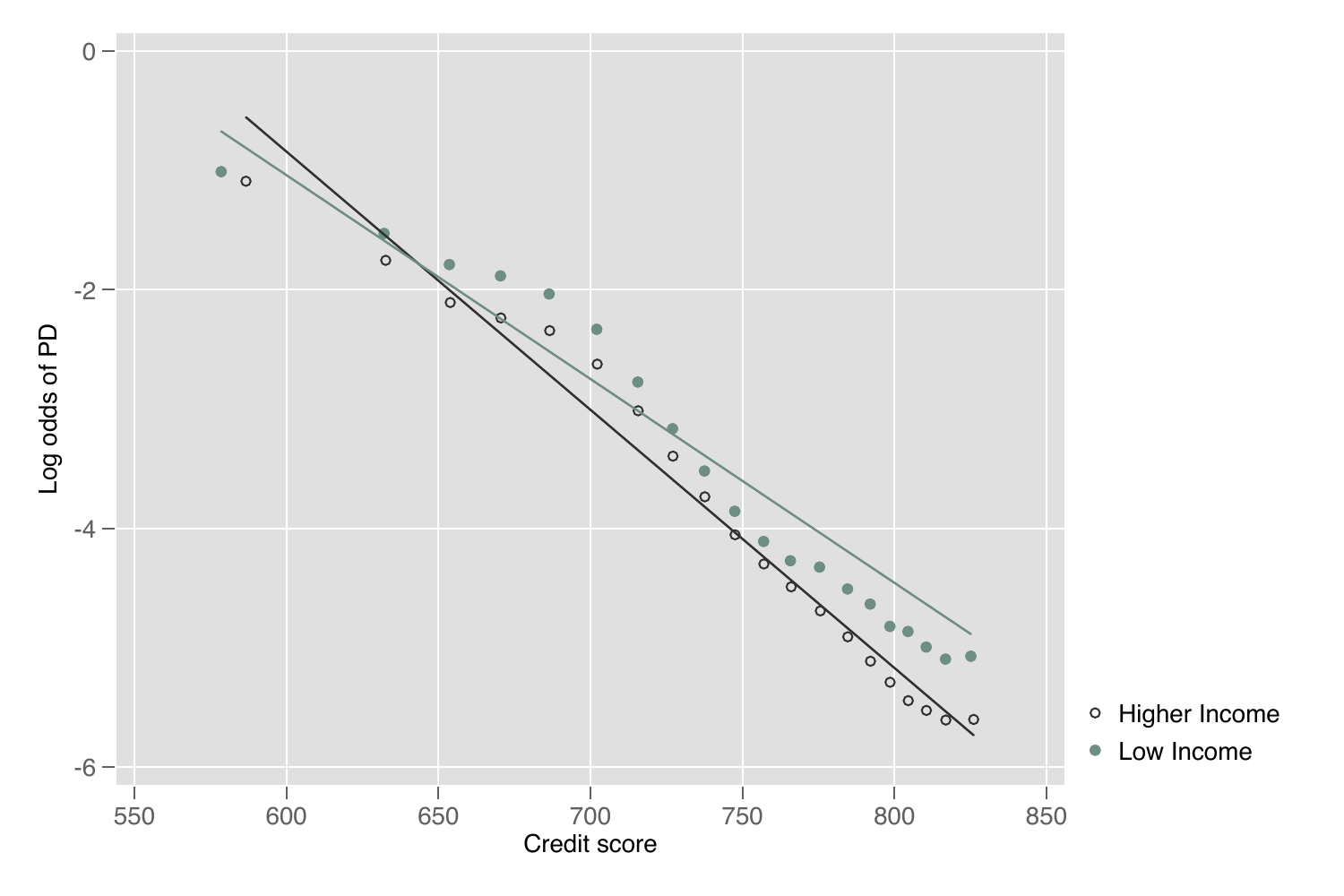}\caption{Approved Sample By Income Status}
     \end{subfigure} 
    \begin{subfigure}[t]{0.45\textwidth}
    \includegraphics[width=1\textwidth]{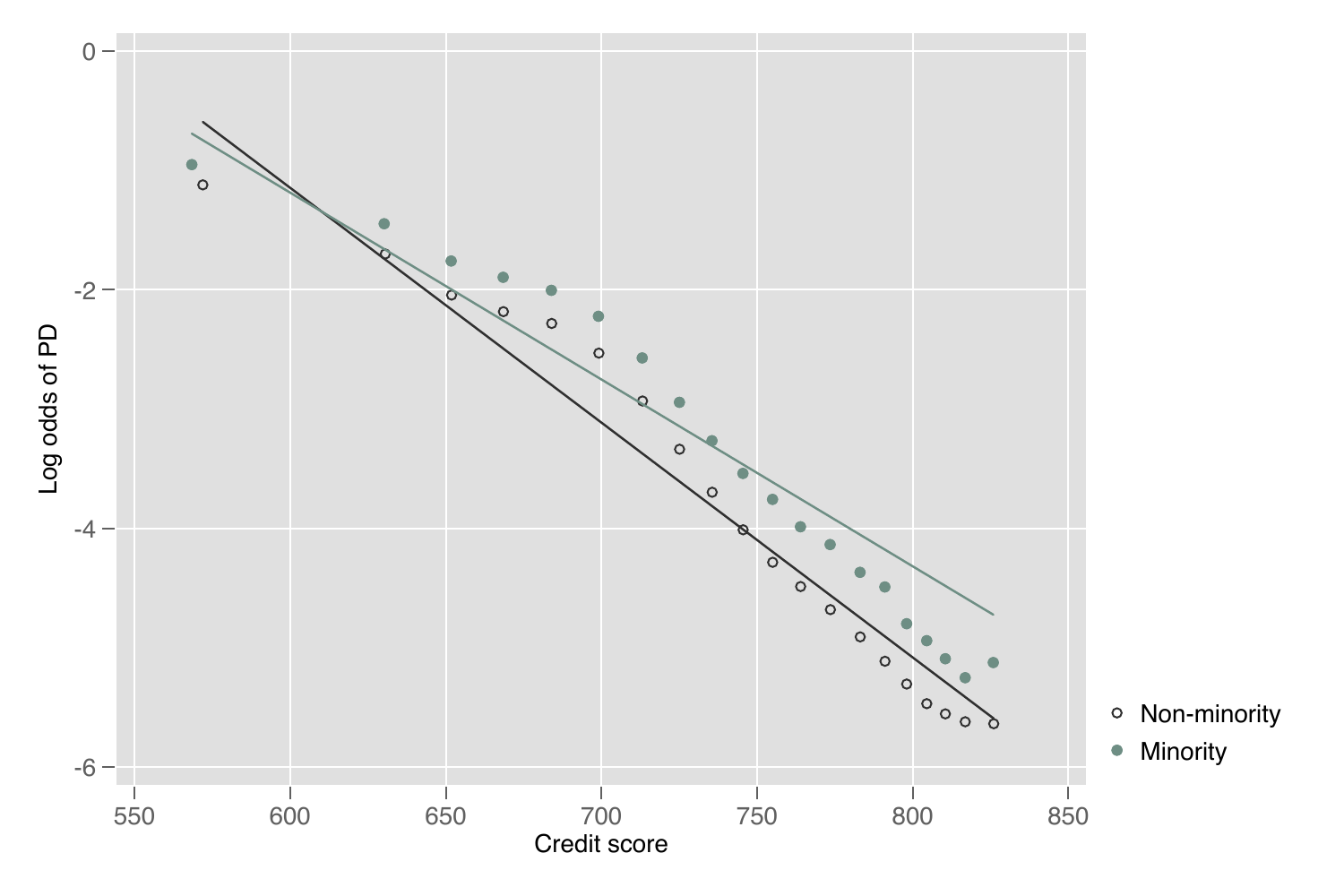}\caption{Approved Sample by Minority Status}
    \end{subfigure}
 
    \begin{flushleft}
    \begin{spacing}{1.1}
    \small{Note: This figure shows the relationship between the log odds of future default and credit score at mortgage application in the samples of mortgage applicants and mortgage originations.  Log odds is defined as log(pd/(1-pd)) where the probability of default (pd) is defined as the fraction borrowers who default. Our measure of default is a non-mortgage delinquency of at least 90 days 24 months after the application date. The graph is a binned scatter plot. For slope coefficients and significant tests, see Table \ref{attenuationtable} in Appendix \ref{appendix_more}. }
    \end{spacing}
    \end{flushleft}
\end{figure}

\captionsetup{justification=centering}


\begin{figure}[H]
\caption{Bias and Noise in Simulated Credit Score \label{bias_noise_sim}}
\centering
    \begin{subfigure}[t]{0.45\textwidth}
   \includegraphics[width=1\textwidth]{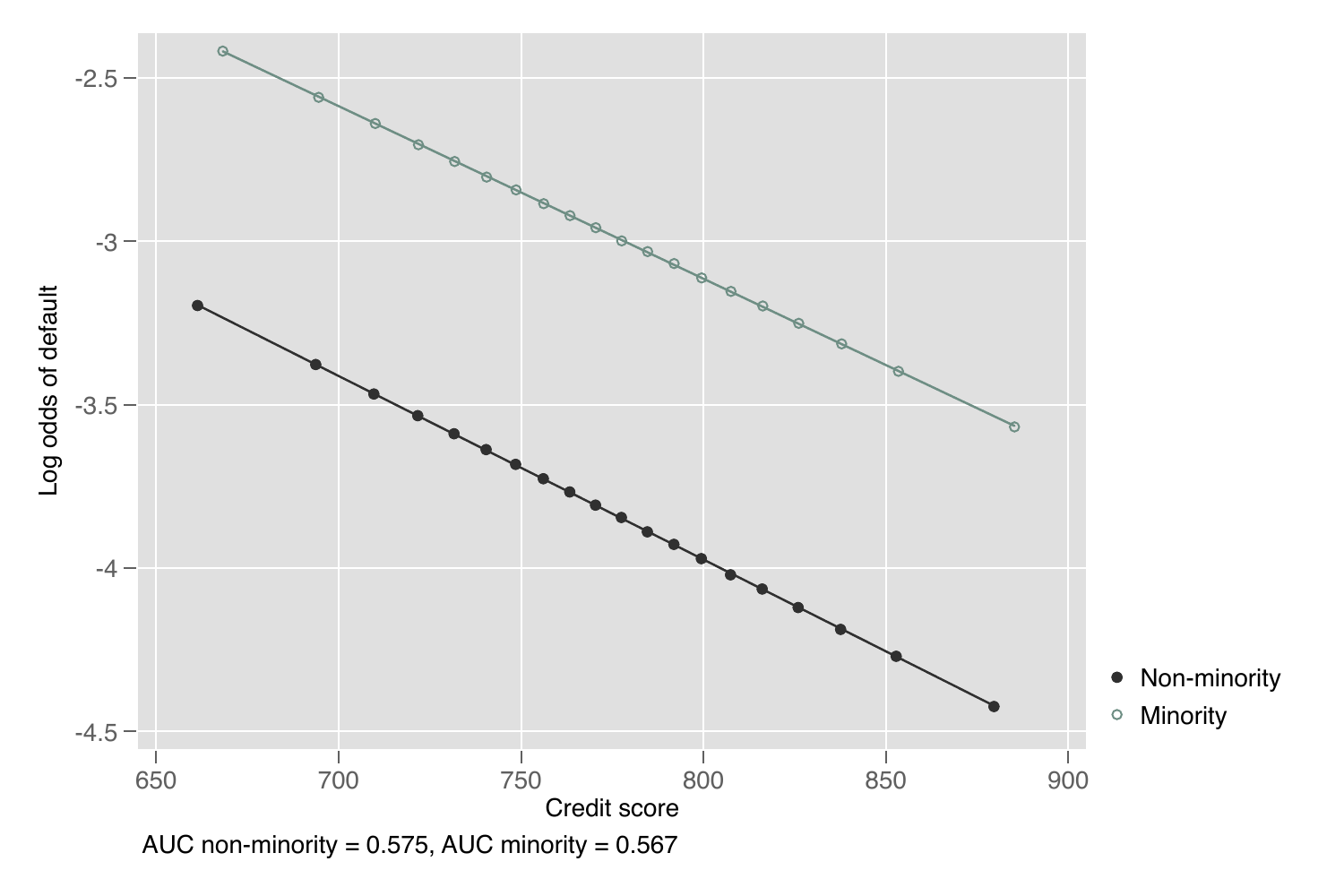}\caption{Only Bias}
     \end{subfigure} \\
    \begin{subfigure}[t]{0.45\textwidth}
    \includegraphics[width=1\textwidth]{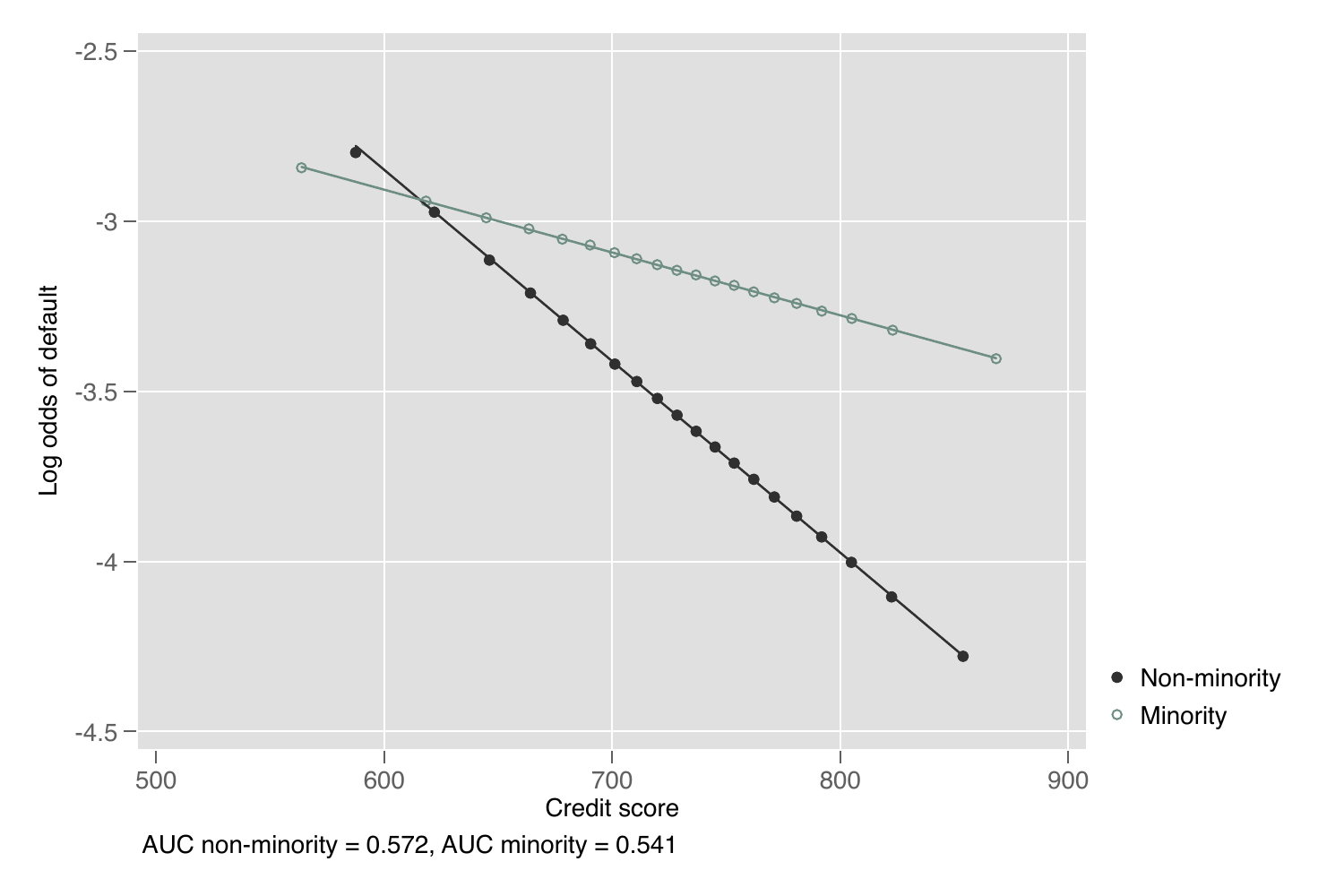}\caption{Only Noise: Different Risk Distributions}
    \end{subfigure}
        \begin{subfigure}[t]{0.45\textwidth}
   \includegraphics[width=1\textwidth]{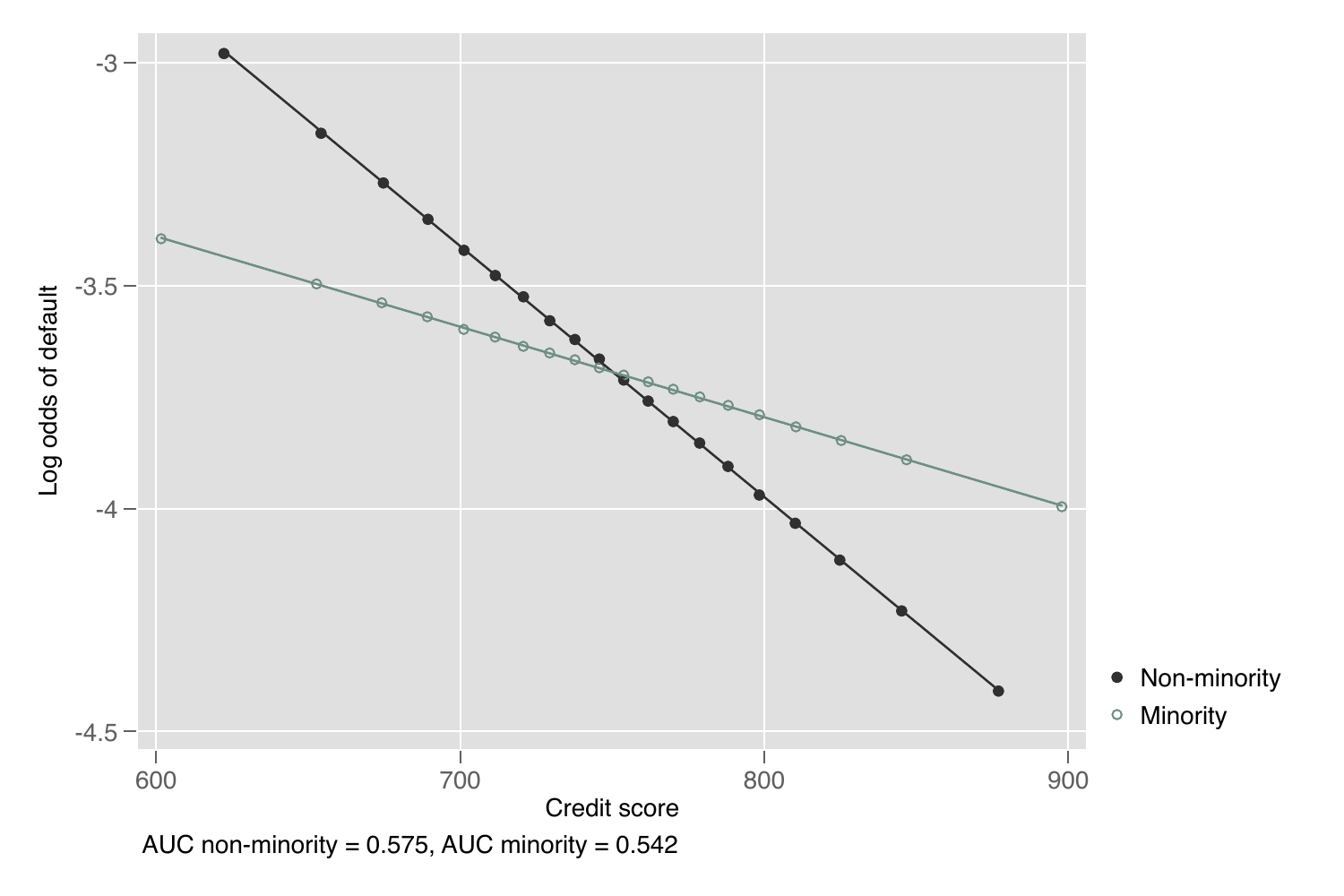}\caption{Only Noise: Same Risk Distribution}
     \end{subfigure} \\
         \begin{subfigure}[t]{0.45\textwidth}
    \includegraphics[width=1\textwidth]{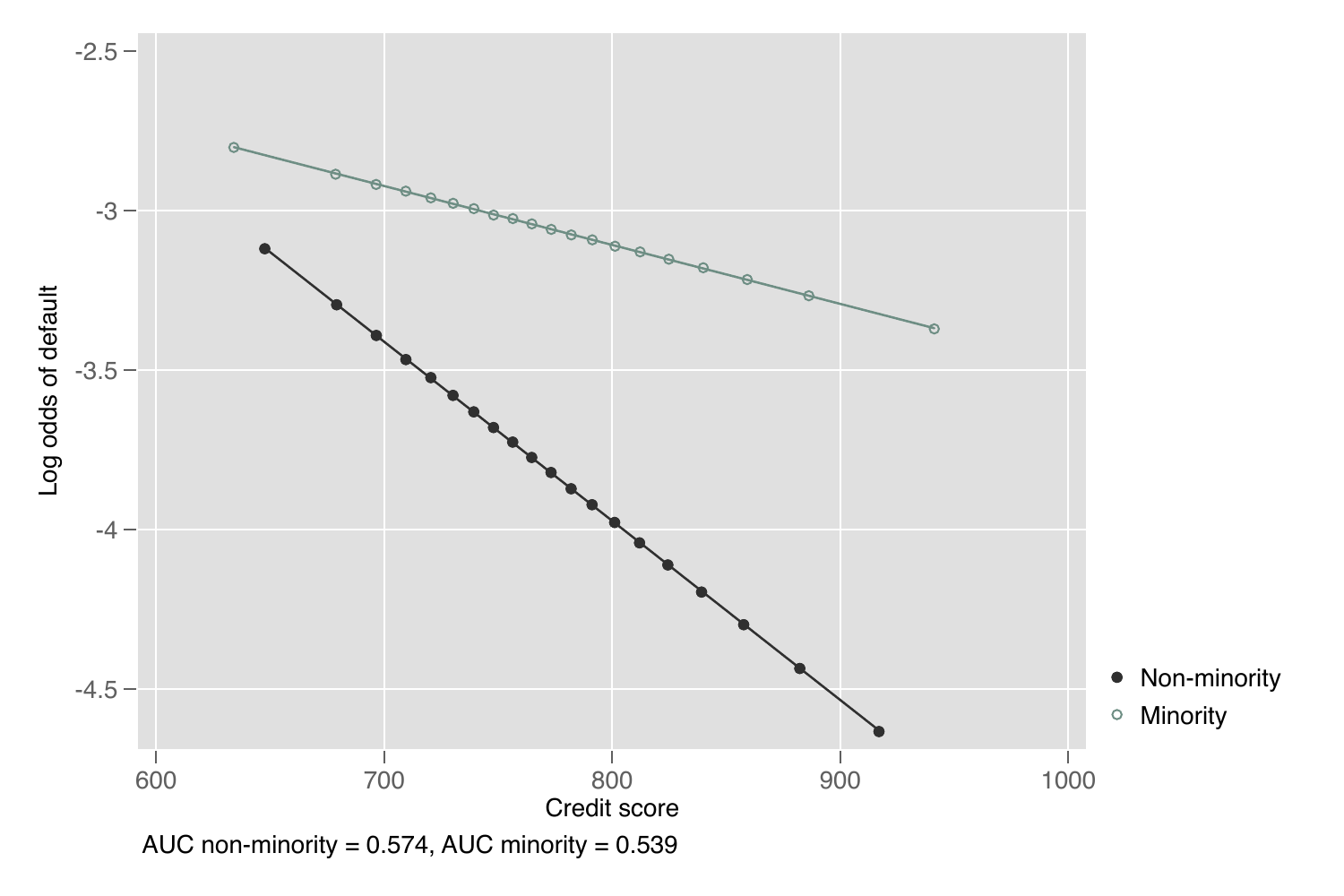}\caption{Bias and Noise: Different Risk Distributions}
    \end{subfigure}
        \begin{subfigure}[t]{0.45\textwidth}
   \includegraphics[width=1\textwidth]{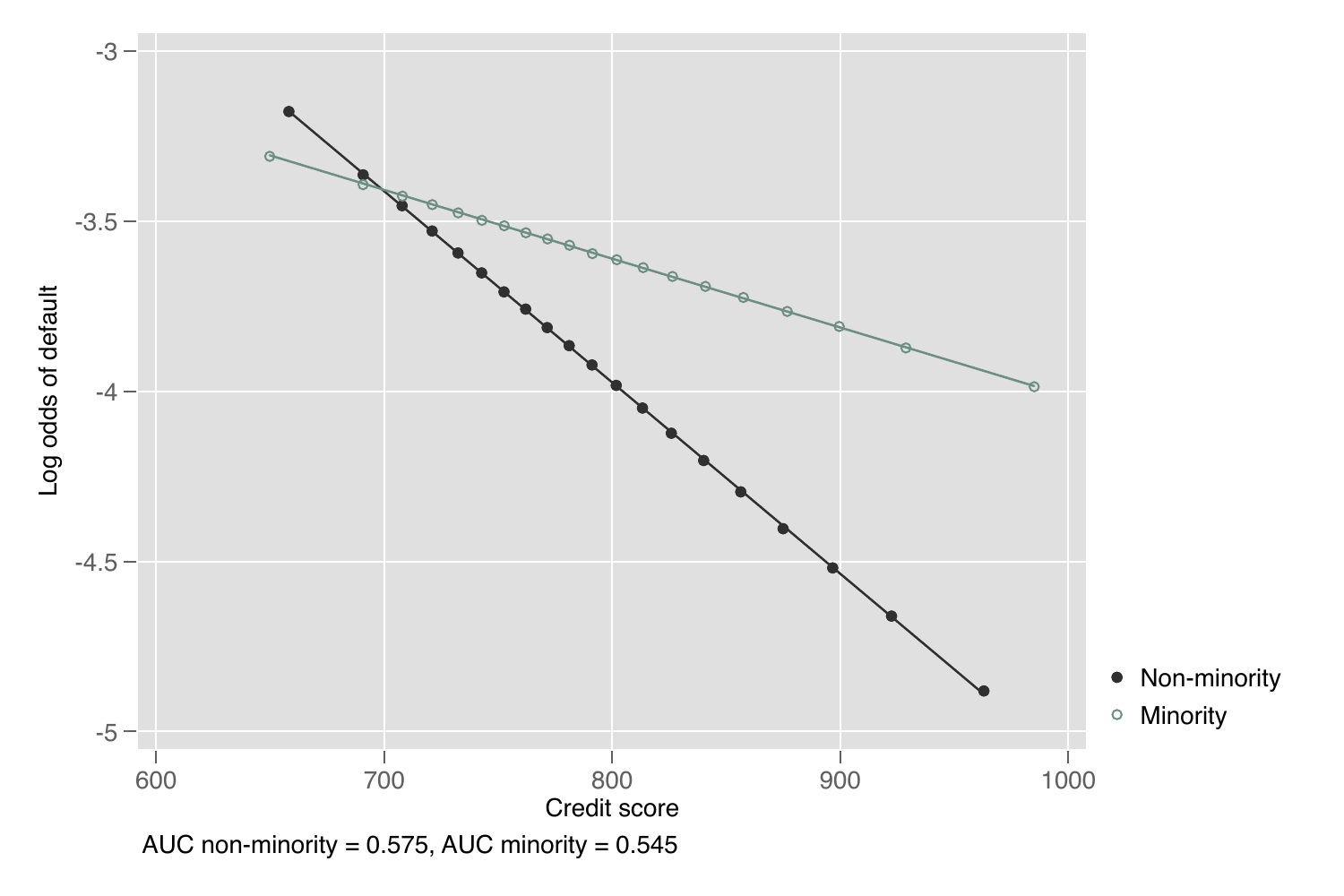}\caption{Bias and Noise: Same Risk Distribution}
     \end{subfigure} \\

    \begin{flushleft}
    \begin{spacing}{1.1}
    \small{Note: The figure shows the relationship between log odds of default and credit score in simulated data from the data-generating process in our structural  model (see Section \ref{sec:model}). Each panel shows a binned scatter plot. When a panel includes ``bias," this refers to the introduction of mean-nonzero signal noise in credit score signal realizations. ``Risk distribution" refers to the distribution of types $\theta$, and ``noise" refers to the variance of credit score signal noise $u_k$.}
    \end{spacing}
    \end{flushleft}
\end{figure}


\begin{table}[H] \centering
\caption{Mortgage Acceptance Rate by Origination Horizon}\label{table_orig_horizon}


\begin{tabular}{lccccc}\toprule
& \multicolumn{4}{c}{ Horizon for origination to occur } & \\ 
Application time & 4 quarters & 3 quarters & 2 quarters & 1 quarter & N \\\midrule
& & & & & \\
1st quarter & 0.457 & 0.433 & 0.404 & 0.354 & 32,870,196 \\
& & & & & \\
2nd quarter & 0.429 & 0.406 & 0.374 & 0.314 & 42,922,951 \\ \bottomrule
\end{tabular}

    \bigskip
    
    \begin{flushleft}
    \small{Note: The table shows acceptance rates of mortgage applications as we vary the time horizon we allow for a mortgage to be opened after the application date. We always include the quarter in which the application was filed plus 1, 2, 3, 4 additional quarters. The first row shows mortgage applications in the first three months of the semiannual panel waves received from TransUnion. The second row shows applications from the second half of each semiannual panel wave.}
    
    \end{flushleft}
\end{table}


\centering
\captionsetup{justification=centering}

\begin{table}[H]
    \caption{Summary Statistics: Applicants}
    \label{table_summarystats_applicants}
    
    \centering
    
    \resizebox{0.6\textwidth}{!}{

\begin{tabular}{lcc}\toprule
& Raw applicant sample & Final applicant sample \\ \midrule
& & \\
Credit score & 688 & 689  \\
& (104.62)& (105.63)  \\
& & \\
Minority share  && 0.29  \\
& & \\
Income (000s) &97& 97 \\
&(59.6) & (59.37)  \\
& & \\
Share & & \\
\multicolumn{1}{r}{Refinance} & 0.45&0.46  \\
\multicolumn{1}{r}{Purchase} &0.55 & 0.54 \\
& & \\
Approval rate & & \\
\multicolumn{1}{r}{Refinance} & 0.50 &0.52 \\
\multicolumn{1}{r}{Purchase} & 0.40&0.41  \\

& & \\
N &47,604,833& 39,331,066  \\ \bottomrule
\end{tabular}
    }
    \bigskip
    
      \begin{flushleft}
      \small{Note: The table shows descriptive statistics for mortgage applicants in our sample between 2009-2015. The first column shows all applicants in our sample of credit bureau records. The second column shows our main applicant sample obtained after conditioning on the applicant having a current address (in Infutor), non-missing information on minority/income status, a credit score at the time of application, and non-missing default outcomes 24 months after the application date.}
    \end{flushleft}
\end{table}


    
    



\begin{table}[H]
\caption{Attenuation in Credit Scores by Group}\small \label{attenuationtable} \centering

\begin{tabular}{lllll} \toprule
Log odds of default &            &              &            &               \\
                    & Minority   & Non-minority & Low Income & Higher Income \\ \midrule
                    &            &              &            &               \\
Credit Score        & -0.010***  & -0.015***    & -0.011***  & -0.020***     \\
                    & (0.000) & (0.000)   & (0.000) & (0.000)    \\
                    &            &              &            &               \\
Constant            & 4.377***   & 7.845***     & 5.341***   & 11.05***      \\
                    & (0.002)    & (0.000)      & (0.001)    & (0.001)       \\ \midrule
Observations        & 9,452,286  & 35,814,724    & 11,935,212 & 34,934,643    \\
R-squared           & 0.635      & 0.791        & 0.803      & 0.951        \\ \bottomrule
\end{tabular}
    \bigskip
    
    \begin{flushleft}
    \small{Note: The table shows regression results from regressing the log odds of default on the VantageScore 3.0 credit score at the time of a mortgage application. We first aggregate the log odds of default by one-point credit score bins.}
    
    \end{flushleft}

\end{table}


\begin{landscape}
\begin{table}[H]
\centering
    \caption{Model Tuning: Random Search Results}
    \label{table_ml_params_mc}
   \resizebox{1.3\textwidth}{!}{

\begin{tabular}{lcccccccc}\toprule
\multicolumn{9}{c}{ Panel a: Random Search Results - XGBoost } \\ \toprule
XGBoost Classifier & & & & & & & & \\
Hyperparameter & maxdepth & minchildweight & nestimators & learning rate & gamma & maxdeltastep & subsample & colsamplebytree \\
Search space & [1,2,3,...,9,10] & [1,2,3,..,9,10] & [100,500,1000] & [0.01, 0.1, 0.3] & [0, 0.1] & [0, 10, 20] & [0.5, 0.75, 1] & [0.3, 0.5, 0.75, 1] \\\toprule
& & & & & & & & \\
Tuned Models (score = neg. LogLoss) & & & & & & & & \\
Baseline: Pooled sample & 10 & 5 & 1000 & 0.01 & 0.1 & 0 & 0.75 & 0.3 \\
Different models: Different sample size (Minority) & 10 & 5 & 1000 & 0.01 & 0.1 & 0 & 0.75 & 0.3 \\
Different models: Different sample size (Non-minority) & 10 & 5 & 1000 & 0.01 & 0.1 & 0 & 0.75 & 0.3 \\
Different models: Same sample size (Non-minority) & 9 & 2 & 1000 & 0.01 & 0 & 20 & 0.5 & 0.5 \\
Different models: Different sample size (Low Income) & 10 & 5 & 1000 & 0.01 & 0.1 & 0 & 0.75 & 0.3 \\
Different models: Different sample size (Higher Income) & 10 & 5 & 1000 & 0.01 & 0.1 & 0 & 0.75 & 0.3 \\
Different models: Same sample size (Higher Income) & 10 & 5 & 1000 & 0.01 & 0.1 & 0 & 0.75 & 0.3 \\
Re-weight training data (Minority/Non-minority) & 10 & 5 & 1000 & 0.01 & 0.1 & 0 & 0.75 & 0.3 \\
Re-weight training data (Income) & 10 & 3 & 500 & 0.01 & 0 & 10 & 0.5 & 0.3 \\
& & & & & & & &  \\
\multicolumn{7}{c}{ Panel b: Random Search Results - Random Forest } & & \\\toprule
Random Forest Classifier & & & & & & & & \\
Hyperparameter & n stimators & max features & min samples eaf & min samples split & max depth & & & \\
Search space & [500] & [10, 20, 40, 80, 120, 160, 200] & [1, 2, 4, 6, 8] & [2, 4, 6, 8] & [10, 30, 50, 70, 90, None] & & & \\\toprule
& & & & & & & & \\
Tuned Models (score = neg. MSE) & & & & & & & & \\
VantageScore 3.0  & & & & & & & & \\
Baseline: Pooled sample & 500 & 40 & 1 & 2 & 50 & & & \\
Different models: Different sample size (Minority) & 500 & 20 & 1 & 4 & 70 & & & \\
Different models: Different sample size (Non-minority) & 500 & 80 & 1 & 4 & 50 & & & \\
Different models: Same sample size (Non-minority) & 500 & 120 & 2 & 8 & 50 & & & \\
Different models: Different sample size (Low Income) & 500 & 80 & 1 & 4 & 50 & & & \\
Different models: Different sample size (Higher Income) & 500 & 40 & 1 & 2 & 50 & & & \\
Different models: Same sample size (Higher Income) & 500 & 80 & 1 & 8 & 80 & & & \\
Re-weight training data (Minority/Non-minority) & 500 & 80 & 1 & 4 & 50 & & & \\
Re-weight training data (Income) & 500 & 80 & 1 & 4 & 50 & & & \\
 \\\bottomrule 
\end{tabular}
   }
    \bigskip 

    \begin{flushleft}
        \small{Note: The table shows the hyperparameters chosen by our tuning procedures (5-fold cross-validation with random search) for both the Random Forest and XGBoost. The top row of each panel shows the hyperparameter search space. }
    \end{flushleft}
    
\end{table}
\end{landscape}


\begin{table}[H]
\caption{Predicting Minority/Income Status}\small \label{tab:predict_group} \centering
\begin{tabular}{ccccc}\toprule
\multicolumn{5}{c}{ XGBoost }\\\midrule
& \multicolumn{2}{c}{ Minority } & \multicolumn{2}{c}{ Low Income } \\
\cmidrule(lr){2-3}\cmidrule(lr){4-5}
 & Predicted: 0 & Predicted: 1 &  Predicted: 0 & Predicted: 1 \\
True Label: 0 & 131528 & 4944 & 124873 & 7545 \\
True Label: 1 & 32787 & 5962 & 10117 & 32686 \\
AUC & 0.727 & & 0.853 & \\
Precision & 0.547 & & 0.812 & \\
Recall & 0.154 & & 0.764 & \\\midrule
& &  & & \\
\multicolumn{5}{c}{ Random Forest }\\\midrule
& \multicolumn{2}{c}{ Minority } & \multicolumn{2}{c}{ Low Income } \\
\cmidrule(lr){2-3}\cmidrule(lr){4-5}
 & Predicted: 0 & Predicted: 1 & Predicted: 0 & Predicted: 1 \\
True Label: 0 & 133684 & 2788 & 119402 & 13016 \\
True Label: 1 & 35008 & 3741  & 9542 & 33261 \\
AUC & 0.712 & & 0.839 & \\
Precision & 0.573 & & 0.719 & \\
Recall & 0.097 & & 0.777 & \\\midrule
& & & & \\
\multicolumn{5}{c}{ Logit }\\\midrule
& \multicolumn{2}{c}{ Minority } & \multicolumn{2}{c}{ Low Income } \\
\cmidrule(lr){2-3}\cmidrule(lr){4-5}
 & Predicted: 0 & Predicted: 1 & Predicted: 0 & Predicted: 1 \\
True Label: 0 & 134420 & 2052 & 125746 & 6672 \\
True Label: 1 & 36599 & 2150 & 30397 & 12406 \\
AUC & 0.661 & & 0.754 & \\
Precision & 0.512 & & 0.650 & \\
Recall & 0.055 & & 0.290 & \\ \bottomrule
\end{tabular}


    \bigskip
    
    \begin{flushleft}
    \small{Note: The table shows confusion matrices and predictive performance for a set of machine learning models that predict minority and low income status, respectively. Performance is evaluated on the same test data set as in Table \ref{table_ml_applicant_mc}. See the note to Table \ref{tab:default_confusionmatrix} for definitions of Precision, Recall, and Accuracy. Logit refers to a 20-variable logit model.}
    
    \end{flushleft}

\end{table}


\begin{table}[]
    \caption{First Stage Regressions (Refinance loans, weighted)}
    \label{table_firststage_refi_mc}
    
    \centering
    
    \resizebox{0.65\textwidth}{!}{
    
    \begin{tabular}{lcccc} \toprule
                   Number of loans                                               & {Non-minority} & {Minority} & {Low Income} & {Higher Income} \\ \midrule
    &   & \\
    Bank Exam                                               & 0.956***              & 1.454***   & 0.470***            & 1.650***        \\
                                                                  & (0.167)              & (0.452)    & (0.190)           & (0.321)       \\
    
    Constant                                                      & 6.277***              & 8.225***    & 3.862***            & 10.673***      \\
                                                                  & (0.020)             & (0.045)   & (0.020)           & (0.038)       \\
                                                                  &                       &          & &         \\
     \textit{Controls}                                             &                       &            & &       \\
     Bank x Tract FE                                               & X                     & X     & X                     & X            \\
    Tract x Time FE                                               & X                     & X      & X                     & X           \\
                                                                  &                       &           & &        \\ \midrule

    N                                                             & 130,871                & 64,294       & 67,170            & 139,441      \\
    First-stage F-stat & 32.82 & 10.37 & 26.45 & 6.11 \\
    R-sq                                                          & 0.793                 & 0.845       & 0.821            & 0.821      \\ \bottomrule
    
    & & & & \\ 
    \end{tabular}
    
    } 
    
    \bigskip
    
    \begin{flushleft}
    \small{Note: The table shows results from estimating the first-stage regression of loan origination volumes on CRA exam activity in eligible geographies, as described in equation \ref{eq:firststage}. Group membership is described in the text. Regressions are weighted by a bank's origination share in a given geography. Standard errors (in parentheses) are clustered at the geography level.}
    
    \end{flushleft}

\end{table}


\begin{table}[H]
    \caption{First Stage Regressions - Robustness (Refinance, unweighted)}
    \label{table_firststage_uw_robustness_refi_mc}
    
    \centering
    
    \resizebox{0.65\textwidth}{!}{
    

\begin{tabular}{lcccc} \toprule
                   Number of loans                                               & Non-minority & Minority & Low Income & Higher Income \\ \midrule
    &   & \\
    Bank Exam                                               & 0.363***              & 0.354**   & 0.074           & 0.473***        \\
                                                                  & (0.067)              & (0.158)    & (0.064)           & (0.098)       \\
    
    Constant                                                      & 2.953***              & 3.348***    & 2.161***            & 3.421***      \\
                                                                  & (0.009)             & (0.019)   & (0.008)           & (0.013)       \\
                                                                  &                       &          & &         \\
     \textit{Controls}                                             &                       &            & &       \\
     Bank x Tract FE                                               & X                     & X     & X                     & X            \\
    Tract x Time FE                                               & X                     & X      & X                     & X           \\
                                                                  &                       &           & &        \\ \midrule

    N                                                             & 51,687               & 18,053       & 16,233         & 54,297      \\
            First-stage F-stat & 29.060 & 5.024 & 23.530 & 1.371 \\

    R-sq                                                          & 0.684             & 0.701      & 0.676          & 0.699     \\ \bottomrule

    \end{tabular}
    
    } 
    
    \bigskip
    
    \begin{flushleft}
    \small{Note: The table shows results from estimating the first-stage regression of loan origination volumes on CRA exam activity in eligible geographies, as described in equation \ref{eq:firststage}. Group membership is described in the text. Standard errors (in parentheses) are clustered at the tract level. The regression specification is unweighted.}
    \end{flushleft}

\end{table}


\begin{table}[H]
    \caption{Second Stage Regressions - Robustness (Refinance, unweighted)}
    \label{table_secondstage_uw_robustness_refi_mc}
    
    \centering
    
    \resizebox{\textwidth}{!}{
    

 \begin{tabular}{lcccccccc} \toprule
                                                                  & \multicolumn{2}{c}{{Non-minority}}   & \multicolumn{2}{c}{{Minority}}    & \multicolumn{2}{c}{{Low Income}}   & \multicolumn{2}{c}{{Higher Income}}   \\\cmidrule(lr){2-5}\cmidrule(lr){6-9}
            Number of loans in default                                                     & {Marginal}  & {Average}     & {Marginal}    & {Average}   & {Marginal}    & {Average} & {Marginal}    & {Average}   \\ \midrule
            & & & & & & & & \\
    Originated Loans                                              & 0.053***           & 0.039***           & 0.069        & 0.056***   &   0.270      & 0.068***          & 0.037**          & 0.039***        \\
                                                                  & (0.019)            & (0.003)           & (0.046)             & (0.004)  & (0.250)            & (0.007)           & (0.015)             & (0.003)            \\
                                                                  &                      &                      &                      &   &                      &                      &                      &                   \\
    \textit{Controls}                                                      &                      &                      &                      & &                      &                      &                      &                      \\
    Bank x Tract FE                                               & X                    & X                     & X                    &  X & X                    & X                     & X                    &  X                   \\
    Tract x Time FE                                               & X                    & X                     & X                    &  X    & X                    & X                     & X                    &  X                \\
                                                                  &                      &                      &                      &   &                      &                      &                      &                    \\ \midrule
              First-stage F-stat & 29.060 & - &  5.024 &-&  23.530 & - & 1.371 & -  \\

    N                                                             & 51,687             & 51,687              & 18,053    & 18,053  & 16,233  & 16,233    & 54,297            & 54,297              \\
    R-sq                                                          & 0.092              & 0.475             & 0.207    & 0.610  & -0.768    & 0.515   & 0.167    & 0.553          \\ \bottomrule
                          
    \end{tabular}
   
    } 
    
    \bigskip
    
    \begin{flushleft}
    \small{Note: The table shows results from estimating the second-stage regression of loan default volumes (any delinquency in 24 months post-origination) on loan origination volumes instrumented by variation in CRA exam activity, as described in equation \ref{eq:2sls}. Averages come from OLS regressions including the same fixed effects as the second stage. Group membership is described in the text. Standard errors (in parentheses) are clustered at the tract level. The regression specification is unweighted.
    }
    \end{flushleft}
    
\end{table}


\begin{table}[H]
    \caption{Second Stage Regressions - Robustness (Refinance, 90+ delinquency)}
    \label{table_secondstage_90plus_robustness_refi_mc}
    
    \centering
    
    \resizebox{\textwidth}{!}{
    

\begin{tabular}{lcccccccc} \toprule
                                                             Number of loans in default     & \multicolumn{2}{c}{{Non-minority}}   & \multicolumn{2}{c}{{Minority}}    & \multicolumn{2}{c}{{Low Income}}   & \multicolumn{2}{c}{{Higher Income}}   \\\cmidrule(lr){2-5}\cmidrule(lr){6-9}
            (90+ days delinquency)                                                     & {Marginal}  & {Average}     & {Marginal}    & {Average}   & {Marginal}    & {Average} & {Marginal}    & {Average}   \\ \midrule
            & & & & & & & & \\
    Originated Loans                                              & 0.014           & 0.014***           & 0.025*          & 0.014***  & 0.063      & 0.017***          & 0.008           & 0.011***        \\
                                                                  & (0.011)            & (0.003)           & (0.013)             & (0.004)  & (0.047)            & (0.005)           & (0.008)             & (0.002)            \\
                                                                  &                      &                      &                      &   &                      &                      &                      &                   \\
    \textit{Controls}                                                      &                      &                      &                      & &                      &                      &                      &                      \\
    Bank x Tract FE                                               & X                    & X                     & X                    &  X & X                    & X                     & X                    &  X                   \\
    Tract x Time FE                                               & X                    & X                     & X                    &  X    & X                    & X                     & X                    &  X                \\
                                                                  &                      &                      &                      &   &                      &                      &                      &                    \\ \midrule
  First stage F-stat & 32.380 & - & 10.370 & - & 6.111 & - & 26.450 & - \\
    N                                                             & 130,871               & 130,871               & 64,294                & 64,294 & 67,170               & 67,170               & 139,441              & 139,441                  \\
    R-sq                                                          & 0.064               & 0.486              & 0.030              & 0.576    & -0.160              & 0.524              & 0.067            & 0.566          \\ \bottomrule

    \end{tabular}

    } 

    \bigskip
    
    \begin{flushleft}
    \small{Note: The table shows results from estimating the second-stage regression of loan default volumes (delinquency of 90+ days in 24 months post-origination) on loan origination volumes instrumented by variation in CRA exam activity, as described in equation \ref{eq:2sls}. Averages come from OLS regressions including the same fixed effects as the second stage. Regressions are weighted by a bank's origination share in a given geography. Group membership is described in the text. Standard errors (in parentheses) are clustered at the tract level.
    }
    \end{flushleft}

\end{table}

\FloatBarrier\newpage \pagebreak

\justify
\section{Appendix: Data Construction} \label{appendix_datamerge}

\setcounter{figure}{0} \renewcommand{\thefigure}{B.\arabic{figure}} 
\setcounter{table}{0} \renewcommand{\thetable}{B.\arabic{table}} 
 
\subsection{Merging Infutor and Property Records}
We first match addresses from Infutor with addresses from tax assessor’s
records of property characteristics and housing transaction data from CoreLogic. Details of this merge are provided in Appendix \ref{sec:addressmerge}. Next, we construct an address-year panel with property owner names of each address in each year using both tax records and transactions data from CoreLogic. More details for creating this owner panel are provided in Appendix \ref{sec:ownershippanel}. Finally,
we compute the Jaro-Winkler string distance between the last names of individuals from Infutor and the last names of property owners from CoreLogic in each year. An individual is classified as a homeowner in a given year if the string distance between her last name and any last name of
the owners at his address in that year is greater or equal to 0.9. See Appendix Section \ref{sec:cutoff} for robustness to other cutoffs. 

For sake of full attribution, we also emphasize: \textbf{some of the data construction was performed jointly with \cite{qian2020effects} and we have jointly written parts of this data appendix.}

\subsubsection{Infutor-CoreLogic Address Merge}\label{sec:addressmerge}
We first convert all distinct addresses in both Infutor and CoreLogic into clean, standardized versions. Note that addresses in CoreLogic are uniquely identified by the APN, county FIPS, and APN sequence number.  We create 5-digit zip codes and 2-digit state codes, we remove all special characters and spaces from street names, split apartment numbers  into numeric and non-numeric parts, and standardize city names using common abbreviations suggested by USPS.  We also create a second city variable based on the preferred spelling of a city name by USPS (preferred city). We then keep all unique addresses in Infutor and CoreLogic respectively. We then merge the two address data sets using the following iterative procedure (at each step we remove uniquely matched observations). 
\begin{itemize}
    \item All address variables (street name and number, apartment number, zip code, city, state);
    \item All variables but only numeric portion of apartment;
    \item All variables using preferred city name;
    \item All variables using preferred city name but only numeric portion of apartment; \item  All variables leaving out zip code;
    \item All variables leaving out zip and using preferred city names;
    \item All variables leaving out zip but only numeric portion of apartment;
    \item All variables but dropping apartment number if and only if there is no apartment number in CoreLogic. This step captures cases where the CoreLogic address is the whole apartment complex but Infutor shows addresses for individual apartments.
    \item Same as previous but using preferred city. 
\end{itemize}

A note on apartment numbers. We observe cases where multi-unit buildings have a single owner in CoreLogic, and hence a missing apartment number, but we observe multiple renters of the individual units in Infutor. If we forced a match on apartment number, we would not match the multi-unit buildings since apartment number is missing in one dataset but not the other. This would imply that we would potentially miss out on classifying a large number of renters in Infutor. For this reason, we allow a final merge  round that does not condition on apartment number. However, this implies that we generate a number of non-unique matches since now one CoreLogic address is associated with multiple Infutor addresses. This non-uniqueness however does not pose a problem for our owner-renter imputation (see details below).

Table \ref{tab:addressmerge} provides the fraction of Infutor addresses that we successfully match to CoreLogic by state. We get low match rates (as \% of Infutor addresses matched) for a handful of states. We confirmed with CoreLogic customer relations representatives that this is driven by low CoreLogic coverage for these states as the deed records in these states have systems in place that make the digitization of the records more difficult. 

\begin{figure}[hbt!]\caption{Homeownership Rate Time Series \label{fig:homeowner_rate}}
   \centering
    \includegraphics[width=0.7\textwidth]{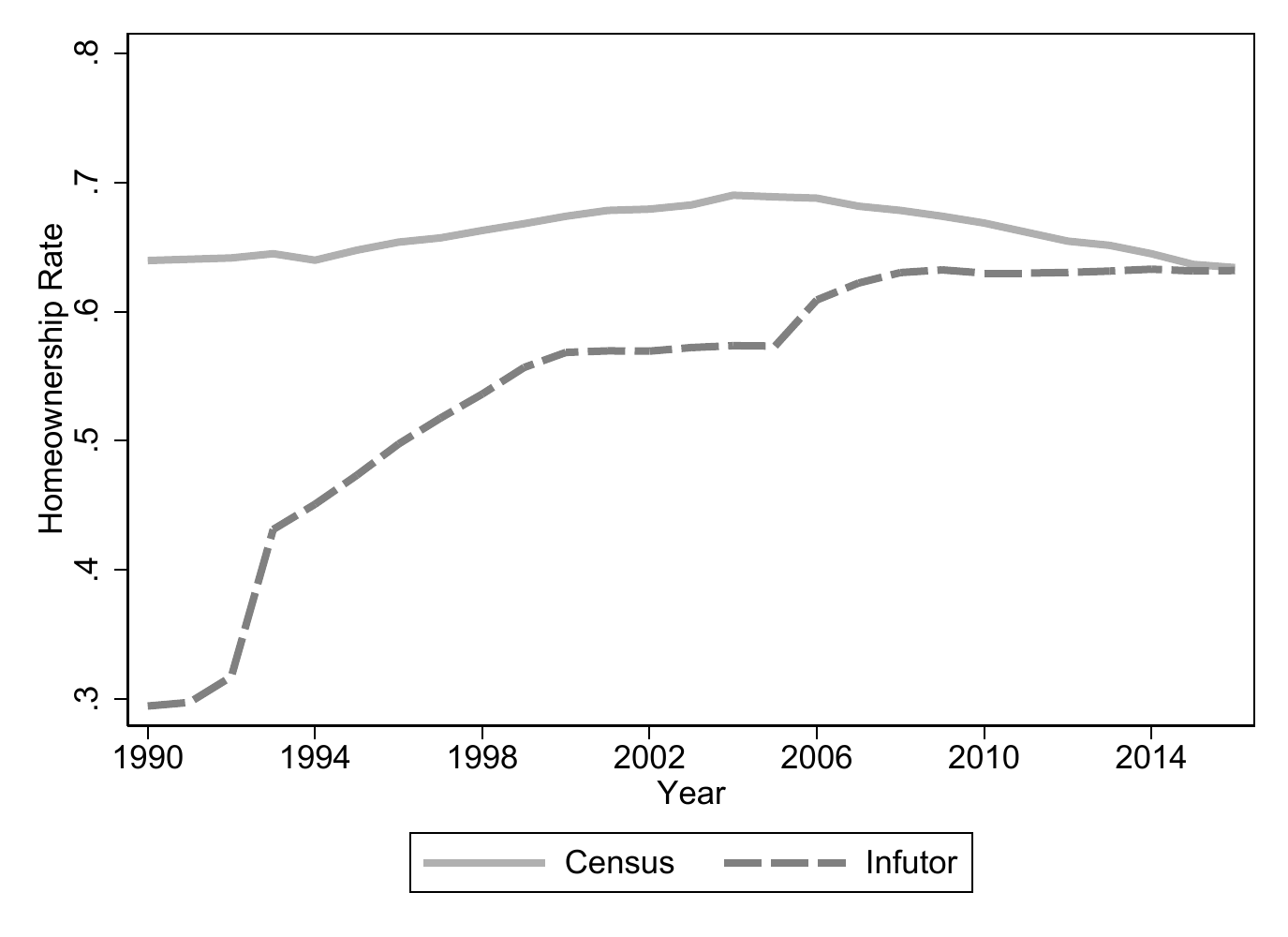}
    \begin{flushleft}
    \small{Note: This figure compares the time series of homeownership rate at the national level from the U.S. Census Bureau and imputed from our Infutor-CoreLogic linked dataset. The Census homeownership rate is the annual average at the national level without seasonal adjustment. The Infutor homeownership rate is calculated using the full Infutor-CoreLogic linked dataset.}
    \end{flushleft}
\end{figure}
\begin{figure}[hbt!]\caption{Validation of Infutor-CoreLogic Homeownership Status \label{fig:homeowneryear}}
   \centering
    
       \begin{subfigure}[h]{0.35\textwidth}
    \includegraphics[width=1.2\textwidth]{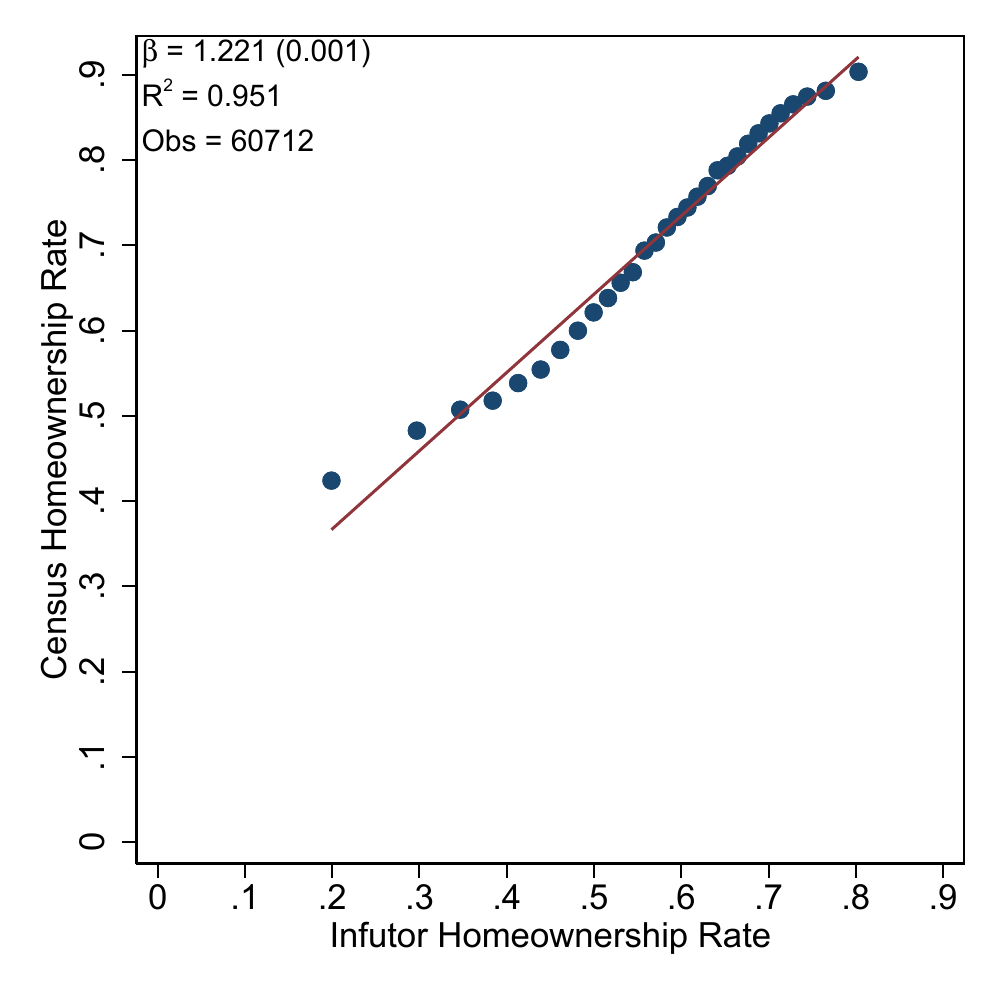}\caption{2000}
     \end{subfigure} 
    \begin{subfigure}[h]{0.35\textwidth}
    \includegraphics[width=1.2\textwidth]{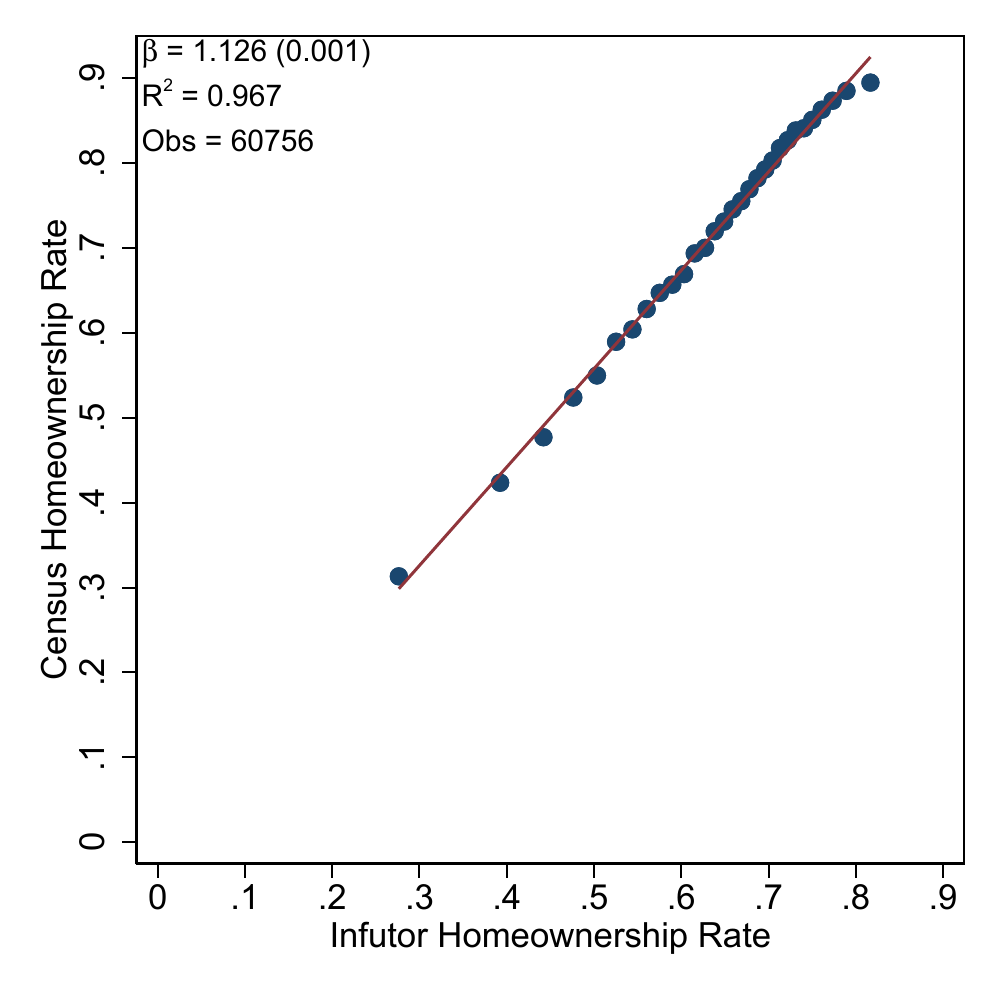}\caption{2010}
    \end{subfigure}\\
       \begin{subfigure}[h]{0.35\textwidth}
    \includegraphics[width=1.2\textwidth]{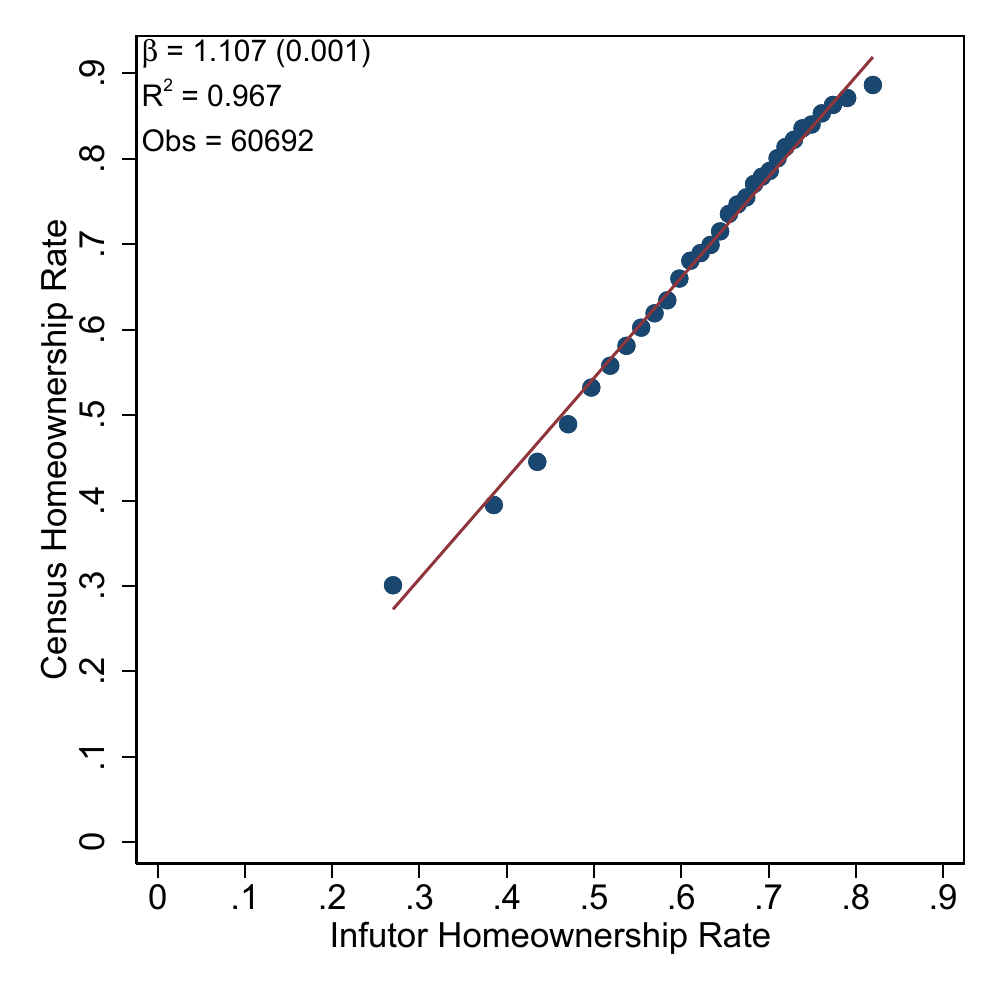}\caption{2015}
    \end{subfigure}\\
  
    \begin{flushleft}
    \begin{spacing}{1.1}
    \small{Note: This figure presents binscatter plots that validate the home ownership imputation at tract level. For the three panels, we keep tracts more than 10 percentile-level observations for which we can impute a home ownership status. We divide the sample into 30 bins and plot the average value for each bin. The x-axis is the Infutor home ownership rate. The y-axis is the Census home ownership rate. The regression is weighted by observations in each tract for which a home ownership status can be imputed
    . 
    In Panel (a), we compare with the home ownership rate from the 2000 Decennial Census using 2010 Census Tract boundaries. In Panel (b), we compare with the home ownership rate from the 2008–2012 5-year ACS. In Panel (c), we compare with the home ownership rate from the 2013–2017 5-year ACS.}
    \end{spacing}
    \end{flushleft}
\end{figure}


\subsubsection{Construction of Ownership Panel}\label{sec:ownershippanel}
To construct the ownership variable, we first use CoreLogic transaction and tax records to establish the owners for each address-year combination. We find the historical tax records to be very incomplete, so we rely on 2017 tax records and reconstruct ownership based on past transactions. We proceed in the following steps:

\begin{itemize}
    \item Create a bookend based on the most recent transaction and tax entry. Note that for transactions, we use the buyer in the last transaction in that year. That is, our imputed owner variable will be the owner ``as of the end of the year". If there is a tax and transaction entry in the same year, the transaction data takes precedence.
    \item  Fill in forward: If the most recent bookend is in a year prior to 2017 (the end of our CoreLogic sample), we assume that owner remains the same until 2017.
    \item Fill in backward: First, we create additional data points from prior transactions. Every time there is a transaction, we say that the last buyer in that year is the owner for the year. We then fill in data between the transactions: If we have a transaction in year t, then the owner at t-1 is the seller of the first transaction at t. For t-2, t-3 etc. we then assume the owner from t-1 remains the owner (until we hit the next transaction). For tax-only entries, we just assume that the owner at t remains the owner. 
\end{itemize}

For robustness, we use an alternative fill-in procedure that only uses transaction data: 
\begin{itemize}
    \item    We find the last buyer in the earliest transaction for that address. That's our first owner.
    \item  That person remains the owner until there is another transaction. Then the last buyer in that year becomes the new owner and remains the owner until there is another transaction.
\end{itemize}

When matching against names in Infutors, we match against owner names from both fill-in procedures if they conflicting information. We repeat this procedure for the property type. 

In the final step, we compare whether the name in Infutor matches the owner name on the deed record. We consider both first and second owners. We first clean  names in both data sets and remove any special characters and spaces. We then compute the Jaro-Winkler string distance between the last names of individuals from Infutor and the last names of property owners from CoreLogic in each year. An individual is classified as a homeowner in a given year if the string distance between her last name and any last name of the owners at his address in that year is greater or equal to 0.9.

Figure \ref{fig:homeowner_rate} shows the time series comparison between the home ownership rate from Censuses (retrieved from FRED) and our imputed home ownership rate from the Infutor-CoreLogic linked data. We consistently under predict home ownership, but over time, the gap between our home ownership rate and the Census home ownership rate becomes smaller. Figure \ref{fig:homeowneryear} shows we can capture
over 95\% of the variation in home ownership rate at the Census Tract level in different years.

\subsubsection{Robustness to Distance Cut-off}\label{sec:cutoff}
We also explore using string distance cutoff values of 0.8, 0.85 and 0.95. We check a random sample of 300 cases that are inconsistent for the different cutoffs. The majority of cases that are a match for 0.8 but not a match for 0.85 are not true matches. Cases that are a match for 0.85, but not for
0.9 are noisy, with about 50\% of them being true matches. Cases that are a match for 0.9, but not for 0.95 are mostly true matches. Only 0.88\% of all cases have inconsistencies using different cutoff values of 0.85, 0.9, and 0.95. In addition, when an Infutor last name has more than 3 characters
and is fully contained in CoreLogic owner last name and vice versa, we adjust the string distance to be 1 and call it a match. This deals with cases where an individual’s last name is contained in a corporation’s name, such as a trust or an estate. This also deals with the structure of some Hispanic
names, for example, ``Jimenez de Armendariz” is matched with ``Armendariz” but is not matched with simply ``de.”

\subsection*{Infutor and Credit Bureau}
We probabilistically sampled 80m individuals from Infutor between ages 18 and 65. For each individual, we provided TransUnion with full name, a preferred address, and a social security number when available. TransUnion returned 49m matched individuals. For the matched individuals, TransUnion provided semi-annual credit report data (May and November months) between 2009-2016, with an additional data draw for November 2017. TransUnion removed all personal identifying information (name, address, SSN) prior to returning the data.

\pagebreak \FloatBarrier
\begin{table}
\centering

\caption{Infutor-CoreLogic: Address Merge Rates}\label{tab:addressmerge}
\label{table_address_merge}

    \resizebox{0.45\textwidth}{!}{
\begin{tabular}{lllll} \toprule
State & Infutor & CoreLogic & Matches & Match rate \\ \midrule
AK & 549,804 & 577,030 & 178,778 & 0.33 \\
AL & 3,686,368 & 3,381,063 & 1,640,014 & 0.44 \\
AR & 2,264,192 & 2,774,138 & 1,117,671 & 0.49 \\
AZ & 4,084,237 & 5,895,776 & 2,140,958 & 0.52 \\
CA & 21,752,607 & 22,692,583 & 10,271,136 & 0.47 \\
CO & 3,637,375 & 4,882,710 & 1,831,535 & 0.5 \\
CT & 2,254,395 & 501,270 & 219,599 & 0.1 \\
DC & 519,947 & 403,199 & 177,154 & 0.34 \\
DE & 673,260 & 814,861 & 341,318 & 0.51 \\
FL & 13,834,611 & 20,195,793 & 7,597,739 & 0.55 \\
GA & 6,545,877 & 7,749,462 & 3,314,527 & 0.51 \\
HI & 991,644 & 1,189,447 & 389,401 & 0.39 \\
IA & 2,114,242 & 2,559,005 & 1,188,525 & 0.56 \\
ID & 1,178,220 & 1,455,100 & 554,947 & 0.47 \\
IL & 8,479,168 & 9,809,275 & 4,273,576 & 0.5 \\
IN & 4,620,489 & 5,518,307 & 2,369,448 & 0.51 \\
KS & 1,977,176 & 2,217,902 & 1,040,895 & 0.53 \\
KY & 3,390,553 & 3,116,883 & 1,580,044 & 0.47 \\
LA & 3,326,382 & 3,391,884 & 1,641,317 & 0.49 \\
MA & 4,129,508 & 960,665 & 380,885 & 0.09 \\
MD & 3,587,829 & 4,640,438 & 1,890,075 & 0.53 \\
ME & 1,250,021 & 139,026 & 30,400 & 0.02 \\
MI & 7,000,246 & 5,635,832 & 3,728,442 & 0.53 \\
MN & 3,447,233 & 4,290,624 & 1,895,469 & 0.55 \\
MO & 4,307,747 & 4,990,370 & 2,150,686 & 0.5 \\
MS & 2,434,565 & 1,901,738 & 969,713 & 0.4 \\
MT & 801,754 & 1,084,111 & 350,949 & 0.44 \\
NC & 7,072,070 & 9,315,465 & 3,586,847 & 0.51 \\
ND & 522,541 & 483,394 & 200,736 & 0.38 \\
NE & 1,247,222 & 1,394,951 & 643,619 & 0.52 \\
NH & 1,043,455 & 132,807 & 47,363 & 0.05 \\
NJ & 5,509,166 & 714,524 & 233,778 & 0.04 \\
NM & 1,499,615 & 1,471,116 & 625,330 & 0.42 \\
NV & 1,707,378 & 2,348,467 & 907,666 & 0.53 \\
NY & 12,511,117 & 11,627,570 & 4,742,301 & 0.38 \\
OH & 8,292,963 & 10,086,281 & 4,228,038 & 0.51 \\
OK & 2,910,524 & 3,106,146 & 1,253,660 & 0.43 \\
OR & 2,705,486 & 3,220,908 & 1,358,130 & 0.5 \\
PA & 8,807,052 & 10,301,095 & 4,532,221 & 0.51 \\
RI & 685,590 & 313,251 & 127,535 & 0.19 \\
SC & 3,350,214 & 4,473,437 & 1,835,782 & 0.55 \\
SD & 584,486 & 448,048 & 262,306 & 0.45 \\
TN & 4,439,823 & 5,608,419 & 2,379,854 & 0.54 \\
TX & 15,942,543 & 19,570,551 & 7,930,927 & 0.5 \\
UT & 1,706,935 & 2,481,290 & 812,449 & 0.48 \\
VA & 5,216,557 & 6,111,974 & 2,758,078 & 0.53 \\
VT & 581,643 & 59,039 & 3,748 & 0.01 \\
WA & 4,987,250 & 5,584,186 & 2,331,650 & 0.47 \\
WI & 4,433,949 & 5,279,060 & 2,024,908 & 0.46 \\
WV & 1,560,793 & 1,582,449 & 468,113 & 0.3 \\
WY & 439,529 & 512,887 & 195,221 & 0.44 \\
Total & 210,597,351 & 228,995,807 & 96,755,461 & 0.46 \\ \bottomrule
\end{tabular}
}
\begin{flushleft}
\small{Note: This table shows merge rates between unique addresses in Infutor and Core Logic deeds and tax records by state. The Infutor  and CoreLogic columns shows the number of distinct addresses in Infutor and CoreLogic respectively (after cleaning the addresses as described in the text). The match rate refers to the fraction of Infutor addresses matched to CoreLogic.}
\end{flushleft}
\end{table}

\FloatBarrier\newpage\pagebreak
\justify
\section{Appendix: Constructing Refinance/Purchase Classification}\label{appendix_refi}

\setcounter{figure}{0} \renewcommand{\thefigure}{C.\arabic{figure}} 
\setcounter{table}{0} \renewcommand{\thetable}{C.\arabic{table}} 

\begin{figure}[H]
\caption{Refinance Rates in Approved Sample: TransUnion vs HMDA }
\label{refi_validation}
\centering
       \begin{subfigure}[t]{0.45\textwidth}
    \includegraphics[width=1\textwidth]{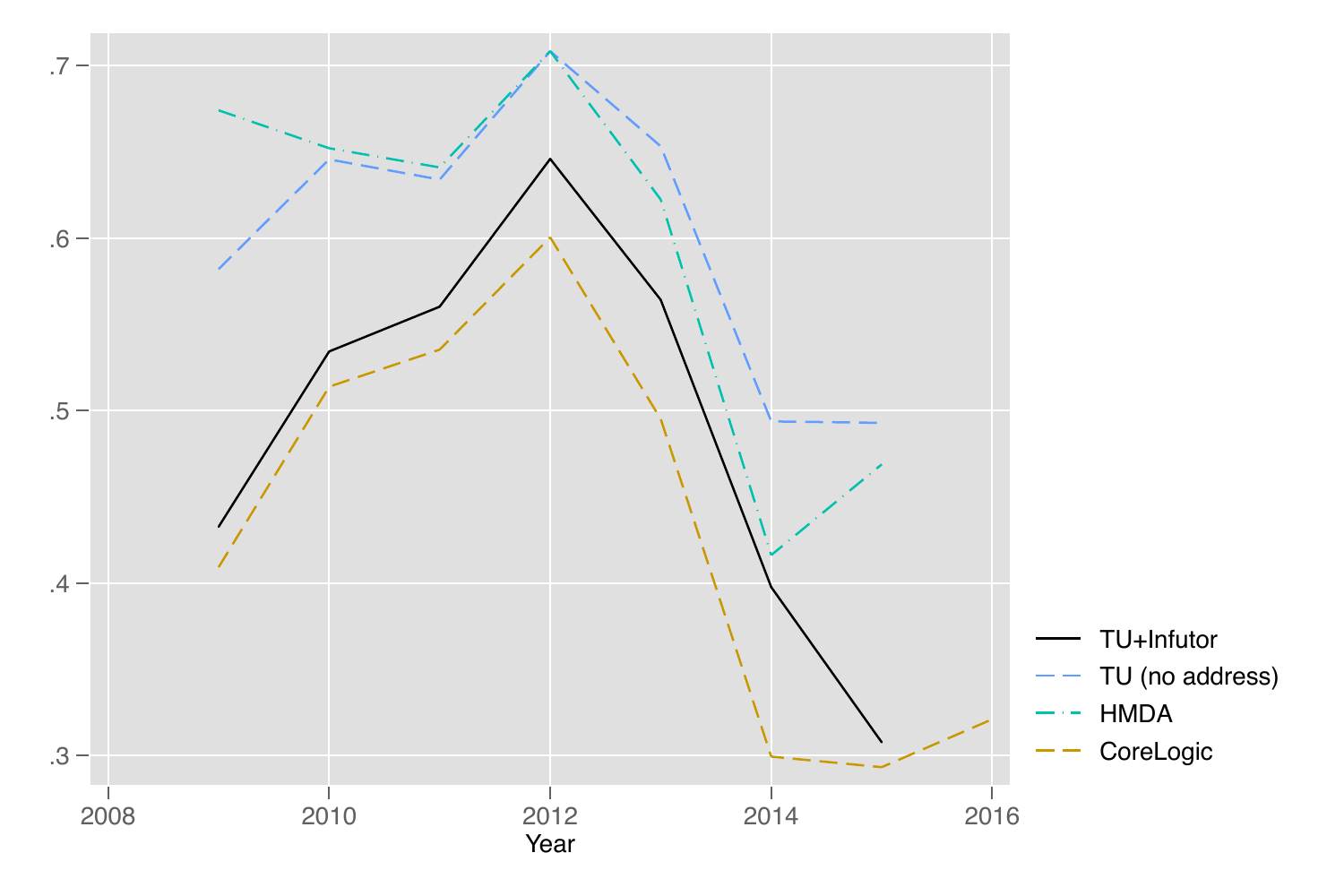}\caption{Originated Mortgages}
     \end{subfigure} 
    \begin{subfigure}[t]{0.45\textwidth}
    \includegraphics[width=1\textwidth]{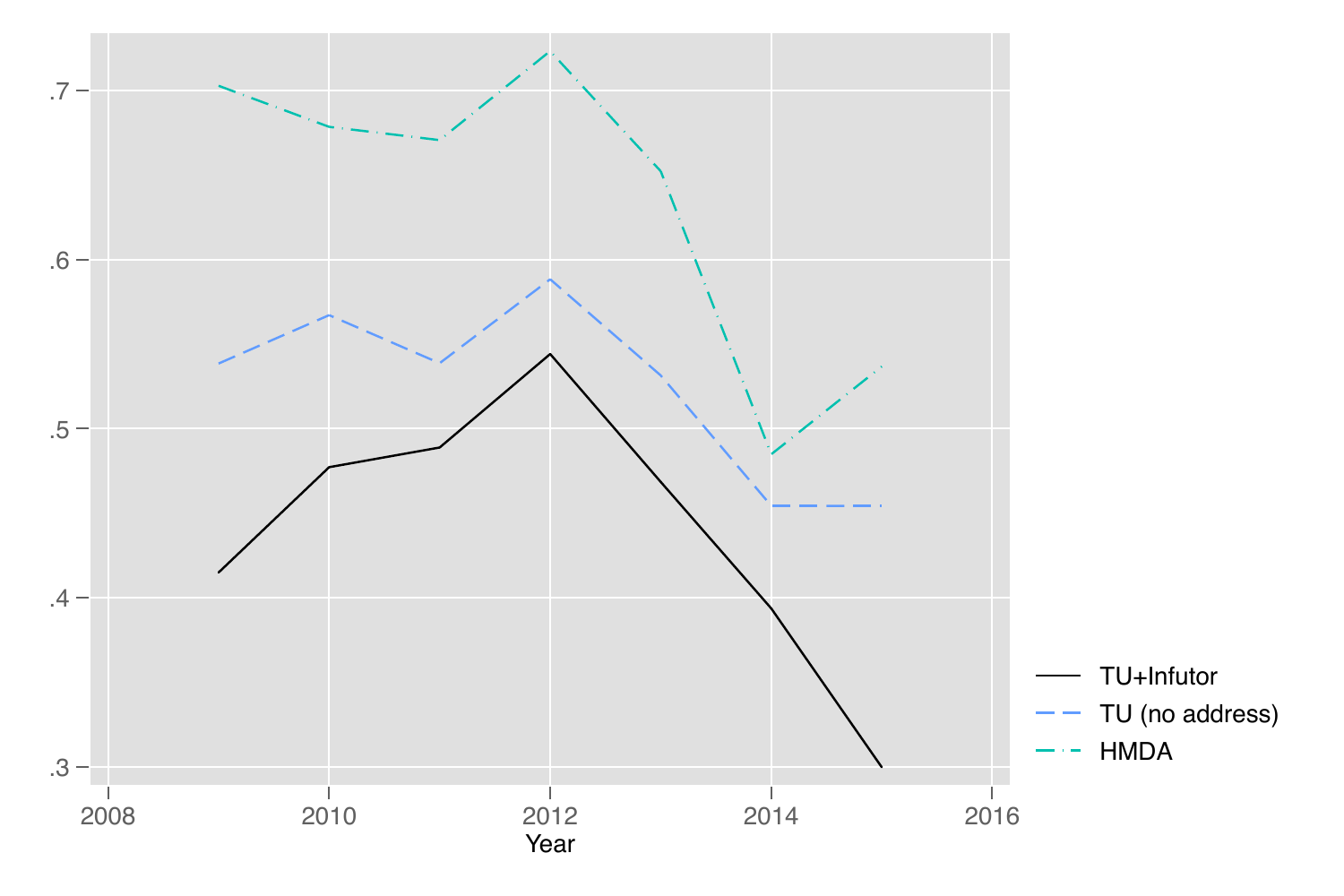}\caption{Mortgage Applications}
    \end{subfigure}
 
    \begin{flushleft}
    \begin{spacing}{1.1}
    \small{Note: This figure shows the share of loans that are refinance loans in the TransUnion sample compared to the universe of HMDA data, as well as CoreLogic Deed Records. TU+Infutor refers to the matched sample of TransUnion and Infutor. TU (no address) refers to a less stringent refi definition that does not impose a constraint on the address remaining the same. CoreLogic refers to a refinance indicator provided in the CoreLogic deeds records. }
    \end{spacing}
    \end{flushleft}
\end{figure}

This appendix describes how we classify both approved mortgages and mortgage applications as refinance or purchase loans, respectively. Our preferred approach combines credit report information and address information from Infutor. We classify a loan or application as a refinance loan if the borrower has at least one open mortgage prior to the application date; the Infutor address indicates that the borrower does not move in a 2-year window around the origination; and the number of open mortgage does not increase (to rule out second homes). We create a second version that drops the address requirement based on HMDA. 
Figure \ref{refi_validation} compares our refi rates to both the refi flag in HMDA and the refi flag provided by CoreLogic in their deeds records (the latter is only available for originated mortgages). Our preferred measure over-predicts refis relative to CoreLogic but underpredicts relative to HMDA. We find that our measure follows very similar dynamics over time as the CoreLogic and HMDA series. The less restrictive version that drops the Infutor address requirement leads to larger refi rates that slightly overpredict refi rates relative to both HMDA and CoreLogic. For this reason, we choose the more restrictive version as our main refi indicator. 

We suspect that we undercount refi applications since credit bureau mortgage inquiries, which we use to define an application, are de-duplicated while HMDA applications are not. That is, if someone applies to three different institutions we observe three applications in HMDA but only one application in the bureau data. If shopping behavior is more pronounced for refis relative to originations, we would expect this to bias our refi application shares downward relative to HMDA. 

\FloatBarrier\newpage\pagebreak

\section{Appendix: Inferring Minority Status} \label{appendix_race}

\setcounter{figure}{0} \renewcommand{\thefigure}{C.\arabic{figure}} 
\setcounter{table}{0} \renewcommand{\thetable}{C.\arabic{table}} 

This appendix provides additional details on the procedure for inferring minority status. We proceed in two steps: We first use first and last names to compute a baseline probability and then update these probabilities using the racial make-up of the census block associated with her or his place of residence. 

Our first step uses two distinct software packages to compute baseline probabilities of race/ethnicity for each individual in our data based on first and last names.  

The first package is a proprietary commercial software (Onolytics) developed by Mateos (2014) which is based on anthropological research on the etymology of first and last names. It is based on a proprietary international database of over 1 million last names and 500,000 first names. We collapse the detailed Onolytics categories to six categories used by the US census.\footnote{These categories are Hispanic, non-Hispanic White, non-Hispanic Black or African American, non-Hispanic Asian/Pacific Islander, non-Hispanic American Indian and Alaska Native, non-Hispanic multi-racial.}

The second package is the ethnicolr package developed by Sood and Laohaprapanon (2018).\footnote{This is Python package is available at \url{https://github.com/appeler/ethnicolr}.} They model the relationship between the sequence of characters in a name and race and ethnicity using Florida Voter Registration data as well as  database of 140k name-race associations from Wikipedia (Ambekar et al. (2009)). 

Our second step updates each individual’s baseline racial/ethnic probabilities with the racial and ethnic characteristics of the census block associated with her or his place of residence using Bayes’ Rule. We compute posterior probabilities based on an individual's 2000 address and 2000 census data on racial and ethnic composition at the block level to create posterior probabilities for the four major racial/ethnic categories used by the US census (Hispanic, non-Hispanic white, non-Hispanic Black or African American, and non-Hispanic Asian/Pacific). 
The posterior probability that an individual with name $s$ residing in geographic area $g$ belongs to race or ethnicity $r$ is then
\begin{equation*}
    \text{Pr}(r|g,s) = \frac{\text{Pr}(r|s)\text{Pr}(g|r)}{\sum_{r'\in R}\text{Pr}(r'|s)Pr(g|r')}
\end{equation*}
where $R$ denotes the set of ethnic categories. 
We then update this posterior again using an individuals 2010 address and the 2010 census data on racial and ethnic composition. 

An individual is assigned to a racial/ethnic category if this category has the highest posterior probability according to both sets of posteriors, and is equal to or above 0.8 on at least one of two sets of posteriors. We make one exception to this rule for non-Hispanic Black or African American. Onolytics has a relatively high mis-classification rate for African-American names who have last names whose etymological roots are European (e.g. ``Washington") and are therefore classified as non-Hispanic white. To address this problem, we assign an individual to the non-Hispanic Black or African American category if the posterior probability of this category based on the ethnicolr baseline is equal to or higher 0.8 even when Onolytics-based posterior places a high probability on the non-Hispanic White category. 

This two step method is similar to methods used by the CFPB to construct race in fair lending analysis. CFPB (2014) and Elliott et al. (2009) show that combining geographic and name-based information outperforms methods using either of these sources of information alone. 

\vskip 10 pt 





\FloatBarrier\newpage\pagebreak

\section{Appendix: Validation Using HMDA} \label{appendix_hmda_income}

\setcounter{figure}{0} \renewcommand{\thefigure}{E.\arabic{figure}} 
\setcounter{table}{0} \renewcommand{\thetable}{E.\arabic{table}} 

This appendix describes in more detail how we use a matched sample with HMDA to validate both our income and minority status measures. 

\subsubsection*{Merging CoreLogic and HMDA}

In what follows, we describe the merge procedure between the CoreLogicand HMDA Loan/Application Registers (LARs) datasets.\footnote{We thank Cody Cook (Stanford GSB) for sharing his initial merge code.} We limit both data sets to the years 2007-2017. Moreover, we merge purchase and refinancing loans separately, splitting observations accordingly before the cleaning steps explained below. The variables in common between the data sets (which we use as merge keys) are the mortgage amount (rounded to nearest thousand), state/county/tract code, year, and lender name.

CoreLogic data includes both Deed Record and Property Tax Record information, which are merged based on county (FIPS) and assessor parcel (APN, APN Sequence Number) identifiers. We drop observations with missing values for any of our merge keys, and convert variables to the appropriate data type for our merge step. Most importantly, we also drop observations that are not uniquely identifiable by our merge key combinations. As a result, we keep 74\% and 65\% of the original Corelogic dataset, respectively for purchase and refinancing. 

HMDA LARs data was obtained through the National Archives online portal. We supplement the HMDA-provided institution names using the publicly available HMDA lender ``Avery" files from Neil Bhutta (FRB), which allows us to link a respondent ID to all possible lender names associated with it. We drop rejected loan applications (based on Action Type) and those with missing values for any of our merge keys. 

Our merge is performed in rounds. We loop over year, state, and counties, merging on census tract (allowing tract code decimal difference due to tract changes in 2010), rounded mortgage amount, and lender name. We increase the rounding of amounts (from one thousand to ten thousand), truncate amounts (instead of rounding), and ``sub-string" lender names (to the first 5 or 7 digits) at each subsequent merge round, allowing for increasingly flexible merge keys. Once an observation is matched to a counterpart from the other data set, the pair is saved and the matched observations are dropped as to not get matched again in subsequent rounds. For each merged pair, we record the respective round in which the merge took place.  Finally, after completing the merge, we drop any observations (from either Corelogic or HMDA) that were matched to more than one counterpart. Table \ref{table_cl_hmda_merge_summary_purchase_mc} includes a summary of the resulting merged dataset in comparison to the two input datasets (post-processing, after keeping only uniquely identified, non-missing, etc).

\begin{table}[H] 
    \caption{CoreLogic-HMDA Merge Summary}
    \label{table_cl_hmda_merge_summary_purchase_mc}

    \centering

    \resizebox{0.85\textwidth}{!}{


\begin{tabular}{lcccccc} \toprule
                       & \multicolumn{3}{c}{Purchase Loans (2007-2017)}                                                                                                                                                 & \multicolumn{3}{c}{Refinancing Loans (2007-2017)}                                                                                                                                              \\ \cline{2-7}
                       & Corelogic  Pre-Processed & HMDA Pre-Processed & Matched & Corelogic Pre-Processed& HMDA Pre-Processed& Matched \\
                       &                                                                                     &                                                                               &                          &                                                                                     &                                                                               &                          \\  \midrule
                       &                                                                                     &                                                                               &                          &                                                                                     &                                                                               &                          \\ 
N                      & 28,260,078                                                                          & 34,747,395                                                                    & 15,951,217               & 51,489,929                                                                          & 44,712,436                                                                    & 23,647,577               \\
                       &                                                                                     &                                                                               &                          &                                                                                     &                                                                               &                          \\
Mortgage amount (000s) & 577.33                                                                              & 234.64                                                                        & 244.33                   & 429.97                                                                              & 227.49                                                                        & 234.79                   \\
                       & (15,185.85)                                                                         & (626.52)                                                                      & (365.09)                 & (25,154.11)                                                                         & (525.63)                                                                      & (336.83)                 \\
                       &                                                                                     &                                                                               &                          &                                                                                     &                                                                               &                          \\
Share minority         & -                                                                                   & 14.82\%                                                                       & 15.23\%                  & -                                                                                   & 11.02\%                                                                       & 11.90\%                  \\
                       &                                                                                     &                                                                               &                          &                                                                                     &                                                                               &     \\ \bottomrule                    
\end{tabular}
} 

\bigskip

    \begin{flushleft}
    \small{Note: The table shows a summary of the Corelogic-HMDA merge results. Corelogic and HMDA counts and amounts refer to the post-cleaning (pre-processed) data, thus only covering observations that had the potential to be merged. Minority status is not present in the CoreLogic data, which is the basis for the merge. A separate merge to Infutor is performed to add imputed race/ethnicity (imputation described separately), and thus minority status, to Corelogic. Note that we de-duplicate CoreLogic observations with identical merge variables prior to the merge.}

    \end{flushleft}
\end{table}

\newpage

\subsubsection*{Income}
Our income variable is based on a credit bureau income estimator provided by TransUnion. The income estimator is available for 91\% of the mortgage applicant sample. 

The income estimator is trained on a 1.2m sample of IRS form 1040 joint income data collected during the mortgage application process from tax years 2008-2012. It predicts total income (not just wage income) which includes investment income, IRA distributions, pensions, and annuities as well as public and private transfer payments. The predicted income range is \$0 to \$1 m. The income estimator is available for consumers who are not deceased and who have one or more trades on file. Some of the key features in the  income estimator model are are spending and payments amounts which provide insight into a consumer's monthly credit card expenditures and payment patterns; historical debt balances and credit utilization; revolving behavior and payment ratios. The data on form 1040 may have been filed jointly or individually. For this reason, the model estimates the probability of an individual filing a joint tax return and, where applicable, bases the joint income estimate on an individual’s credit report.

We validate the income estimator using the merged sample CoreLogic-HMDA sample described above. HMDA contains self-reported income and therefore allows us to compare the bureau income estimator to an independent data source. Note that both income measures are nominal income. Since CoreLogic is linked to our Infutor/credit bureau data set, we can use the above CoreLogic merge to link our credit bureau sample to HMDA. 

\begin{figure}[H] 
 \caption{TransUnion and HMDA Income Distributions}
  \label{hist_income_mc} 
  
  \subcaption*{\large{Purchase Loans}}
  
  \begin{minipage}[b]{0.5\linewidth}
    \centering
    \subcaption{TU income} 
    \includegraphics[width=\linewidth]{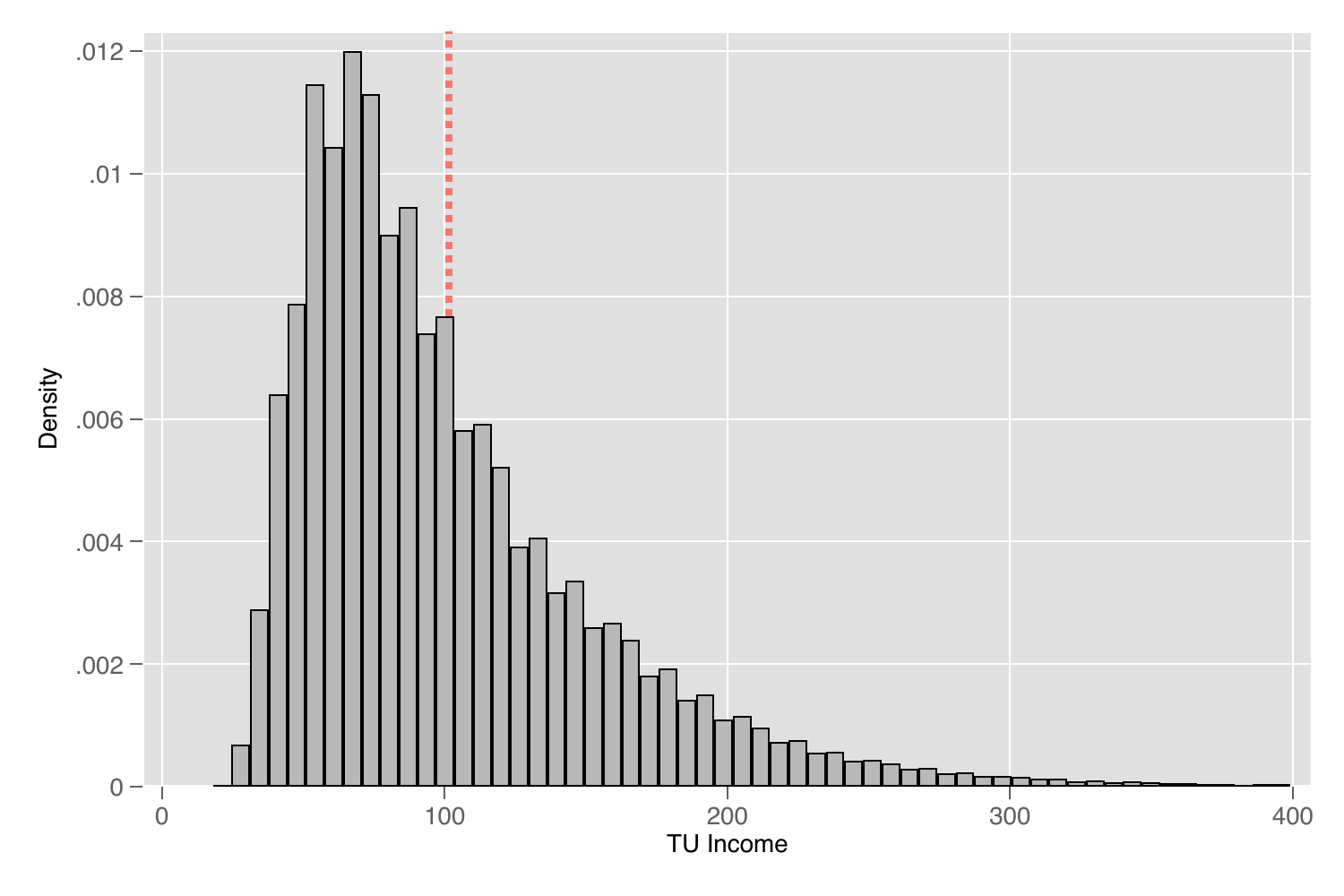} 
    \vspace{1ex}
  \end{minipage}
  \begin{minipage}[b]{0.5\linewidth}
    \centering
    \subcaption{HMDA income} 
    \includegraphics[width=\linewidth]{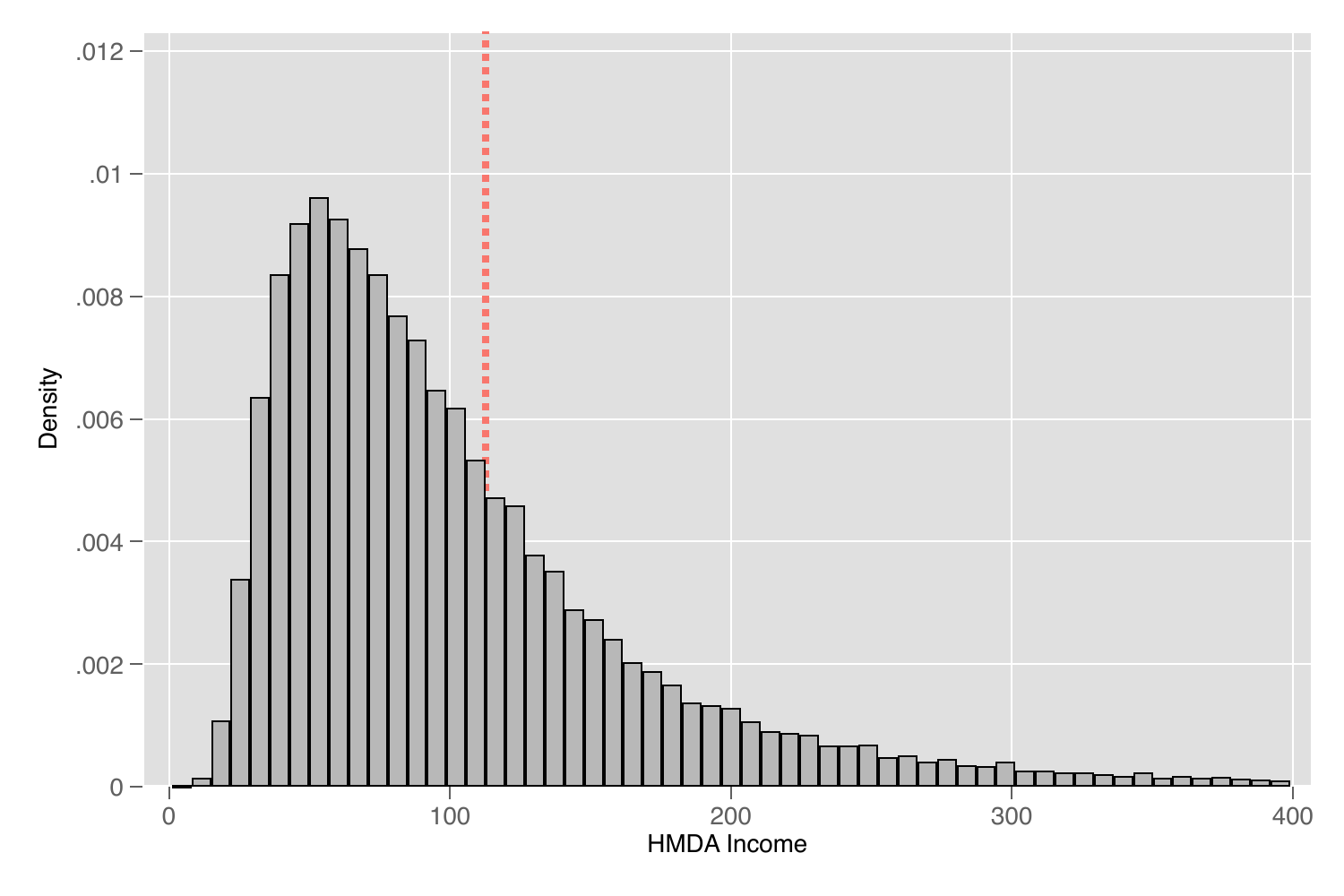} 
    \vspace{1ex}
  \end{minipage} 
  
  \subcaption*{\large{Refinancing Loans}}
  
  \begin{minipage}[b]{0.5\linewidth}
    \centering
    \subcaption{TU income} 
    \includegraphics[width=\linewidth]{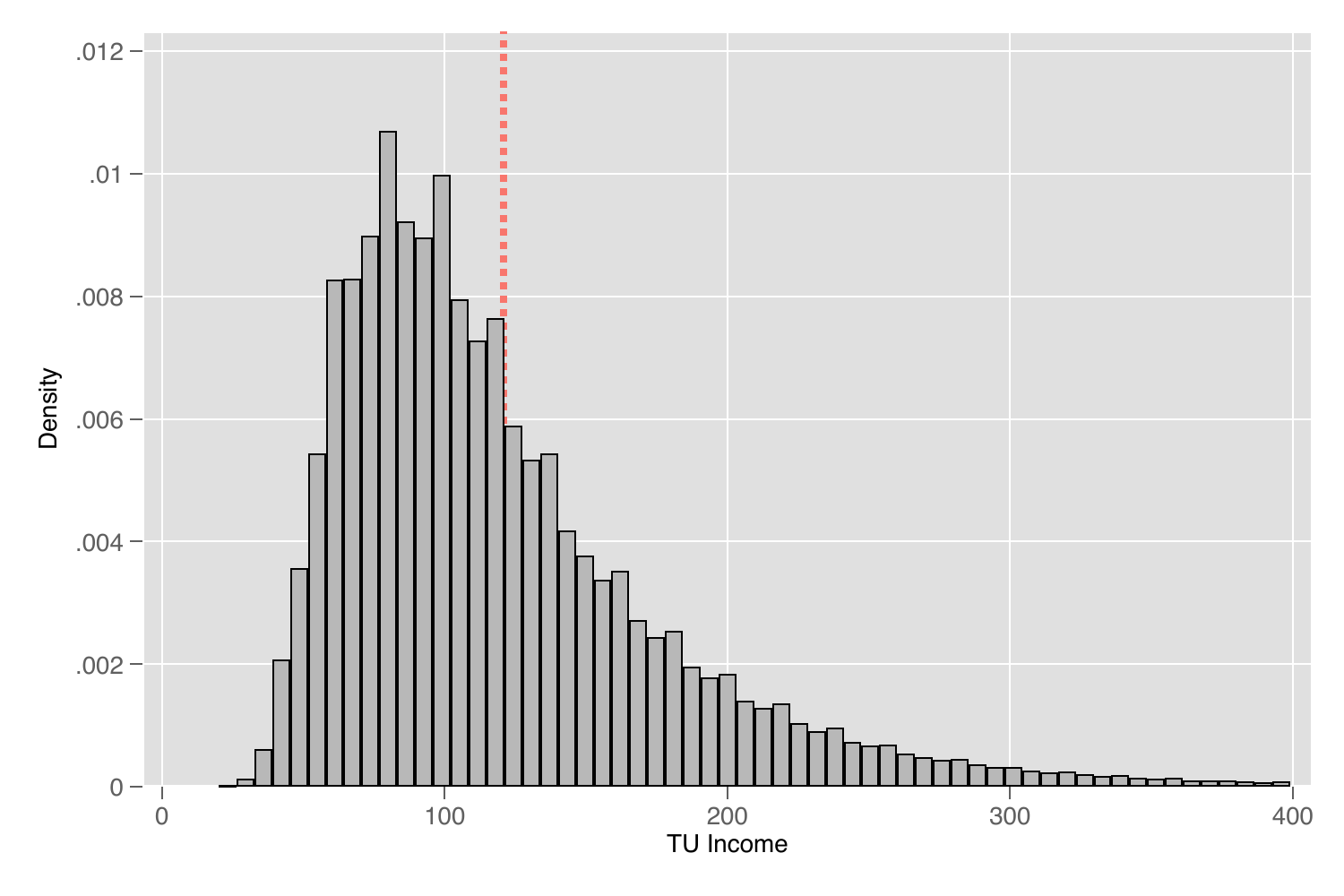} 
    \vspace{1ex}
  \end{minipage}
  \begin{minipage}[b]{0.5\linewidth}
    \centering
    \subcaption{HMDA income}
    \includegraphics[width=\linewidth]{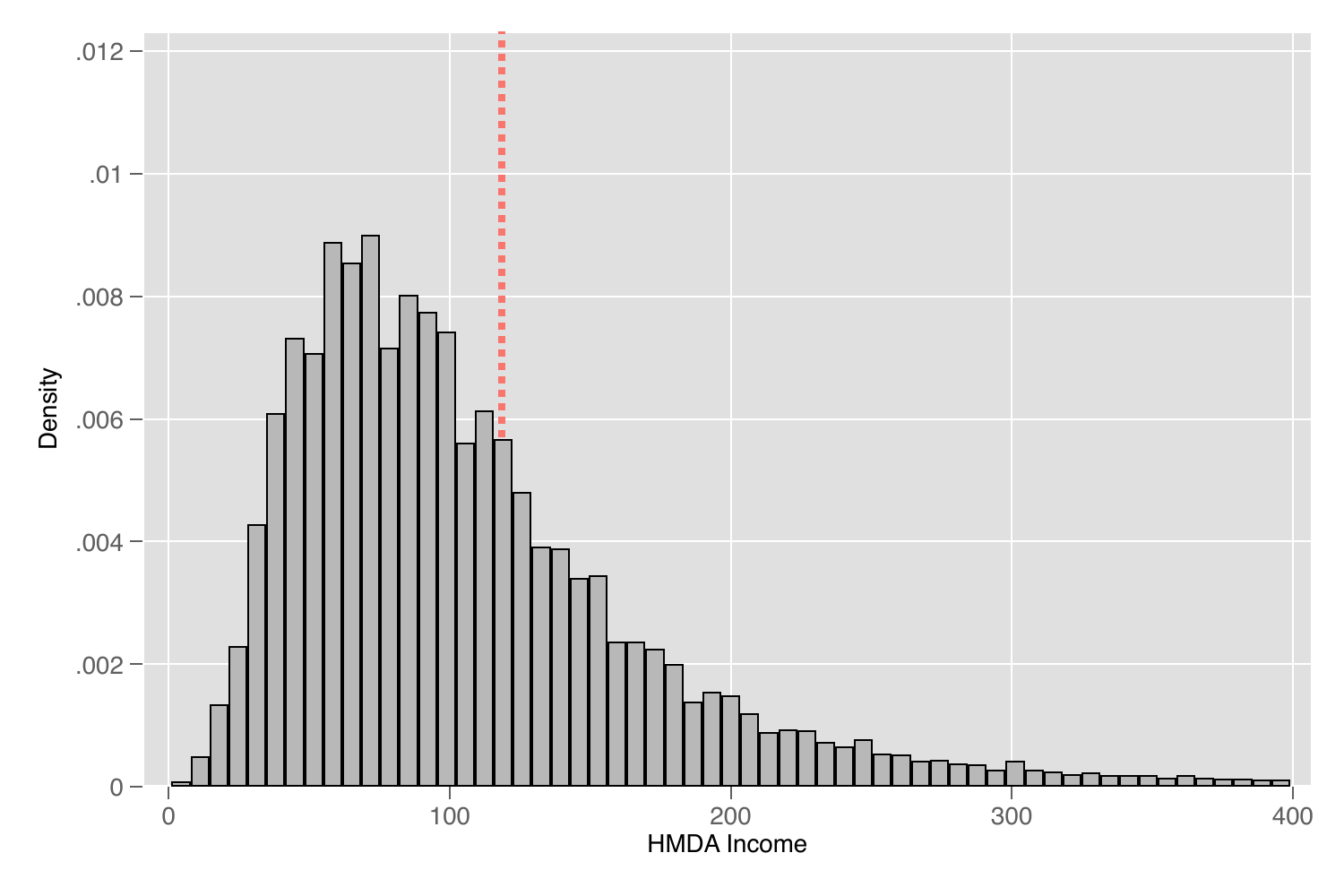} 
    \vspace{1ex}
  \end{minipage} 
  
\bigskip
\begin{flushleft}
\small{Note: The above histograms show the distribution of reported income in HMDA and the income estimator in the TransUnion dataset by transaction type. Sample includes all transactions in the final TransUnion-Corelogic-HMDA merged dataset. Observations more than two standard deviations above mean are not plotted. Red vertical dotted lines indicate mean income. Note that we de-duplicate CoreLogic observations with identical merge variables prior to the merge.}
\end{flushleft}
  
\end{figure}

\begin{figure}[H]
\caption{Income Difference between TransUnion and HMDA (\% HMDA income)}
\label{incomediff_purchase_mc}
\centering
    \begin{subfigure}[t]{0.45\textwidth}
    \includegraphics[width=1\textwidth]{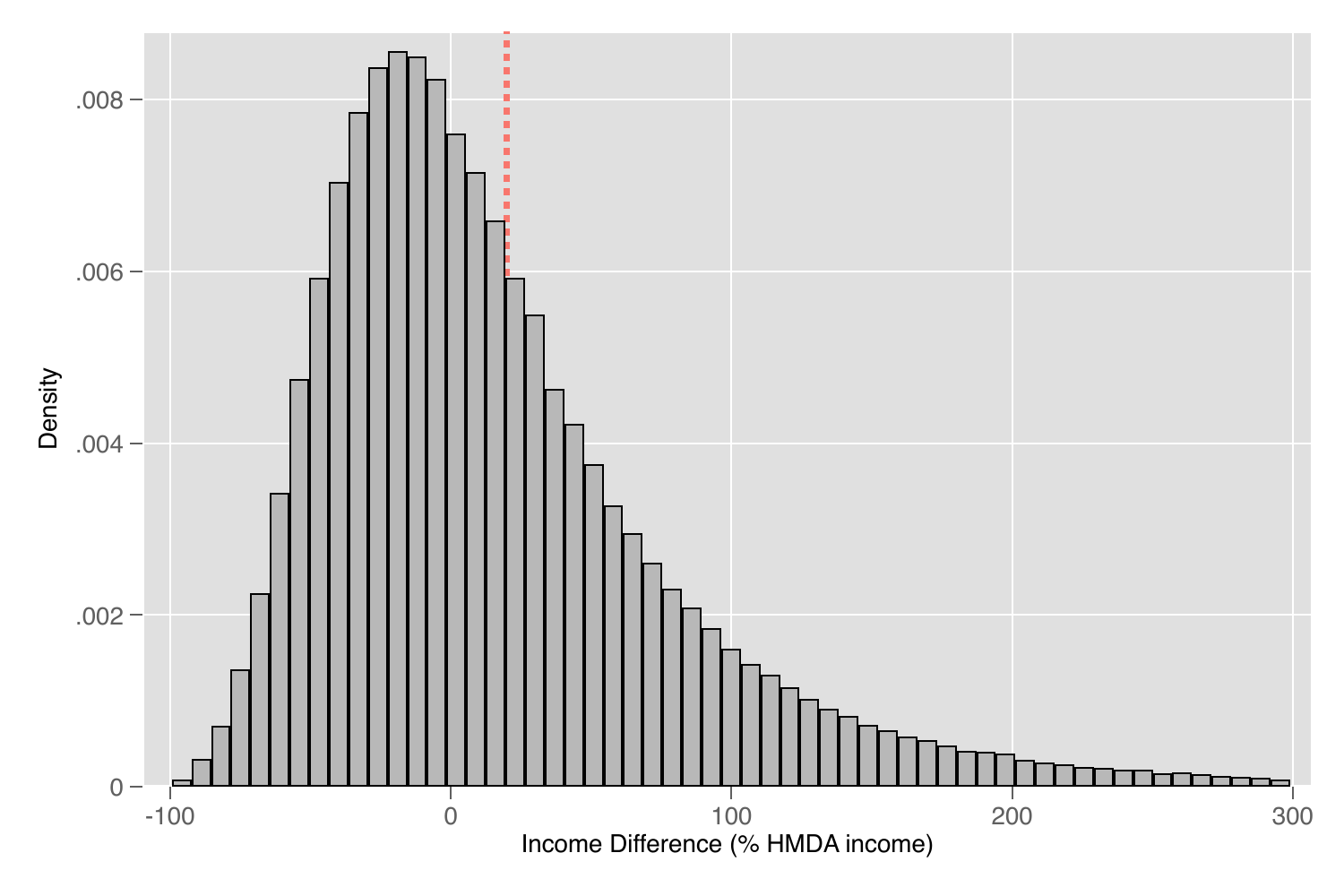}\caption{Purchase Loans}
    \end{subfigure} 
    \begin{subfigure}[t]{0.45\textwidth}
    \includegraphics[width=1\textwidth]{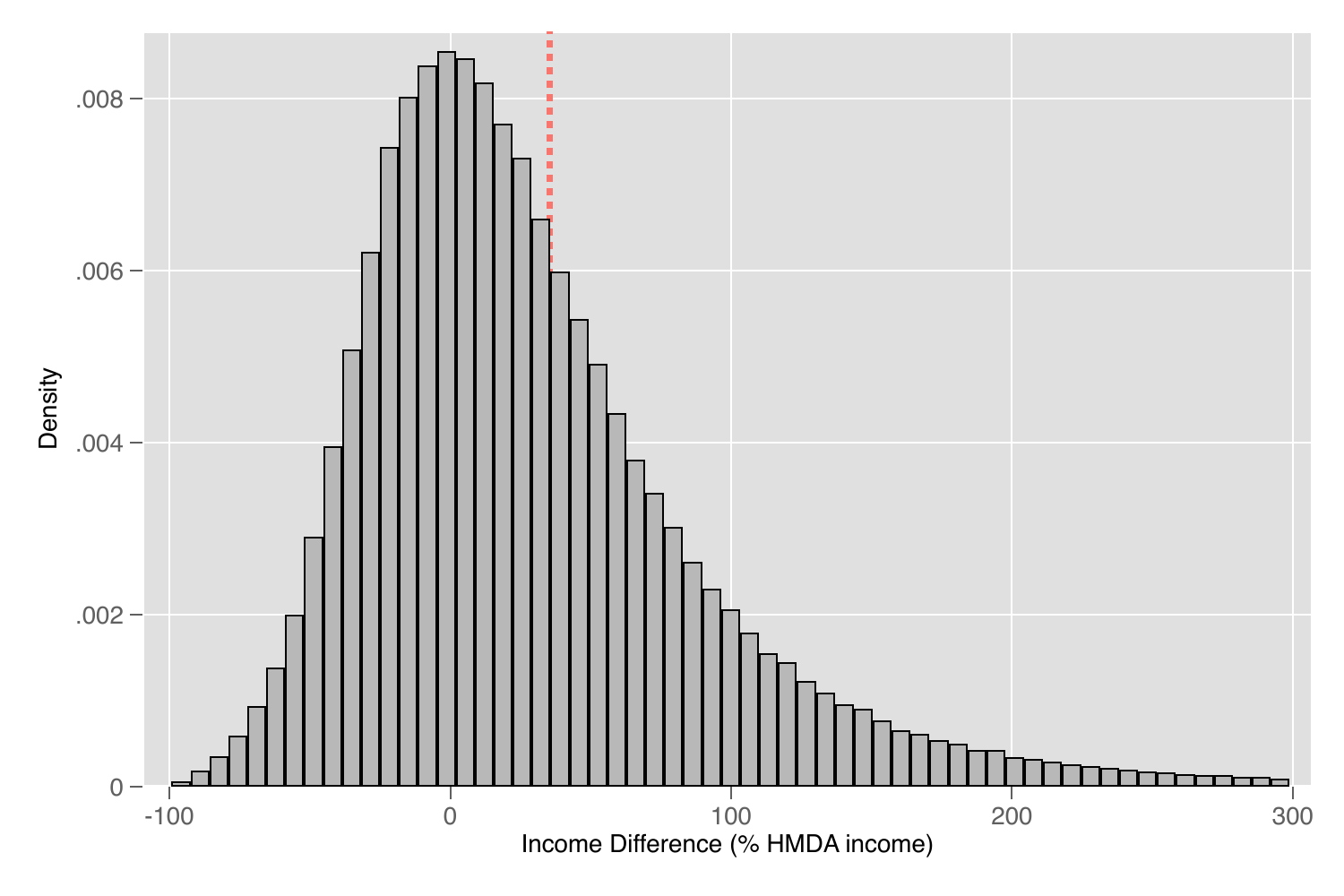}\caption{Refinancing Loans}
    \end{subfigure}
 
    \begin{flushleft}
    
    \small{Note:  Histogram of the difference between the income estimator and HMDA income expressed in percent relative to HMDA income. Outliers are removed from the plot for easier visualization (income difference above 30). Red vertical dotted lines indicate mean income.}

    \end{flushleft}
\end{figure}

 Overall, both income measures are highly correlated with a correlation coefficient of 0.436. The distributions of income are similar (Figure \ref{hist_income_mc}), with differences being in most cases no larger than a few thousand dollars (see Figure \ref{incomediff_purchase_mc}). The median divergence in income is \$2,000, while the average difference is \$11,000. The credit bureau income estimator tends to slightly underestimate income relative to HMDA.

\begin{figure}[H]\caption{Coefficient Plot of Income Difference Regression.
\label{coeffplot_tu_hmda_incomediff}}
\centering
    \includegraphics[width=.95\textwidth]{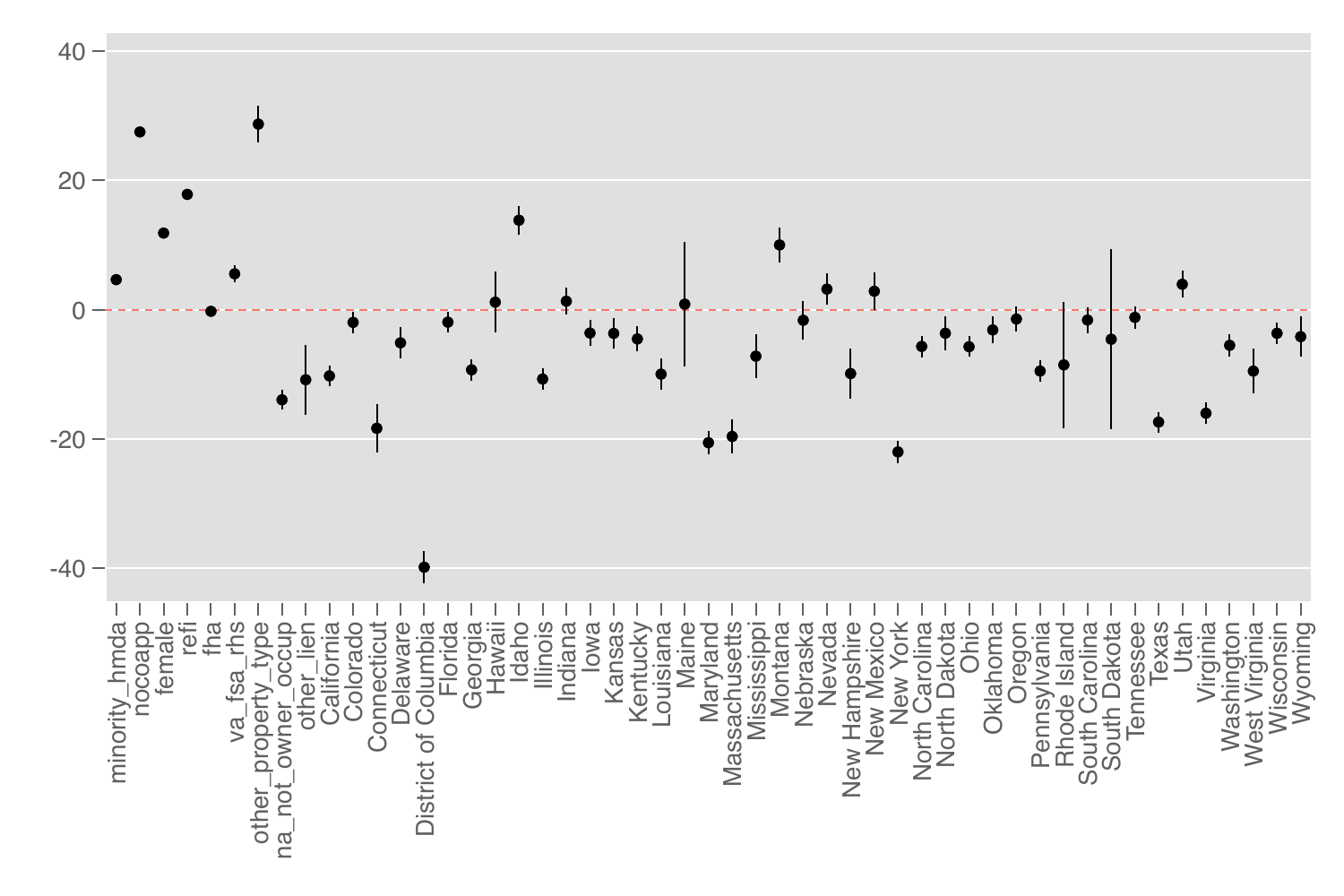}
     \begin{flushleft}
    \begin{spacing}{1.1}
    \small{Note: The figure shows regression coefficients from a regression of the TU-HMDA income difference on observables in HMDA based on our pooled matched sample.}
    \end{spacing}
    \end{flushleft}
\end{figure}

Figure \ref{coeffplot_tu_hmda_incomediff} shows the main drivers of the divergence by regressing the percentage difference between the bureau and HMDA income on several observables including household structure (potential joint filer), gender, race/ethnicity, transaction type (Purchase/Refinancing), loan type (Conventional/FHA/VA), low documentation status, and state dummies. While we observe some significant coefficients in both positive and negative directions, there does not seem to exist clear evidence of systematic mistakes that could invalidate our identification.

To explore robustness of our results, Table \ref{table_income_matrix_mc} presents confusion matrices that indicate how many of our low income individuals would be re-classified as high income if we had used their HMDA reported income instead, and vice-versa.  We use the 25th percentile of TU income as the threshold to split low and high income individuals.

\begin{table}[H]
\renewcommand{\arraystretch}{1.5}
\centering
\caption{Confusion Matrix - HMDA and TU income groups}
\label{table_income_matrix_mc}


\subcaption*{Panel A: Purchase Loans}

        \begin{tabular}{c|c|c} \toprule
        \diagbox{\small{TU}}{\small{HMDA}} & Higher income & Lower income \vspace*{0.01cm} \\ \midrule
       Higher income & 415,707      & 103,623   \vspace*{0.01cm}  \\ \midrule
        Lower income     &  54,650        & 62,486    \\ \bottomrule
        \end{tabular}

\bigskip

\subcaption*{Panel B: Refinancing Loans}

        \begin{tabular}{c|c|c} \toprule
        \diagbox{\small{TU}}{\small{HMDA}} & Higher income & Lower income \vspace*{0.01cm} \\ \midrule
         Higher income & 794,071      &  157,041   \vspace*{0.01cm}  \\ \midrule
        Lower income     &  32,815       &  41,647 \\ \bottomrule
        \end{tabular}

\bigskip
\begin{flushleft}
\small{Note: The table shows confusion matrices for the low income definition using the matched CoreLogic-Infutor-HMDA sample. We use the 25th percentile of the TransUnion income estimates to create higher and low income groups.}
\end{flushleft}

\end{table}

\clearpage

\subsubsection*{Minority Status}

We validate our BISG minority imputation using HMDA. We link our matched Infutor-credit bureau sample that contains BISG-imputed race to HMDA using the CoreLogic-HMDA merge described above. HMDA contains self-reported race and ethnicity. Table \ref{table_race_matrix_mc} shows that overall our minority indicator is highly correlated with the minority indicator derived from HMDA self-reported categories (correlation coefficient of 0.573). HMDA allows mortgage applicants to report race (American Indian/Alaska Native, Asian, Black/African American, Native Hawaiian/Pacific Islander, or White) and ethnicity (Hispanic/Latino or not). HMDA also has the option not to report race/ethnicity. We define minority as African-American or Hispanic/Latino, and non-minority as Non-Hispanic/Latino White. Other groups are not included in the analysis.

\begin{figure}[H]
\caption{Misclassification and group (BISG) probability thresholds}
\label{bisg_misclassfication}
\centering
    \begin{subfigure}[t]{0.45\textwidth}
    \includegraphics[width=\textwidth]{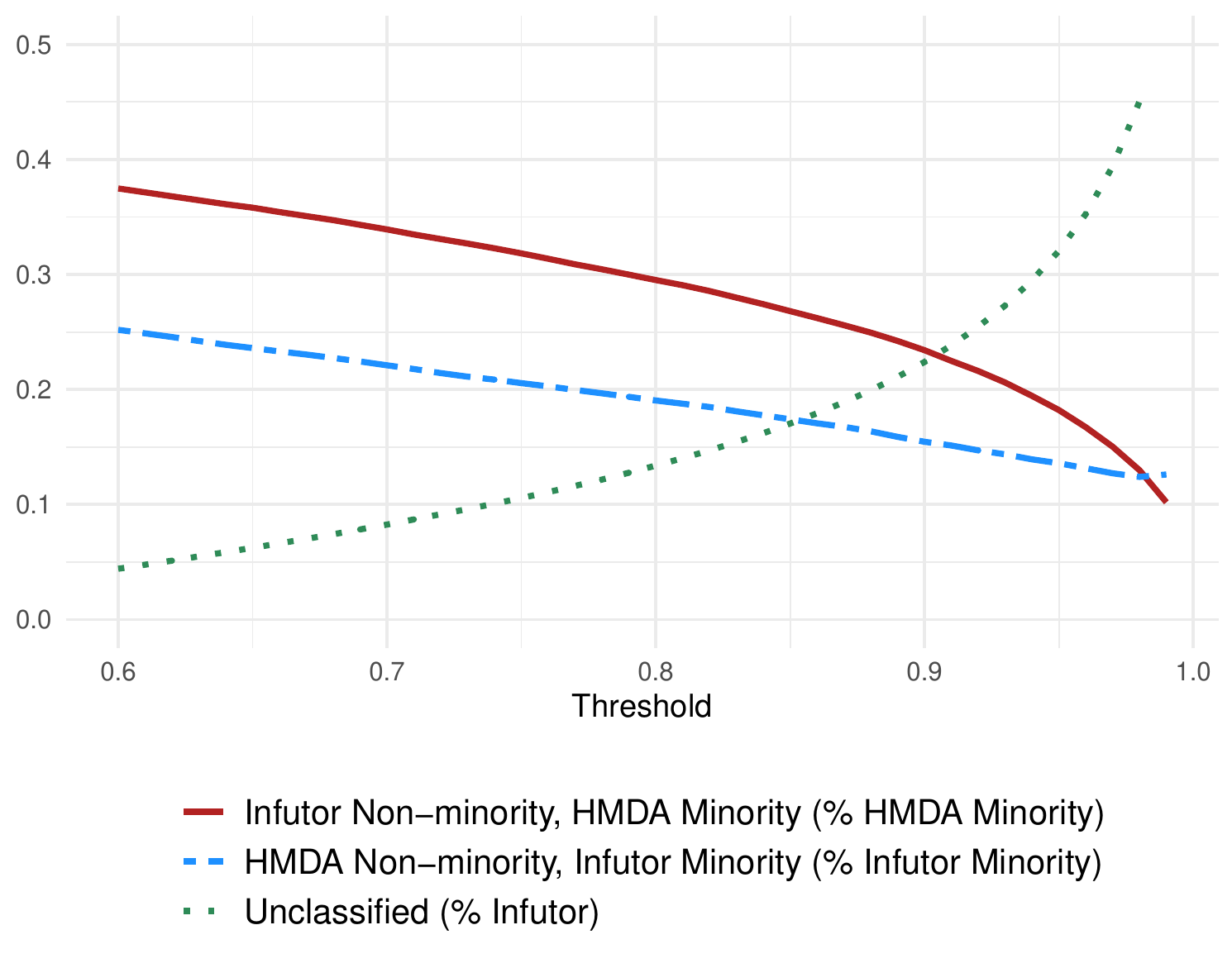}\caption{Purchase Loans}
    \end{subfigure} 
    \begin{subfigure}[t]{0.45\textwidth}
    \includegraphics[width=\textwidth]{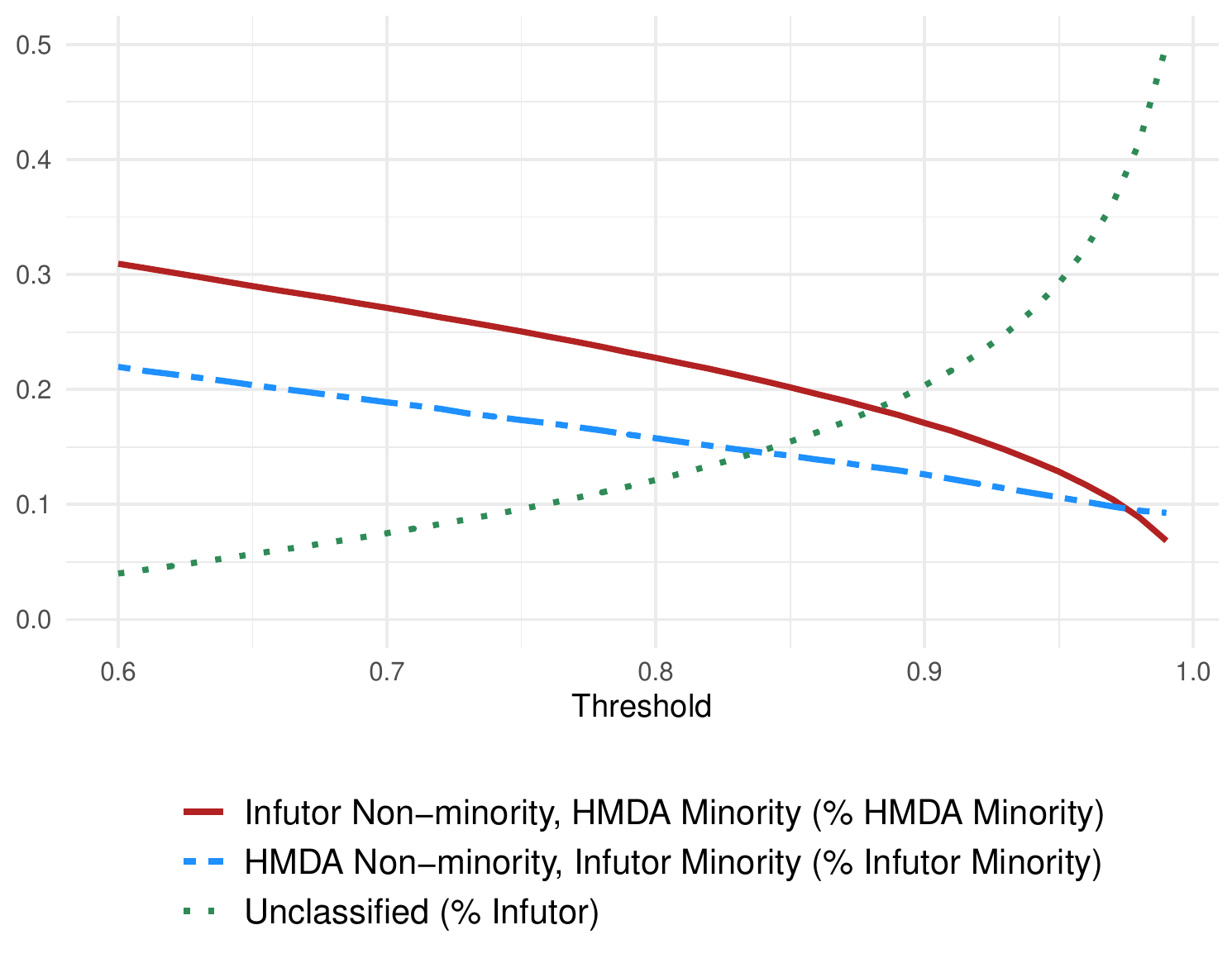}\caption{Refinancing Loans}
    \end{subfigure}
 
    \bigskip
    \begin{flushleft}
    
    \small{Note: The plot traces out the classification errors by the probability threshold that the Infutor-based imputed race has to satisfy to be classified - those below the threshold are left with no assigned race (unclassified). The red continuous line indicates the share of those reported as minority in HMDA that are mis-classified as non-minority  by the Infutor-based imputation procedure. This share decreases as we increase the probability threshold (keeping only those imputed with a relatively high probability). The blue dashed line indicates the share of those classified as minorities based on Infutor that are instead reported as non-minority in HMDA.}

    \end{flushleft}
\end{figure}

Figure \ref{bisg_misclassfication} shows the percentage of individuals we incorrectly label as minority according to HMDA, as well as the percentage of individuals we incorrectly label as non-minority, as a function of the probability threshold that we use to classify someone as belonging to a particular racial/ethnic group. Changing this threshold allow us to check the robustness of our results, as we do in Table \ref{table_race_matrix_mc} (Panels (b) and (d)).

\begin{figure}[H]
\centering
\caption{Coefficient Plot of Race Regressions}
\begin{subfigure}[b]{0.9\textwidth}
    \caption{HMDA minority, Infutor non-minority}
    \includegraphics[width=.95\textwidth]{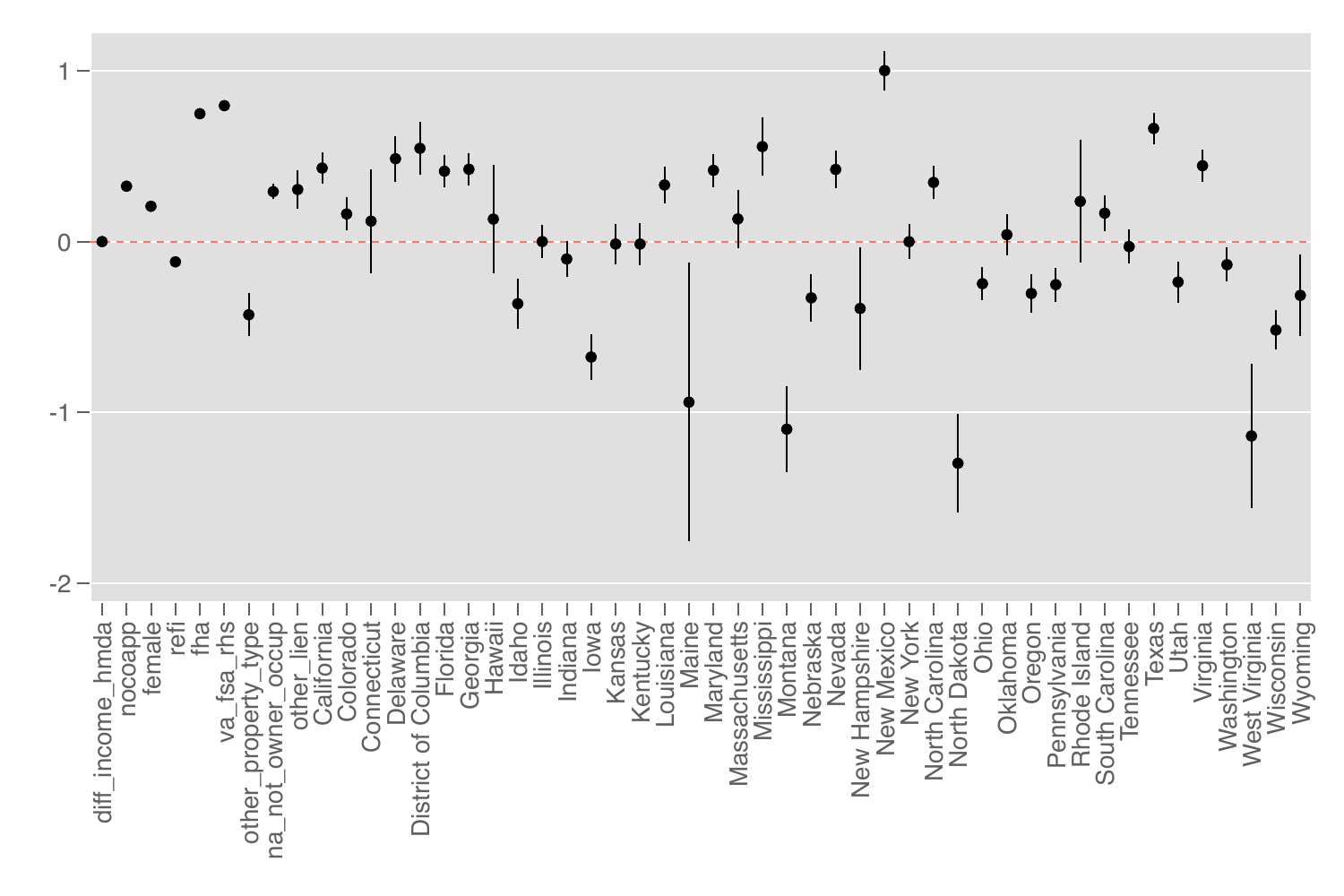}
    \label{fig:coeffplot_race_reg1_mc} 
\end{subfigure}

\begin{subfigure}[b]{0.9\textwidth}
    \caption{Infutor minority, HMDA non-minority}
    \includegraphics[width=.95\textwidth]{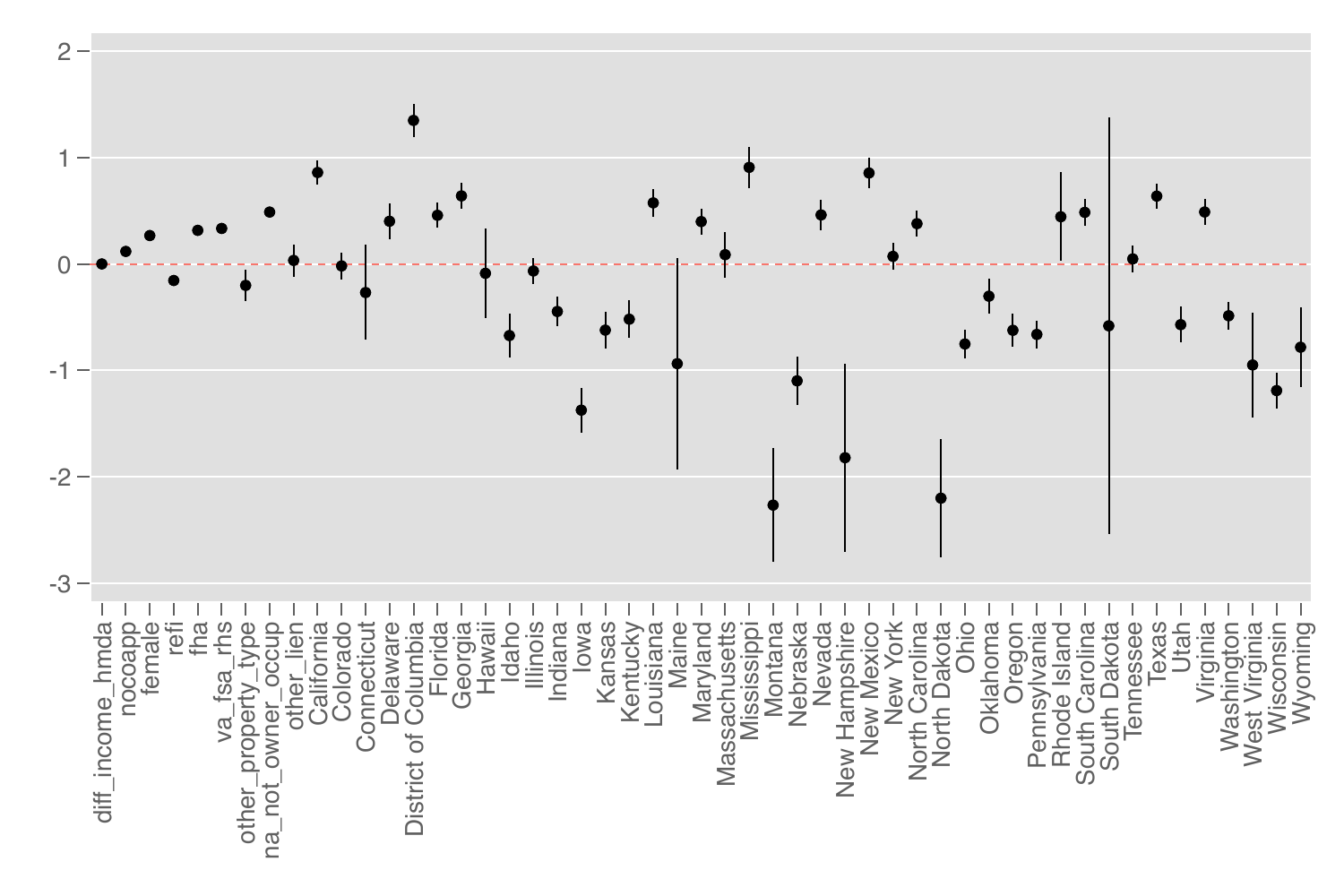}
    \label{fig:coeffplot_race_reg2_mc}
\end{subfigure}
\begin{flushleft}
 \small{Note: The figure shows regression coefficients from a regression of divergence in minority status difference in HMDA/Corelogic-Infutor on observables in HMDA based on our pooled matched sample.}
 \end{flushleft}
\end{figure}

Finally, we also fit logistic regressions to  indicators of mismatch between the two minority indicators (HMDA and BISG-based imputation) to analyze the source of our imputation errors. Figure \ref{fig:coeffplot_race_reg1_mc} plots the coefficients of the regression where the outcome is a mismatch when an individual is reported in HMDA as a minority, but imputed as as non-minority. On the other hand, Figure \ref{fig:coeffplot_race_reg2_mc} covers the case when an individual is reported in HMDA as a non-minority, but imputed as as minority. Our covariates include household structure (potential joint filer), gender, race/ethnicity, transaction type (Purchase/Refinancing), loan type (Conventional/FHA/VA), low documentation status, and state indicators. While we observe some significant coefficients in both positive and negative directions, there does not seem to exist clear evidence of systematic mistakes that could invalidate our identification.

\begin{table}[H]
\renewcommand{\arraystretch}{1.5}
\centering
\caption{Confusion Matrix - HMDA and Infutor-Based Race/Ethnicity}
\label{table_race_matrix_mc}

  \vspace{4ex}
  

 \subcaption*{\large{Purchase Loans}}

  \begin{minipage}[b]{0.5\linewidth}
    \centering
    \subcaption{Baseline} 
    
        \begin{tabular}{ccc} \toprule
        \diagbox{\small{HMDA}}{\small{Infutor}} & Non-minority & Minority \vspace*{0.01cm} \\ \midrule
         Non-minority & 435,524      & 19,799  \vspace*{0.01cm}  \\ \midrule
        Minority     & 32,409       & 43,630  \\ \bottomrule
        \end{tabular}
   
  \end{minipage}
  \begin{minipage}[b]{0.5\linewidth}
    \centering
    \subcaption{Robustness Check ($p>0.9$)} 
        \begin{tabular}{ccc} \toprule
        \diagbox{\small{HMDA}}{\small{Infutor}} & Non-minority & Minority \vspace*{0.01cm} \\ \midrule
        Non-minority & 365,485      & 4,615 \vspace*{0.01cm}  \\ \midrule
        Minority     & 17,427       & 24,776 \\ \bottomrule
        \end{tabular}
  
  \end{minipage} 
  
  \vspace{4ex}
  \subcaption*{\large{Refinancing Loans}}
  
  \begin{minipage}[b]{0.5\linewidth}
    \centering
    \subcaption{Baseline} 
   \begin{tabular}{ccc} \toprule
        \diagbox{\small{HMDA}}{\small{Infutor}} & Non-minority & Minority \vspace*{0.01cm} \\ \midrule
         Non-minority & 709,438      & 25,393 \vspace*{0.01cm}  \\ \midrule
        Minority     & 36,967      & 65,574 \\ \bottomrule
        \end{tabular}
  
  \end{minipage}
  \begin{minipage}[b]{0.5\linewidth}
    \centering
    \subcaption{Robustness Check ($p>0.9$)} 
    \begin{tabular}{ccc} \toprule
        \diagbox{\small{HMDA}}{\small{Infutor}} & Non-minority & Minority \vspace*{0.01cm} \\ \midrule
         Non-minority & 608,987      & 5,383 \vspace*{0.01cm}  \\ \midrule
        Minority     & 17,113       & 37,381 \\ \bottomrule
        \end{tabular}
    
  \end{minipage}

\vspace{4ex}
\bigskip

\begin{flushleft}
\small{Note: The table shows confusion matrixes between imputed minority status and HMDA self-reported race/ethnicity status. Minority includes African-American and Hispanic/Latino individuals, while non-minority includes Non-Hispanic/Latino White individuals. We exclude from the table those belonging to other groups or with no reported race in HMDA. Our baseline version does not apply any probability threshold to BISG/Infutor-based race imputation. Robustness version excludes observations whose BISG/Infutor-based imputed race had a probability below 0.9.}

\end{flushleft}
\end{table}

\begin{table}[H]
\caption{Cross-Tabulation between Income and Minority Status}\label{tab:minority_income_crosstab}
\centering

\begin{tabular}{ccccccc}
\toprule
&&& &&&\\
Mortgage applicant sample&\multicolumn{3}{c}{TU sample} & \multicolumn{3}{c}{HMDA} \\ \midrule
&& \multicolumn{2}{c}{Low income } &  & \multicolumn{2}{c}{Low income } \\
& & 0 & 1 & & 0 & 1 \\
Minority & 0 & 0.626 & 0.172 && 0.684 & 0.181\\
& 1 & 0.116 & 0.085 &&0.090 &0.046 \\
\midrule
&&& &&&\\
Approved mortgage sample& & \multicolumn{2}{c}{TU sample} &  & \multicolumn{2}{c}{HMDA}\\\midrule

& & \multicolumn{2}{c}{Low income } &  & \multicolumn{2}{c}{Low income }\\
& & 0 & 1 && 0&1\\
Minority & 0 & 0.796 & 0.092&&0.699&0.186 \\
& 1 & 0.089 & 0.023&&0.076&0.039\\
\bottomrule
\end{tabular}

\begin{flushleft}
\small{Note: The table shows cross-tabs between minority and low income status in both our samples of approved mortgages and mortgage applicants. Numbers are percentages of the total number of observations.}
\end{flushleft}
\end{table}



\FloatBarrier\newpage\pagebreak

\section{Appendix: Default Measures}\label{appendix_default}

\setcounter{figure}{0} \renewcommand{\thefigure}{F.\arabic{figure}} 
\setcounter{table}{0} \renewcommand{\thetable}{F.\arabic{table}} 

\subsection*{Default Measures}
This appendix describes how we define default outcomes in our data. There are several types of variables that describe default in credit bureau data. First, we use credit bureau variables that indicate the severity of delinquency, or Manner of Payment (MOP), in the past 24 months. Second, we use information on charge-offs, collections, bankruptcies, foreclosures and repossessions. The MOP categories are: 30 days overdue (MOP 2), 60 days overdue (MOP 3), 90 days overdue (MOP 4), 120 days (MOP 5). After MOP 5, an account is typically charged-off and potentially sold to a third-party collector who will then try to collect on the outstanding debt. For this reason we flag as default if there was a charge-off or collection on the account. In addition, we count any instances of bankruptcies as default. Finally, for mortgages and auto loans, foreclosures and repossessions (respectively) are also counted as default outcomes. 

One potential concern is that our measurement of default in other loans works differentially well across groups. Table \ref{default_confusionmatrix_bygroup} shows the confusion matrix separately by group. While overall accuracy is lower (the share of observations correctly predicted), we find that precision/recall are higher for the disadvantaged group. These trends reflect that default rates are generally higher for the disadvantaged group. 

\subsection*{A Different Approach to Constructing Mortgage ROC Curves}

We rely on the relationship between mortgage and non-mortgage default to directly compute the mortgage ROC. To understand this approach, let $A = {0,1}$ denote an accepted application, let $Y^*$ denote whether the applicant would have defaulted on the mortgage and let $Y$ denote whether applicant would have defaulted on the non-mortgage credit product. We define $Y=1$ as the non-default event to make the exposition easier. Let $\hat{Y}$ be the default prediction given by the credit score. At a particular threshold in the ROC curve, the true positive rate (TPR) for the set of approved applicants is given by $E(Y^*|A=1,\hat{Y}=1)$  which is observable. While we cannot observe the TPR for the group of rejected applicants, $E(Y^*|A=0,\hat{Y}=1)$, we can use the fact that we observe default on a non-mortgage product. In particular, by Bayes rule, $E(Y^*|A=0,\hat{Y}=1) = \sum_y E(Y^*|Y=y,  A=0, \hat{Y}=1) P(Y=y|A=0, \hat{Y}=1).$ If we assume that the relationship between mortgage and non-mortgage default is the same for accepted and rejected applicants, that is $E[Y^* | Y, \hat{Y}=1, A=0] = E[Y^*| Y, \hat{Y}=1, A=1]$, we can compute the TPR for rejected applicants as
\begin{equation*}
    E(Y^*|A=0,\hat{Y}=1) = \sum_y E(Y^*|Y=y,  A=1, \hat{Y}=1) P(Y=y|A=0, \hat{Y}=1).
\end{equation*}

To illustrate, consider a VantageScore 3.0 credit score cut-off of 620. Using the non-mortgage default for the entire sample of applicants, we obtain true positive rates of 0.90 for the majority and 0.76 for the minority. Among the accepted sample, the mortgage TPRs are 0.97 and 0.92, respectively.  We can reconstruct the TPRs for the rejects by first computing the mortgage TPR for the subset of approvals who have a default on their non-mortgage product, which are 0.78 and 0.72, respectively. Among the group of approvals without a non-mortgage default, the TPRs are 0.97 and 0.93. Given that the share of non-mortgage defaults among the rejects, is 22\% for majority and 32\% for the minority group, we can combine the two TPRs for overall `reject' TPRs of 0.93 and 0.86, respectively. We can then further combine the TPR rate for the rejected sample with that of the accepted sample to obtain our final mortgage TPR. The procedure for FPR is analogous.


\begin{figure}[H]
\caption{Receiver Operating Characteristic Curves by Consumer Type \label{rocgraph_alt}}
\centering
       \begin{subfigure}[t]{0.45\textwidth}
    \includegraphics[width=1\textwidth] {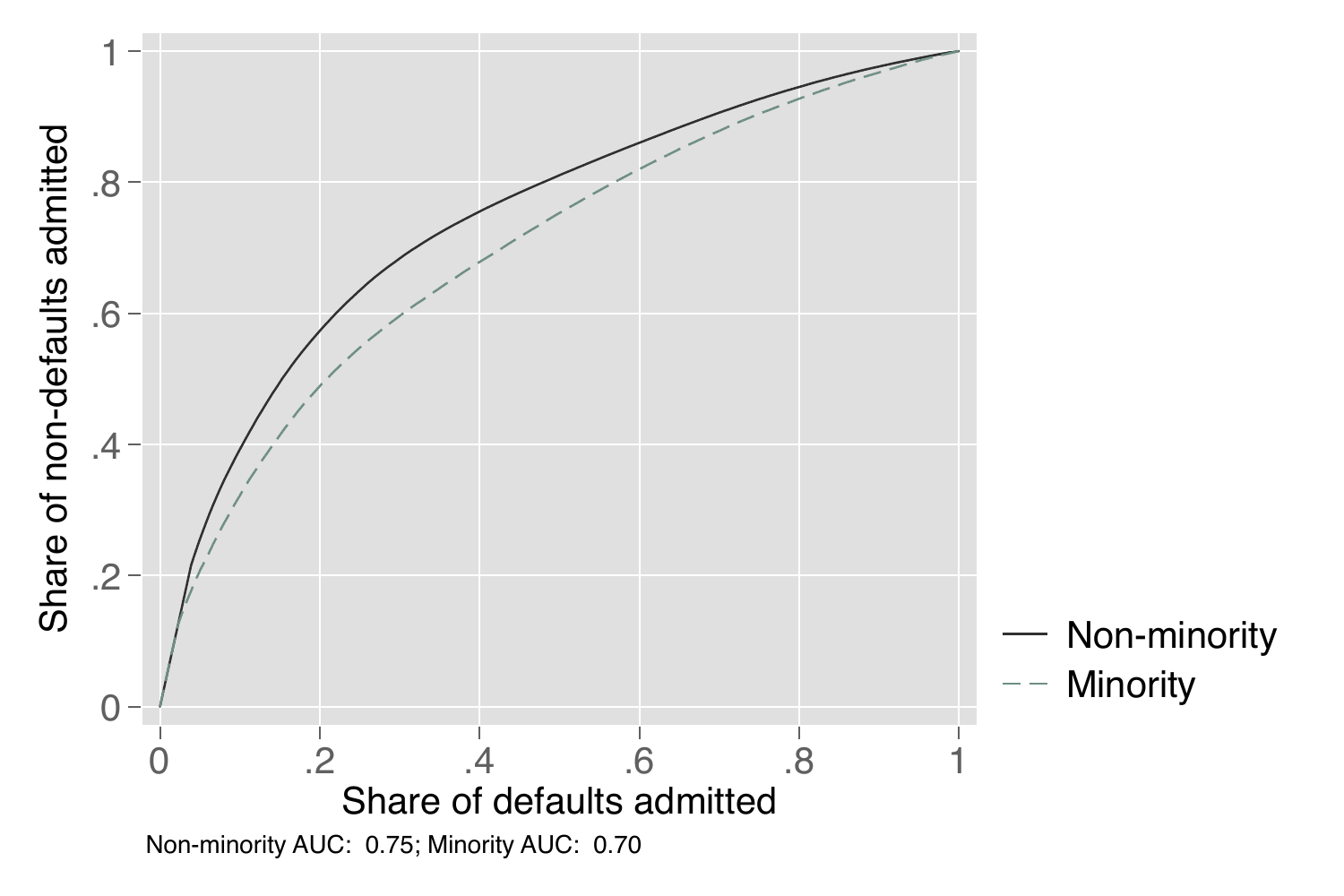}\caption{Mortgage Default: Minority}
     \end{subfigure} 
    \begin{subfigure}[t]{0.45\textwidth}
   \includegraphics[width=\textwidth]{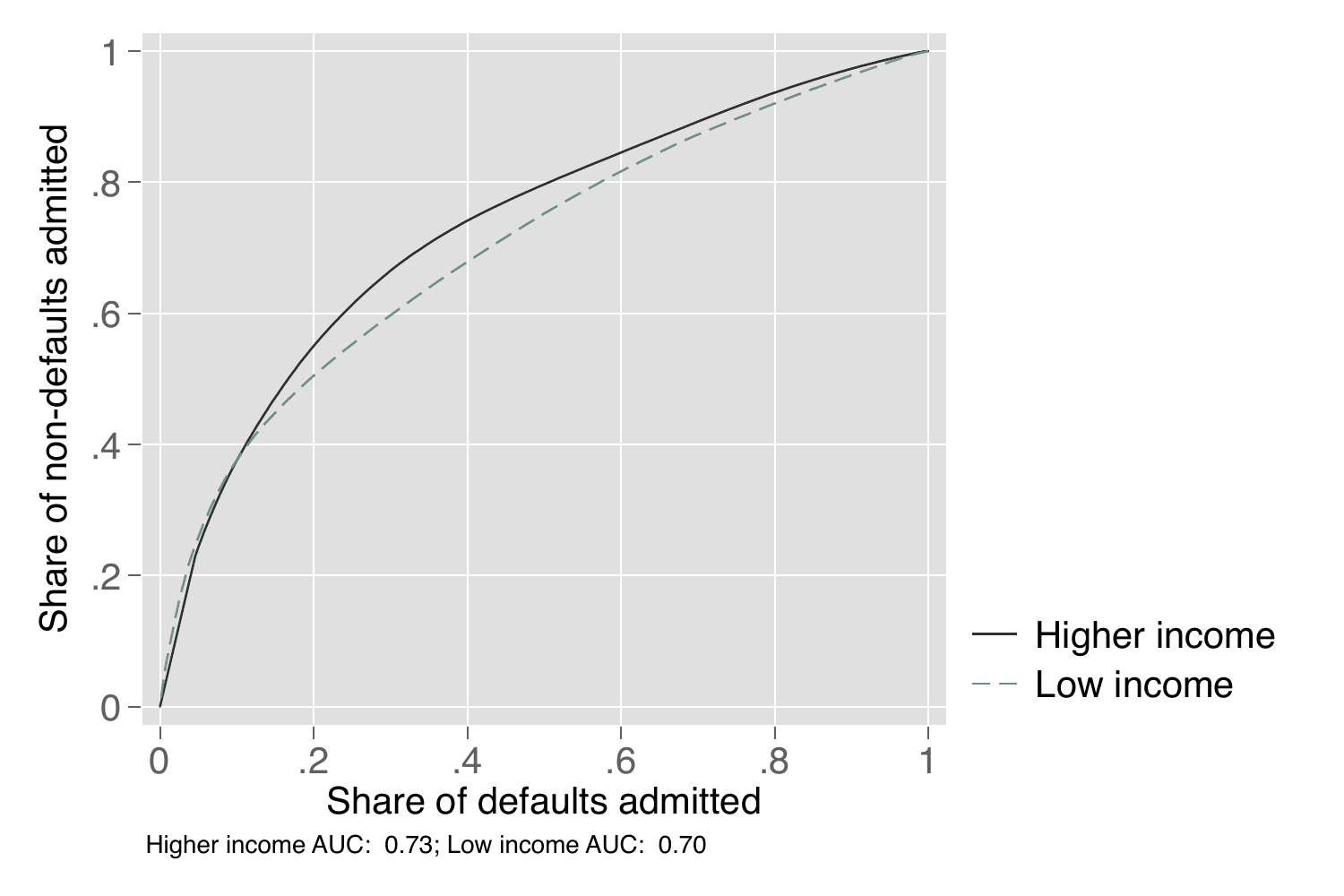}\caption{Mortgage Default: Low income }
    \end{subfigure}\\
    \begin{flushleft}
    \begin{spacing}{1.1}
    \small{Note: This figure shows Receiver Operating Characteristic (ROC) curves for the VantageScore 3.0 credit score in a sample of mortgage applicants between 2009-2016. ROC curves plot the fraction of non-defaults (True Positive Rate) admitted for a given score cutoff against the fraction of defaults admitted (False Positive Rate). We use any mortgage default 24 months after application, reconstructing reject outcomes using the Bayes rule procedure described in this appendix. The group classifications are described in the main text. }
    \end{spacing}
    \end{flushleft}
\end{figure}



\begin{table}[]
    \caption{Confusion Matrices for Default Measures by Group}
    \label{default_confusionmatrix_bygroup}
    
    \centering
    
    \resizebox{\textwidth}{!}{

\begin{tabular}{ccccccc}\toprule

 \multicolumn{3}{c}{ Non-minority } & & \multicolumn{3}{c}{ Minority } \\
 \cmidrule(lr){1-3}\cmidrule(lr){5-7}
& \multicolumn{2}{c}{ Mortgage delinquency } & & & \multicolumn{2}{c}{ Mortgage delinquency } \\

 Non-mort. delinq. (90+ DPD) & 0 & 1 & & Non-mort. delinq. (90+ DPD) & 0 & 1 \\

0 & 14,269,994 & 512,930 & & 0 &  1,967,429 &  148,419 \\
1 & 389,693 & 273,608 & & 1 & 113,991 & 91,227 \\
& & & & & & \\

\multicolumn{1}{r}{Precision} & 0.41 & & & \multicolumn{1}{r}{Precision} & 0.44 & \\
\multicolumn{1}{r}{Recall} & 0.35 & & & \multicolumn{1}{r}{Recall} & 0.38 & \\
\multicolumn{1}{r}{Accuracy} & 0.94 & & & \multicolumn{1}{r}{Accuracy} & 0.89 & \\ \midrule
& & & & & & \\

\multicolumn{3}{c}{Higher income} & & \multicolumn{3}{c}{Low income} \\
\cmidrule(lr){1-3}\cmidrule(lr){5-7} 
& \multicolumn{2}{c}{ Mortgage delinquency } & & & \multicolumn{2}{c}{ Mortgage delinquency } \\

Non-mort. delinq. (90+ DPD) & 0 & 1 & & Non-mort. delinq. (90+ DPD) & 0 & 1 \\

0 & 15,064,234 & 525,255 & & 0 &  2,165,546 &  158,743 \\
1 & 373,562 & 234,516 & & 1 & 145,888 & 136,341 \\
& & & & & & \\

\multicolumn{1}{r}{Precision} & 0.39 & & & \multicolumn{1}{r}{Precision} & 0.48 & \\
\multicolumn{1}{r}{Recall} & 0.31 & & & \multicolumn{1}{r}{Recall} & 0.46 & \\
\multicolumn{1}{r}{Accuracy} & 0.94 & & & \multicolumn{1}{r}{Accuracy} & 0.88 & \\

\bottomrule

\end{tabular}

}

    \bigskip
    
    \begin{flushleft}
    \small{Note: The table shows confusion matrices in the sample of approved mortgage applicants by group (low income and two minority definitions). Mortgage delinquency refers to the ground truth, i.e. whether an approved applicant ended up defaulting on the mortgage 24 months after origination. We show how well this is predicted by the non-mortgage delinquency measure (non-mort. delinq.) conditioning on at least 90 days past due (DPD). See the note to Table \ref{tab:default_confusionmatrix} for definitions of Precision, Recall, and Accuracy.}
    
    \end{flushleft}

\end{table}
\FloatBarrier\newpage\pagebreak
\section{Appendix: Model Estimation Details}\label{appendix_model}

\setcounter{figure}{0} \renewcommand{\thefigure}{G.\arabic{figure}} 
\setcounter{table}{0} \renewcommand{\thetable}{G.\arabic{table}} 

This appendix discusses techniques used to improve performance of the minimization routine to estimate model parameters. 

The normal-normal parameterization that we specify for empirical tractability has some difficulty matching the thickness of the tails of the empirical credit score distribution. To address this difficulty, we truncate (i.e., trim) the empirical credit score distribution before calculating empirical moments, and then apply the same truncation to the model moments. To select which truncation to use, we perform a grid search over possible truncation points and minimize the weighted sum of, first, a Cramér–von Mises test statistic that is modified to compare the truncated empirical distribution to a truncated normal distribution, and second, a penalty term for how much mass has been truncated from the empirical distribution. The benchmark normal distribution used for this exercise has its two moments determined by the median and the 84th percentile (one standard deviation from the median) of the empirical distribution. The weighting of the penalty term is calibrated based on the zero-truncation value of the Cramér–von Mises statistic.

After choosing these truncation levels, we normalize parameter scaling and use a standard gradient descent method alternating with one-parameter gradient descent (line search). Given possible sensitivity to fitting tail masses as discussed above, we explore robustness across different moment weighting schemes, which even in a just-identified setting can affect which moments are fit most closely. While parameter estimates are broadly similar across weighting schemes, we use estimates from weighting schemes that generate the lowest average deviation in percentage terms between model moments and target moments. 
\FloatBarrier\newpage\pagebreak
\section{Appendix: Additional Results on Data Bias} \label{appendix_data_bias}

\setcounter{figure}{0} \renewcommand{\thefigure}{H.\arabic{figure}} 
\setcounter{table}{0} \renewcommand{\thetable}{H.\arabic{table}} 

This appendix includes additional results discusssed in Section \ref{sec:databias}.



\begin{figure}[hbt!]
\caption{AUC and Compositional Differences in Credit Files  \label{fig:auc_comp1}}
\centering

    \begin{subfigure}[t]{0.45\textwidth}
    \includegraphics[width=1\textwidth]{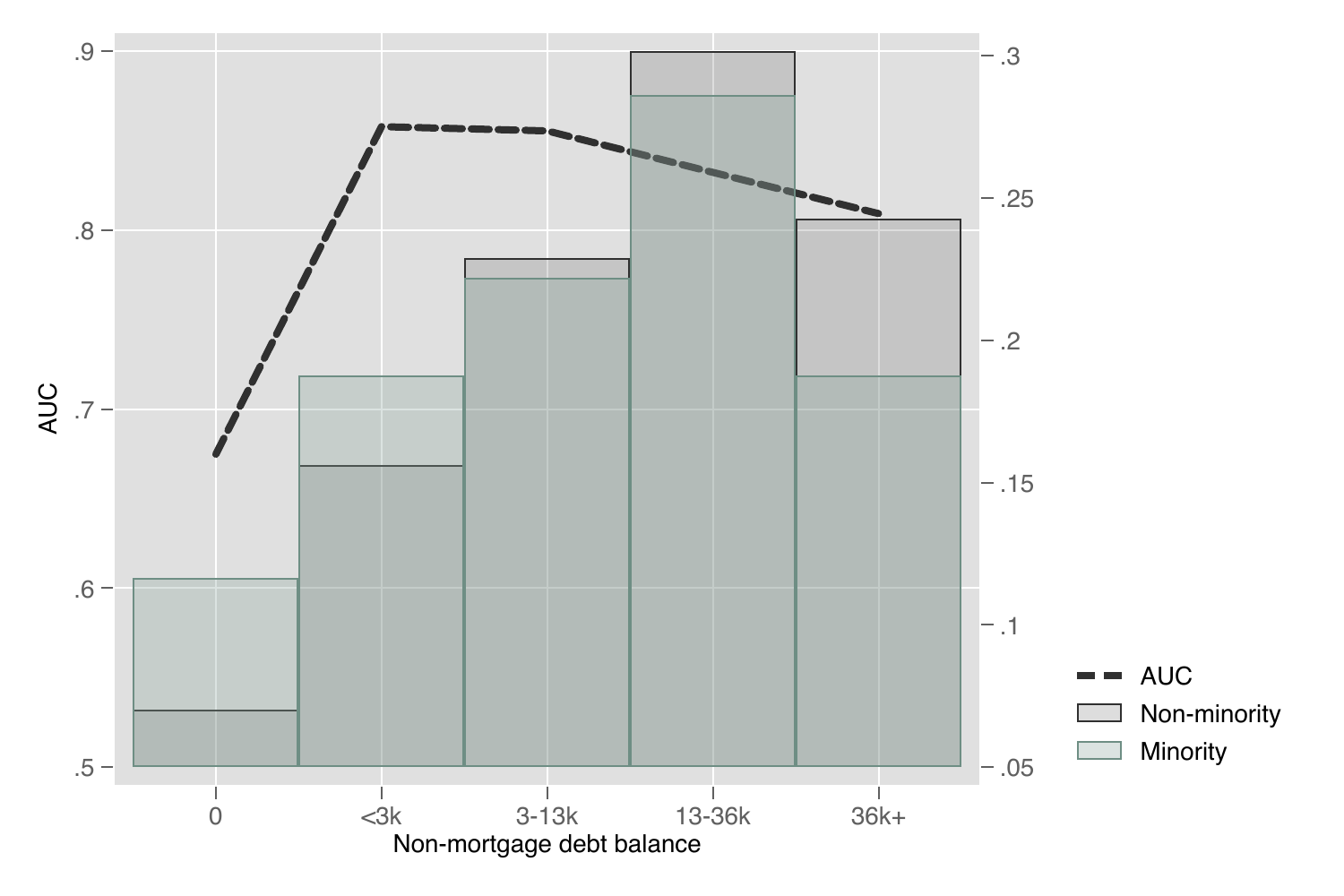}\caption{Debt balance by Minority Status}
     \end{subfigure} 
    \begin{subfigure}[t]{0.45\textwidth}
    \includegraphics[width=1\textwidth]{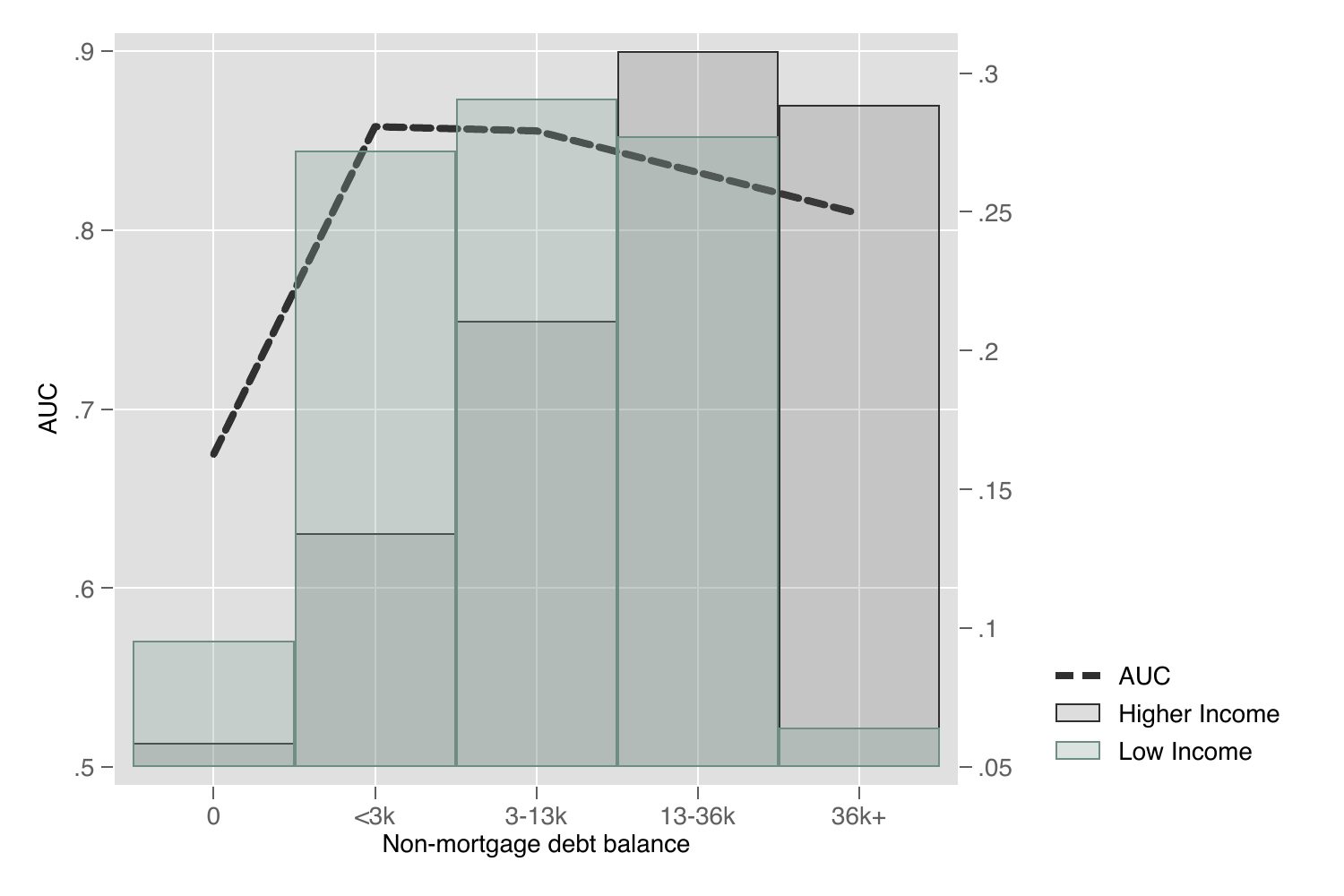}\caption{Debt balance by Income}
    \end{subfigure}\\
            \begin{subfigure}[t]{0.45\textwidth}
    \includegraphics[width=1\textwidth]{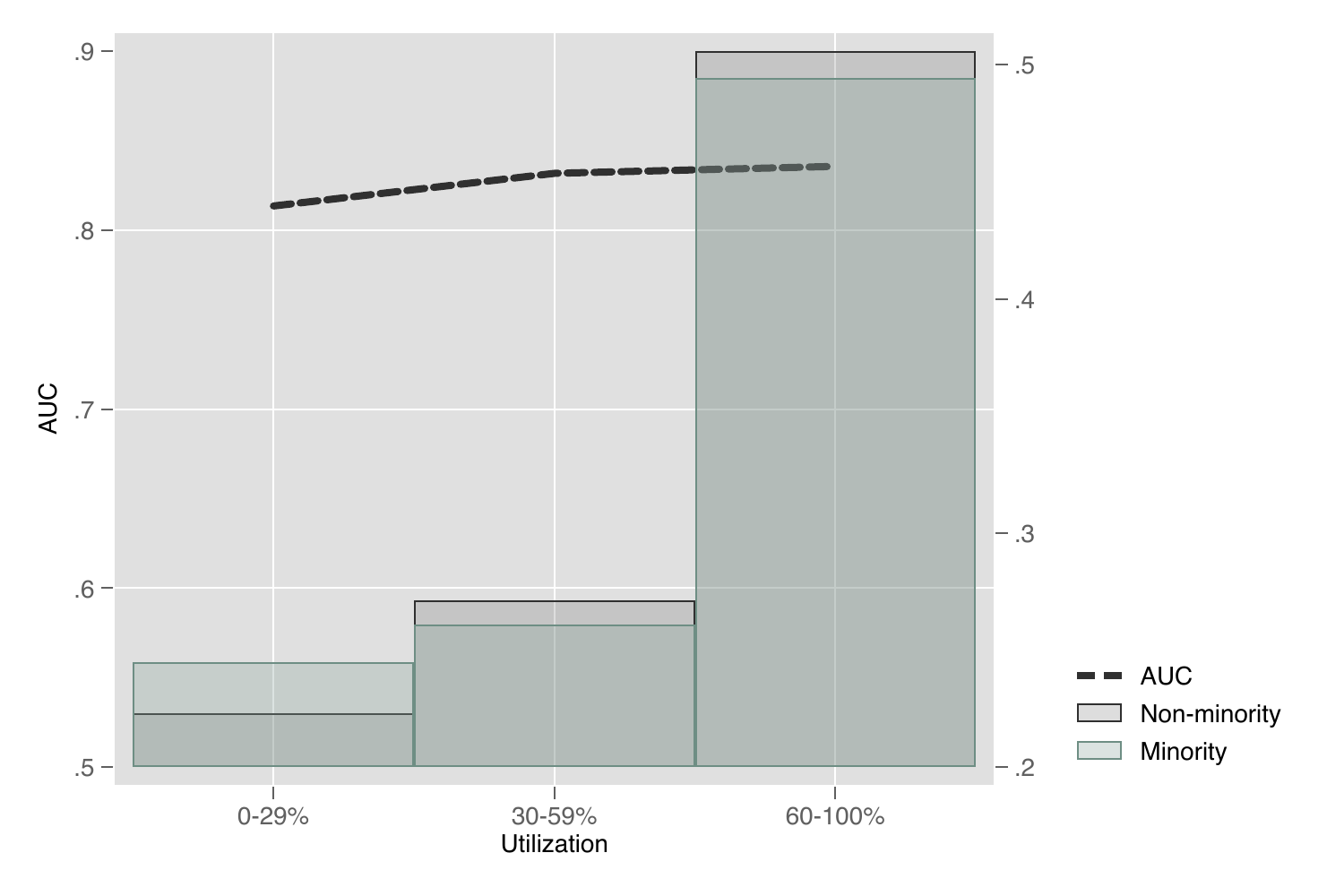}\caption{Utilization by Minority Status}
     \end{subfigure} 
    \begin{subfigure}[t]{0.45\textwidth}
    \includegraphics[width=1\textwidth]{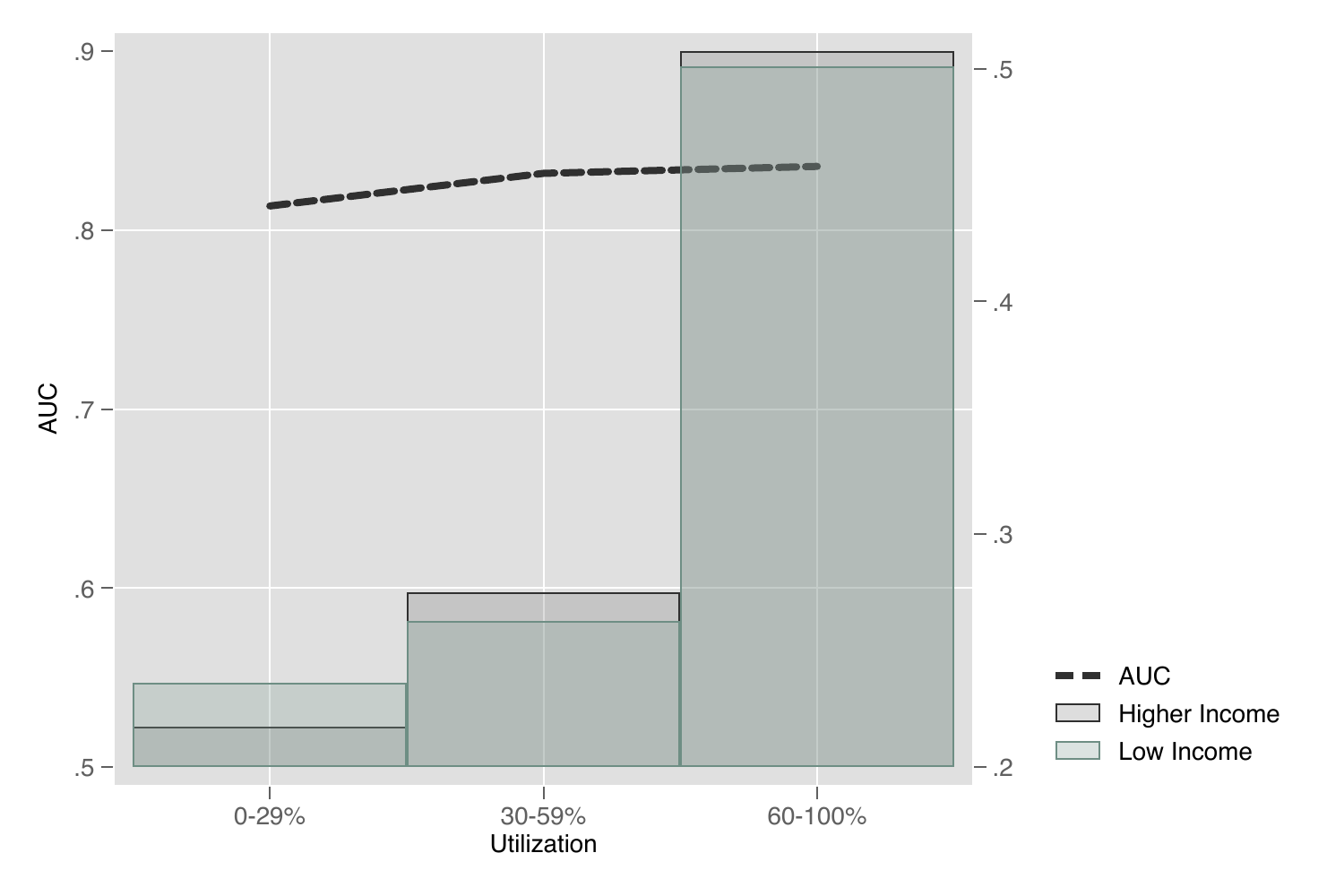}\caption{Utilization by Income}
    \end{subfigure}\\
            \begin{subfigure}[t]{0.45\textwidth}
    \includegraphics[width=1\textwidth]{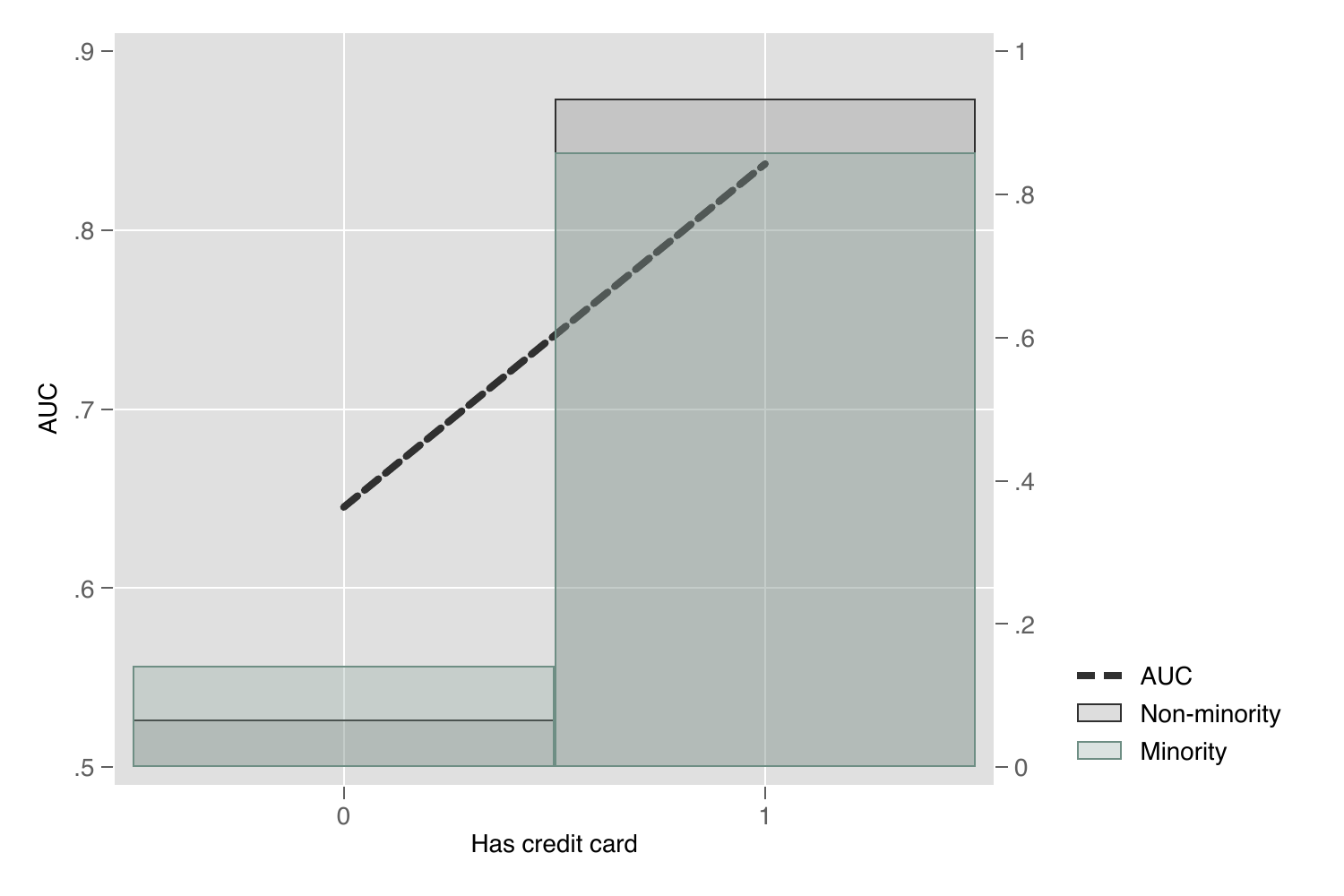}\caption{Bank Card by Minority Status}
     \end{subfigure} 
    \begin{subfigure}[t]{0.45\textwidth}
    \includegraphics[width=1\textwidth]{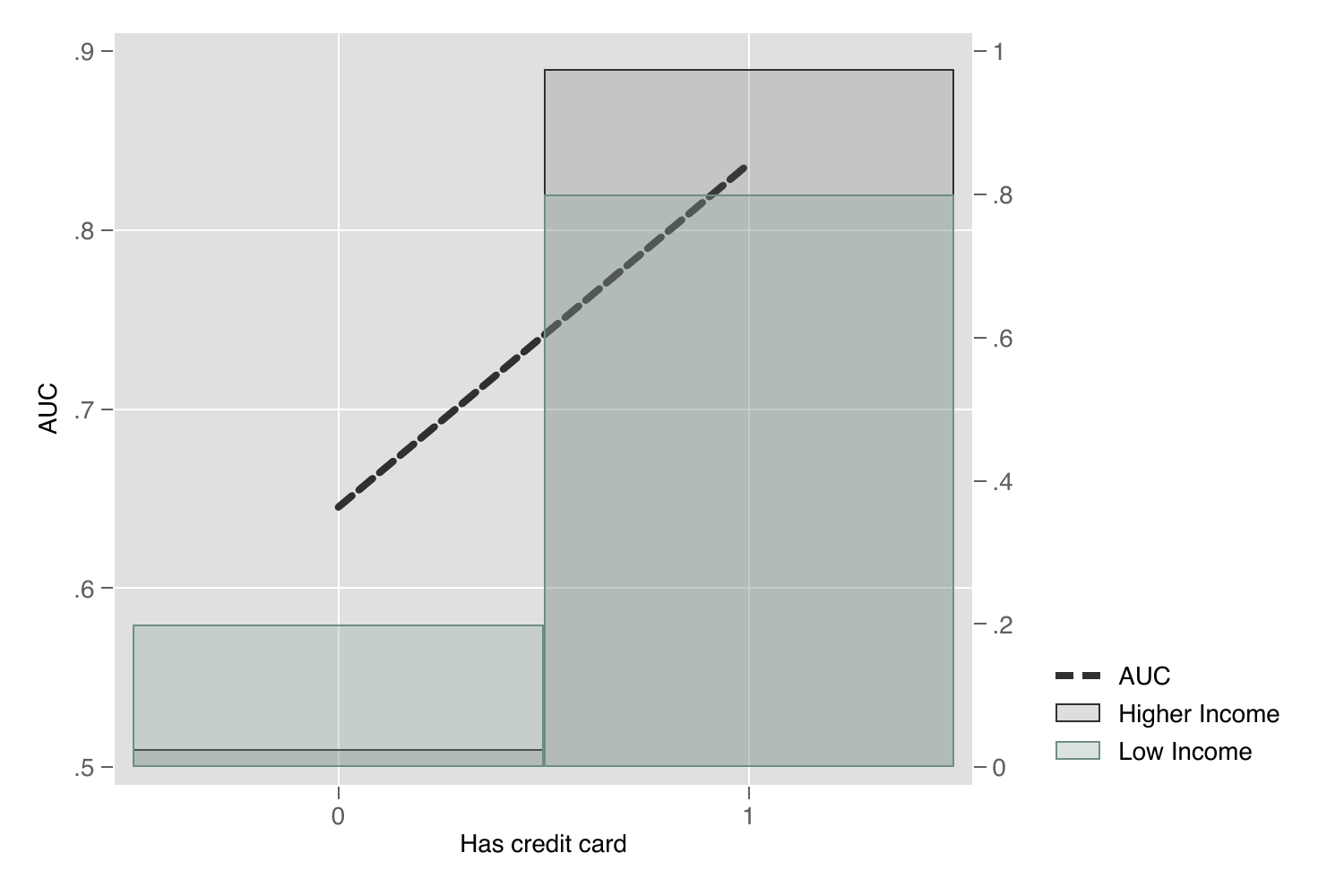}\caption{Bank Card by Income}
    \end{subfigure}\\
    
    \begin{flushleft}
    \begin{spacing}{1.1}
    \small{Note: Each panel of the figure highlights a different characteristic of credit report data (e.g., file thickness) for two complementary consumer groups (e.g. minority and non-minority applicants). Different subsamples defined by a given characteristic of credit report data are arrayed on the x-axis (e.g., different levels of file thickness). The y-axes of the plot then show subsample-specific Area Under the Curve (AUC) for the VantageScore 3.0 credit score (left axis), and group-specific distributions across these subsamples (y axis). All subsamples are drawn from the sample of mortgage applicants. DPD = days past due.}
    \end{spacing}
    \end{flushleft}
\end{figure}

\begin{table}[hbt!]
    \caption{Descriptive Statistics on Default Persistence}
    \label{tab:default_persistence}

    \centering

    \resizebox{0.6\textwidth}{!}{


\begin{tabular}{cccccccc}\toprule
\multicolumn{1}{l}{Panel a: full sample}& & &&&& &  \\ 
& \multicolumn{3}{c}{Non-minority } & & \multicolumn{3}{c}{Minority} \\ \cline{2-4} \cline{5-8}
\diagbox{Current default}{Past default}& 0 & 1 & 2 & & 0 & 1 & 2 \\
0 & 0.97 & 0.07 & 0.03 & & 0.95 & 0.06 & 0.03 \\
1 & 0.02 & 0.87 & 0.02 & & 0.02 & 0.84 & 0.02 \\
2 & 0.01 & 0.06 & 0.95 & & 0.02 & 0.1 & 0.95 \\
& & & & & & & \\ \midrule
\multicolumn{1}{l}{Panel b: full sample }& & &&&& &  \\ 
& \multicolumn{3}{c}{Higher income} & & \multicolumn{3}{c}{Low income} \\ \cline{2-4} \cline{5-8}
\diagbox{Current default}{Past default}& 0 & 1 & 2 & & 0 & 1 & 2 \\
0 & 0.98 & 0.07 & 0.03 & & 0.95 & 0.05 & 0.03 \\
1 & 0.02 & 0.88 & 0.02 & & 0.02 & 0.81 & 0.01 \\
2 & 0.01 & 0.05 & 0.94 & & 0.03 & 0.13 & 0.96 \\
& & & & & & & \\\midrule
\multicolumn{1}{l}{Panel c: has mortgage}& & &&&& &  \\ 
& \multicolumn{3}{c}{Non-minority } & & \multicolumn{3}{c}{Minority} \\\cline{2-4} \cline{5-8}
\diagbox{Current default}{Past default}& 0 & 1 & 2 & & 0 & 1 & 2 \\

0 & 0.97 & 0.07 & 0.03 & & 0.95 & 0.06 & 0.04 \\
1 & 0.02 & 0.86 & 0.02 & & 0.02 & 0.82 & 0.01 \\
2 & 0.01 & 0.07 & 0.95 & & 0.03 & 0.11 & 0.95 \\
 & & & & & & & \\\midrule
 \multicolumn{1}{l}{Panel d: no mortgage}& & &&&& &  \\ 
& \multicolumn{3}{c}{Non-minority} & & \multicolumn{3}{c}{Minority } \\\cline{2-4} \cline{5-8}
\diagbox{Current default}{Past default}& 0 & 1 & 2 & & 0 & 1 & 2 \\

0 & 0.97 & 0.07 & 0.02 & & 0.95 & 0.06 & 0.01 \\
1 & 0.02 & 0.88 & 0.03 & & 0.04 & 0.86 & 0.02 \\
2 & 0.01 & 0.05 & 0.95 & & 0.01 & 0.08 & 0.97 \\
 & & & & & & & \\\bottomrule
\end{tabular}

} 

\bigskip

    \begin{flushleft}
    \small{Note: The table shows descriptive statics on the persistence of default states. We define three default states. No past default on file (0), some delinquency of less than 90 days past due on file (1), and some severe delinquency of at least 90 days past due  on file (2). We show transition matrices into future default states (rows) based on past default status (columns). Panel c and d break down panel a by whether or not the borrower has an active mortgage account on their credit file. The table is based on panel data of a random 1m sample of our main sample for computational feasibility.  } 

    \end{flushleft}

\end{table}















\begin{table}[hbt!]
    \caption{Credit Score Performance with FactorTrust Attributes}
    \label{tab:ml_ft}

    \centering

    \resizebox{.7\textwidth}{!}{


\begin{tabular}{llcc} \toprule
                                          &                     & TU-only Features & TU+FT Features \\
XGBoost Classifier                        & Sample                    & AUC       & AUC     \\ \midrule
Baseline (Pooled)              & Low Income          & 0.814            & 0.814          \\
                                          & Higher Income       & 0.891            & 0.891          \\
                                          & {Difference} & {0.077}   & {0.077} \\ 
     & Minority            & 0.841            & 0.840          \\
                                          & Non-minority        & 0.888            & 0.887          \\
                                          & {Difference} & {0.047}   & {0.047} \\ \midrule
Re-weighted Training Data                   & Low Income          & 0.814            & 0.812          \\
                                          & Higher Income       & 0.890            & 0.888          \\
                                          & {Difference} & {0.076}   & {0.076} \\ 
          & Minority            & 0.840            & 0.840          \\
                                          & Non-minority        & 0.888            & 0.886          \\
                                          & {Difference} & {0.048}   & {0.046} \\ \midrule
Different Models by Group              & Low Income          & 0.814            & 0.814          \\
                                          & Higher Income       & 0.891            & 0.891          \\
                                          & {Difference} & {0.077}   & {0.077} \\ 
      & Minority            & 0.839            & 0.839          \\
                                          & Non-minority        & 0.888            & 0.887          \\
                                          & {Difference} & {0.049}   & {0.048} \\ \bottomrule
\end{tabular}

} 

\bigskip

    \begin{flushleft}
    \small{Note: The table shows AUCs for default prediction models with and without including a set of FactorTrust (FT) variables. All results are for the XGBoost algorithms. Our measure of default is a non-mortgage deliquency of at least 90+ days 24 months after the application date. The table reports Area Under the Curve (AUC) statistics from the test data. See text for more details on the training and test data sets. The re-weighting and ``different model" sub-panels of the table replicate the exercises from Table \ref{table_ml_applicant_mc}, described further in Section \ref{sec:sources}.} 
    \end{flushleft}

\end{table}



\end{document}


\section{Appendix: Results with Alternative Minority Definition}\label{alt_minority}

\setcounter{figure}{0} \renewcommand{\thefigure}{I.\arabic{figure}} 
\setcounter{table}{0} \renewcommand{\thetable}{I.\arabic{table}} 

This appendix provides the main empirical results of the paper if we use a different classification cut-off to define minority status.


\begin{figure}[H]\caption{Receiver Operating Curves by Consumer Type \label{rocgraph_robust}}
   \includegraphics[width=0.6\textwidth]{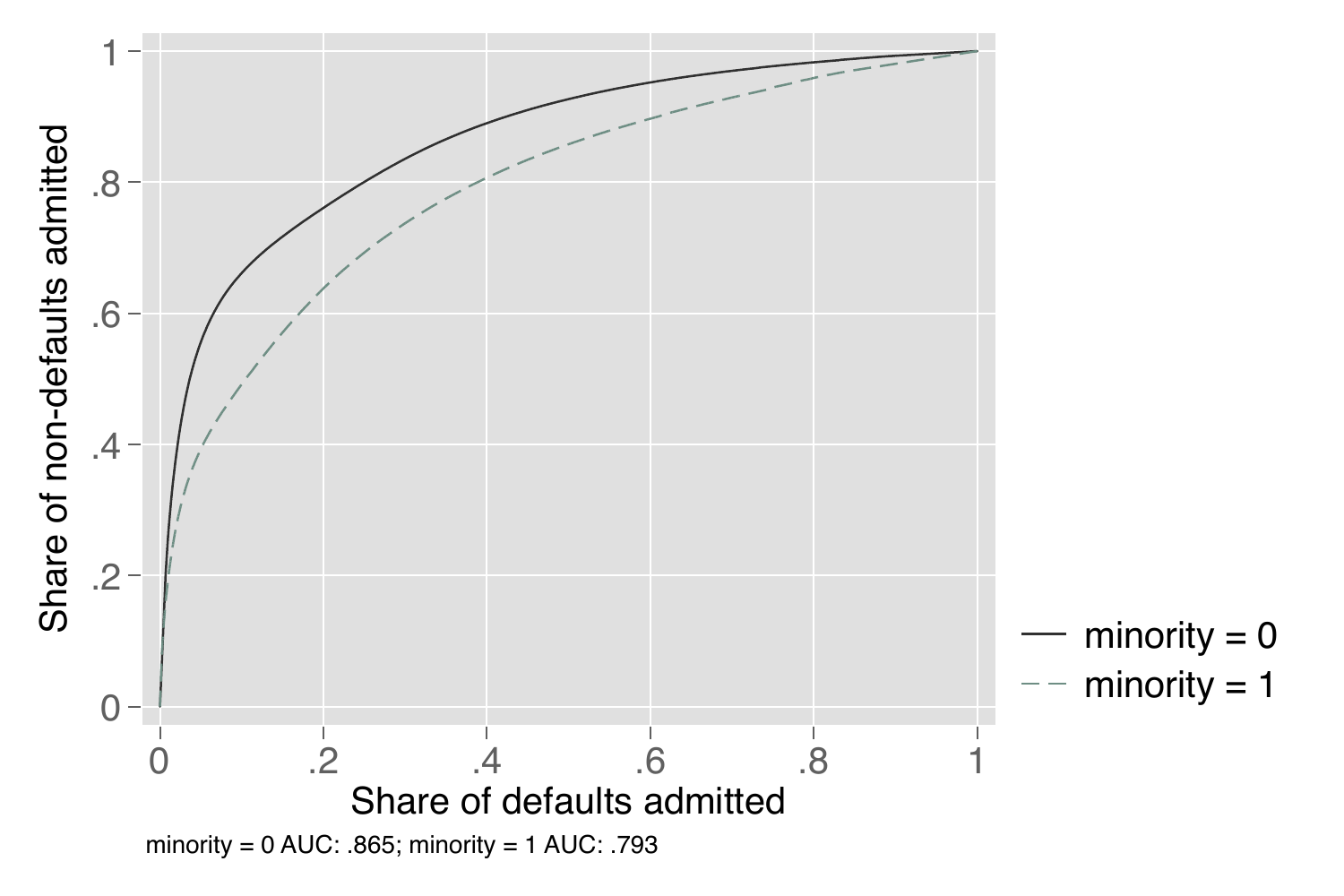}\caption{Minority }

    \begin{flushleft}
    \begin{spacing}{1.1}
    \small{Note: This figure shows Receiver Operating Curves (ROC) for the VantageScore 3.0 credit score in a sample of mortgage applicants between 2009-2016. ROC curves plot the fraction of non-defaults (True Positive Rate) admitted for a given score cutoff against the fraction of defaults admitted (False Positive Rate). Our measure of default is any  delinquency of at least 90+ days 24 months after the application date. We use our alternative definition of minority group. }
    \end{spacing}
    \end{flushleft}
\end{figure}